\documentclass[journal,onecolumn]{IEEEtran}
\usepackage[T1]{fontenc}

\usepackage{url}
\usepackage{ifthen}
\usepackage{braket}
\usepackage{cite}
\usepackage[cmex10]{amsmath} 
\usepackage{longtable}
\usepackage{makecell}
\usepackage{algpseudocode}
\usepackage[linesnumbered,ruled,vlined]{algorithm2e}
\usepackage{color}
\usepackage[normalem]{ulem}
                             
\usepackage{amssymb,amsthm}
\usepackage{mathtools,bbm}  
\usepackage{xifthen,xparse}
\usepackage{hyperref}  % remove when submitting to IEEE Trans.
\usepackage{float}
\usepackage{hhline}
\usepackage{caption}

\usepackage{stmaryrd}

\usepackage{booktabs, tabularx}
\newcolumntype{C}{>{\centering\arraybackslash}X}

\usepackage[table]{xcolor}

\usepackage{tikz,pgfplots}
\pgfplotsset{compat=1.17}

\definecolor{dkgreen}{rgb}{0,0.6,0}
\definecolor{gray}{rgb}{0.5,0.5,0.5}
\definecolor{mauve}{rgb}{0.58,0,0.82}

\usepackage{textcomp}

\allowdisplaybreaks

\newtheorem{theorem}{Theorem}

\newtheorem{remark}[theorem]{Remark}
\newtheorem{lemma}[theorem]{Lemma}
\newtheorem{example}{Example}

\newenvironment{example*}
  {\addtocounter{example}{-1}\example}
  {\endexample}

\newcommand{\ghzmap}[1]{\widehat{#1}}
\newcommand{\dket}[1]{\left\lvert #1 \right\rangle}
\newcommand{\dbra}[1]{\left\langle #1 \right\rvert}
\NewDocumentCommand\dketbra{+m+g}{%
  \IfNoValueTF{#2}
    {\left\lvert #1 \right\rangle \left\langle #1 \right\vert}
  {\left\lvert #1 \right\rangle \left\langle #2 \right\rvert}%
}
\NewDocumentCommand\dbraket{+m+g}{%
  \IfNoValueTF{#2}
    {\left\langle #1 \vert #1 \right\rangle}
  {\left\langle #1 \vert #2 \right\rangle}%
}

\newcommand{\etal}{\emph{et al.~}}

\newcommand{\MCC}{\mathcal{C}}

\newcommand{\syminn}[2]{\langle #1, #2 \rangle_{\text{s}}}
\newcommand{\llbr}{[\![}
\newcommand{\rrbr}{]\!]}

\newif\ifnotes
\notestrue

\def\mathcolor#1#{\@mathcolor{#1}}
\def\@mathcolor#1#2#3{%
  \protect\leavevmode
  \begingroup
    \color#1{#2}#3%
  \endgroup
}
\begin{document}

\title{Distilling GHZ States using Stabilizer Codes}

\author{%
   \IEEEauthorblockN{Narayanan Rengaswamy, Ankur Raina, Nithin Raveendran and Bane Vasi{\'c}}%
   \thanks{%DRAFT DATE: Sep. 13, 2021. 
           %This paper was presented virtually in part at the 2021 Beyond IID in Information Theory workshop.
           %This paper is available on the arXiv preprint server (\url{https://arxiv.org/abs/2109.06248}) and has been submitted to the IEEE Transactions on Information Theory.
           N. Rengaswamy, N. Raveendran and B. Vasi{\'c} are with the 
           Department of Electrical and Computer Engineering,
           University of Arizona, Tucson, Arizona 85721, USA.
           A. Raina is with the Department of Electrical Engineering and Computer Sciences, Indian Institute of Science Education and Research, Bhopal, Madhya Pradesh 462066, India.
           Email: \{narayananr,nithin\}@email.arizona.edu, ankur@iiserb.ac.in, vasic@ece.arizona.edu}%
}

\maketitle

% \begin{abstract}
% When multiple copies of the Bell state are shared between communicating parties such as Alice and Bob, it can be shown that the logical qubits of quantum error correcting stabilizer codes generated locally by them are also entangled in a Bell state.  We show that this idea can be generalized using multipartite GHZ states. 
% \end{abstract}

\begin{abstract}
Entanglement distillation is a well-studied problem in quantum information, where one typically starts with $n$ noisy Bell pairs and distills $k$ Bell pairs of higher fidelity. 
While distilling Bell pairs is the canonical setting, it is important to study the distillation of multipartite entangled states because these can be useful for realizing distributed algorithms on quantum networks. 
In this paper, we study the distillation of GHZ states using quantum error correcting codes (QECCs). 
Using the stabilizer formalism, we begin by explaining the QECC-based Bell pair distillation protocol of Wilde \etal (2007), which relies particularly on the transpose symmetry between Alice's and Bob's qubits in Bell states.
Extending this idea, we show that, given $n$ GHZ states, performing a matrix on Alice's qubits is equivalent to performing a ``stretched'' version of the transpose of the matrix on the qubits of Bob and Charlie. 
We call this mapping to the stretched version of the matrix the \emph{GHZ-map}, and show that it is an algebra homomorphism.
Using this property, we show that Alice projecting her qubits onto an $\llbr n,k \rrbr$ stabilizer code implies the simultaneous projection of Bob's and Charlie's qubits onto an induced $\llbr 2n,k \rrbr$ stabilizer code. 
Guided by this insight, we develop a GHZ distillation protocol based on local operations and classical communication that uses any stabilizer code. 
Inspired by stabilizer measurements on GHZ states, we also develop a new algorithm to generate logical Pauli operators of any stabilizer code and use it in the protocol.
Since quantum codes with finite rate and almost linear minimum distance have recently been discovered, this paper paves the way for high-rate high-output-fidelity GHZ distillation. 
We provide simulation results on the $5$-qubit perfect code to emphasize the importance of the placement of a certain local Clifford operation in the protocol.
\end{abstract}

\section{Introduction}
\label{sec:intro}

% \ankur{}
% \narayanan{}
% \nithin{}
% \bane{}

Entanglement distillation, or purification, is a well-studied problem in quantum information, where 
%the idea is to start with a noisy multipartite entangled state and distill several copies of a higher fidelity entangled state from them.
one typically starts with $n$ copies of a noisy Bell pair, or a general mixed state, and distills $k$ Bell pairs of higher fidelity~\cite{Bennett-prl96}.
Several teams of researchers have worked on this problem, and the contributions range from fundamental limits~\cite{Bennett-prl96,Bennett-pra96,Miyake-prl05,Dur-rpp07,Leditzky-it17,Fang-it19} to simple and practical protocols~\cite{Bennett-pra96,Wilde-isit10,Rozpedek-pra18,Krastanov-quantum19}. 
Some schemes involve two-way communications between the involved parties while others only need one-way communication.
We focus on one-way schemes in this paper.
While distilling (bipartite) Bell pairs is the canonical setting, it is equally important to study the distillation of other multipartite entangled states because these can prove useful for realizing various protocols on quantum networks~\cite{Dur-pra99}.
Of course, if one can distill Bell pairs, then these can be used in sequence to entangle multiple parties, but direct distillation of an entangled resource between all parties is more efficient~\cite{Murao-pra98}.
In this paper, we propose a new protocol to directly distill GHZ (Greenberger-Horne-Zeilinger) states using quantum error correction (QEC).
% Our protocol requires only local operations and classical communications (LOCC), and these are assumed to be noiseless.
Since quantum code families with asymptotically constant rate and almost linear distance have recently been discovered~\cite{Hastings-stoc21,Panteleev-arxiv20,Breuckmann-it21,Breuckmann-arxiv21}, our work paves the way for high-rate high-output-fidelity entanglement distillation.

The connection between one-way entanglement purification protocols (1-EPPs) and quantum error correcting codes (QECCs) was established by Bennett \etal in 1996~\cite{Bennett-pra96}.
They showed that any QECC can be converted into a 1-EPP (and vice-versa) as follows.
Given an $\llbr n,k,d \rrbr$ QECC, they consider Alice generating $k$ perfect Bell pairs, retaining one qubit of each pair, encoding the remaining $k$ qubits into $n$ qubits via the QECC, sending these $n$ qubits to Bob via a teleportation channel induced by noisy shared Bell pairs between Alice and Bob, then Bob performing error correction on the received qubits, and finally Bob decoding his qubits to obtain the $k$ qubits that are entangled with the corresponding qubits of Alice.
This is a natural way to use a QECC for distilling higher fidelity Bell pairs from the noisy Bell pairs shared between Alice and Bob.
In his dissertation, Aschauer~\cite{Aschauer-phd05} showed that 2-EPPs can also be constructed using QECCs, but there the code is only used for error detection.
In~\cite{Dur-rpp07}, D{\"u}r and Briegel provide a comprehensive review of entanglement purification protocols constructed until 2007.
As they show, most of those protocols work on $2$ copies of a mixed state. 
Here, some initial operations are performed to entangle the two states, then certain measurement results on one copy imply that the other (unmeasured) copy now has higher fidelity with respect to the desired entangled resource state.
While these can be generalized to $n$-to-$k$ protocols in several cases~\cite{Dur-rpp07}, such schemes are not always efficient.
Therefore, the framework of QECCs enables systematic $n$-to-$k$ protocols where the rate and average output fidelity are directly a function of the QECC performance when paired with a specific decoder.

In 2007, Wilde \etal \cite{Wilde-isit10} showed that any classical convolutional code can be used to distill Bell pairs via their entanglement assisted 1-EPP scheme.
In the development of this scheme, they mention a potentially different method to use a QECC for performing 1-EPP~\cite[Section II-D]{Wilde-isit10}, compared to the aforementioned connection established initially by Bennett \etal in~\cite{Bennett-pra96}. 
Their method forms the motivation for our work.
Initially, Alice generates $n$ perfect Bell pairs locally, marks one qubit of each pair as `A' and the other as `B', and measures the stabilizers of her chosen $\llbr n,k \rrbr$ code on qubits A.
Due to a property of Bell states, this simultaneously projects qubits B onto an equivalent code (see Section~\ref{sec:bell_state_identity}).
Then, she performs a local Pauli operation on qubits A to fix her obtained syndrome, shares her code stabilizers and syndrome with Bob through a noiseless classical channel, and sends qubits B to Bob over a noisy Pauli channel.
Using the Bell state identity, Bob measures the appropriate code stabilizers on qubits B, and corrects channel errors by combining his syndrome with Alice's syndrome.
Finally, Alice and Bob decode their respective qubits, i.e., invert the encoding unitary for the code (see Section~\ref{sec:stabilizer_codes}), to convert the $k$ Bell pairs on the logical qubits into $k$ physical Bell pairs.
Since the code helps correct some errors, on average the output Bell pairs are of higher fidelity than the initial $n$ noisy Bell pairs (obtained by sending qubits B to Bob).

While the protocol is clear, it is not obvious why the logical qubits of Alice's and Bob's codes must form Bell pairs, assuming that Bob's error correction was successful.
After discussing some background in Section~\ref{sec:background}, we begin our contributions by explaining the reason for the above phenomenon in Section~\ref{sec:bell_distillation}.
We start with a state vector perspective for CSS (Calderbank-Shor-Steane) codes~\cite{Calderbank-physreva96,Steane-physreva96}, and then use the stabilizer formalism~\cite{Gottesman-icgtmp98} to address general stabilizer codes~\cite{Gottesman-phd97,Calderbank-it98}, with the $5$-qubit perfect code~\cite{Laflamme-prl96,Gottesman-phd97} as an example.
In the process, we also highlight the fact that it is convenient to choose the logical Pauli operators for the code in a way that is ``compatible'' with our analysis of the protocol.
These choices determine the respective encoding unitaries to be inverted at the end.
Inspired by stabilizer measurements on Bell/GHZ states, we develop a new algorithm to generate logical Paulis for any stabilizer code (see Algorithm~\ref{algo:logical_paulis_ghz_msmt} and its explanation in Appendix~\ref{sec:logical_paulis_ghz_msmt}), and use that in the protocol.
Our approach using the stabilizer formalism also clarifies that the protocol has a non-zero error rate only because we disallow non-local measurements.
Indeed, if Alice and Bob were allowed to make joint measurements on qubits A and B together, then \emph{any} error is correctable since the initial state is deterministically given by $n$ copies of the known Bell state (see Remark~\ref{rem:Bell_local_msmt}).
Given these insights, we proceed to investigate the distillation of GHZ states.

As in the Wilde \etal protocol, we consider only local operations and classical communications (LOCC), and assume that these are noiseless.
Since we are constructing a new GHZ distillation protocol based on a new insight on GHZ states, we have considered this simple model of noiseless LOCC and noisy qubit communications.
While the problem of noisy local operations is important, and has received attention~\cite{Pan-nature01,Dur-rpp07,Nickerson-ncomms13,Krastanov-quantum19}, we leave these investigations on our protocol for future work.

We begin Section~\ref{sec:ghz_distillation} by exploring an extension of the Bell state matrix identity (``transpose trick'') to GHZ states.
Given $n$ copies of the GHZ state, whose three subsystems are marked `A', `B' and `C', we find that applying a matrix on qubits A is equivalent to applying a ``stretched'' version of the matrix on qubits B and C together (see Lemma~\ref{lem:ghz_state_identity}).
We call this mapping to the stretched version of the matrix the \emph{GHZ-map}, and prove that it is an \emph{algebra homomorphism}~\cite{Dummit-2004}, i.e., linear, multiplicative, and hence projector-preserving.
Recollect that we are interested in measuring stabilizers on qubits A and understanding its effect on the remaining qubits.
Using the properties of the GHZ-map, we show that it suffices to consider only the simple case of a single stabilizer.
With this great simplification, we prove that imposing a given $\llbr n,k,d \rrbr$ stabilizer code on qubits A simultaneously imposes a certain $\llbr 2n, k, d' \rrbr$ stabilizer code jointly on qubits B and C.
By performing diagonal Clifford operations on Charlie's qubits, which commutes with Alice's measurements, one can vary the distance $d'$ of the induced BC code.
Then, we use this core technical result to devise a protocol that distills GHZ states using any stabilizer code.
We perform simulations on the $\llbr 5,1,3 \rrbr$ perfect code and compare the protocol failure rate to the logical error rate of the code on the appropriate Pauli channel. 
In terms of error exponents, we show that it is always better for Bob to perform a local diagonal Clifford operation on Charlie's qubits, rather than Alice doing the same.
We support the empirical observation with an analytical argument on the induced BC code and Charlie's code.
Finally, we finish by showing that the average output $k$-qubit density matrix of the protocol is diagonal in the GHZ-basis, and its fidelity is directly dictated by the protocol's failure rate.

As mentioned above, our noise model here is a Pauli channel for qubit communications, combined with noiseless LOCC.
It is common in the literature to consider distillation of GHZ states when the initial noisy states are diagonal in the GHZ-basis.
However, in~\cite{Wang-oc17}, the authors show that it is possible to distill a perfect GHZ state from two copies of a non-GHZ-diagonal initial state, as long as some conditions are satisfied.
They also show that these states are better than GHZ-diagonal states for distillation of GHZ states.
Using their insight, they are able to apply their scheme for practical noisy environments such as the amplitude-damping channel.
While this work also considers a $2$-to-$1$ protocol, in the future, we plan to integrate our insights with theirs to develop systematic QECC-based $n$-to-$k$ protocols that are operable on more general channels than the simple Pauli channel considered here.

\section{Background and Notation}
\label{sec:background}

% \narayanan{$E(a,b)$ notation, 
% Stabilizer codes and their projectors, 
% Stabilizer formalism for measurements,
% Transpose property of Bell states.}

\subsection{Pauli Matrices}
\label{sec:paulis}

We will use the standard Dirac notation to represent pure quantum states. 
An arbitrary $n$-qubit state will be denoted as a ket $\dket{\psi} = \sum_{v \in \mathbb{F}_2^n} \alpha_v \dket{v}$, where $\{ \alpha_v \in \mathbb{C}\, ; \, v \in \mathbb{F}_2^n \}$ satisfy $\sum_{v \in \mathbb{F}_2^n} |\alpha_v|^2 = 1$ as required by the Born rule~\cite{Wilde-2013}.
Here, $\dket{v} = \dket{v_1} \otimes \dket{v_2} \otimes \cdots \otimes \dket{v_n}$ is a standard basis vector for $v = [v_1,v_2,\ldots,v_n]$, with $v_i \in \mathbb{F}_2 = \{0,1\}$, and $\otimes$ denotes the Kronecker (or tensor) product.
Define $\imath \coloneqq \sqrt{-1}$.
Then, the well-known $n$-qubit Pauli group $\mathcal{P}_n$ consists of tensor products of the single-qubit Pauli matrices
\begin{align}
I = 
\begin{bmatrix}
1 & 0 \\ 0 & 1
\end{bmatrix}, \ 
X = 
\begin{bmatrix}
0 & 1 \\ 1 & 0
\end{bmatrix}, \ 
Z = 
\begin{bmatrix}
1 & 0 \\ 0 & -1
\end{bmatrix}, \ 
Y = \imath XZ =
\begin{bmatrix}
0 & -\imath \\ \imath & 0
\end{bmatrix},
\end{align}
multiplied by scalars $\imath^\kappa, \kappa \in \{0,1,2,3\}$, i.e., 
$\mathcal{P}_n \coloneqq \{ \imath^\kappa E_1 \otimes E_2 \otimes \cdots \otimes E_n, \ E_i \in \{I,X,Z,Y\}, \ \kappa \in \mathbb{Z}_4 = \{0,1,2,3\} \}$.

Given two binary row vectors $a = [a_1,a_2,\ldots,a_n], b = [b_1,b_2,\ldots,b_n] \in \mathbb{F}_2^n$, we will write $E(a,b)$ to denote an arbitrary Hermitian (and unitary) Pauli matrix, where $a$ represents the ``$X$-component'' and $b$ represents the ``$Z$-component''~\cite{Rengaswamy-pra19}:
\begin{align}
\label{eq:Eab}
E(a,b) \coloneqq \left( \imath^{a_1 b_1} X^{a_1} Z^{b_1} \right) \otimes \left( \imath^{a_2 b_2} X^{a_2} Z^{b_2} \right) \otimes \cdots \otimes \left( \imath^{a_n b_n} X^{a_n} Z^{b_n} \right) = \imath^{ab^T} \bigotimes_{i=1}^n (X^{a_i} Z^{b_i}).
\end{align}
For example, $E([1,0,1],[0,1,1]) = X \otimes Z \otimes Y$.
It can be verified that $E(a,b)^2 = I_N = I^{\otimes n}$, where $N \coloneqq 2^n$.
Hence,
\begin{align}
\label{eq:Pauli_group}
\mathcal{P}_n = \{ \imath^\kappa E(a,b) \colon \ a,b \in \mathbb{F}_2^n, \ \kappa \in \mathbb{Z}_4 = \{0,1,2,3\} \}.
\end{align}
Using the properties of the Kronecker product, primarily the identities $(A \otimes B) (C \otimes D) = (AC) \otimes (BD)$ and $(A \otimes B)^T = A^T \otimes B^T$, we can show the following.
We represent standard addition by ``$+$'' and modulo $2$ addition by ``$\oplus$''.

\begin{lemma}
\label{lem:Eab}
For any $a,b \in \mathbb{F}_2^n$, the Pauli matrix $E(a,b)$ satisfies the following properties:
\begin{enumerate}
    
\item[(a)] $E(a,b)^T = (-1)^{ab^T} E(a,b)$;
    
\item[(b)] $E(a,b) \cdot E(c,d) = \imath^{bc^T - ad^T} E(a+c, b+d)$, where the exponent and the sums $(a+c), (b+d)$ are performed modulo $4$ and the definition in~\eqref{eq:Eab} is directly extended
% \footnote{We can write $a+c = a \oplus c + 2(a \ast c)$, where $a \ast b = [a_1 b_1, \ldots, a_n b_n]$, and apply the definition in~\eqref{eq:Eab} to obtain an expression in terms of $E(a \oplus c, b \oplus d)$.}
 to $a,b \in \mathbb{Z}_4^n$;
    
\item[(c)] $E(a,b) \cdot E(c,d) = (-1)^{\syminn{[a,b]}{[c,d]}} E(c,d) \cdot E(a,b)$, where 
\begin{align}
\label{eq:syminn}
\syminn{[a,b]}{[c,d]} \coloneqq ad^T + bc^T \ (\bmod\ 2)
\end{align}
is the symplectic inner product between $[a,b]$ and $[c,d]$ in $\mathbb{F}_2^{2n}$, as indicated by the subscript `s'.
    
\end{enumerate}
\end{lemma}
% \bane{Move \cite{Rengaswamy-pra19} from from lemma to a remark after lemma?}

Hence, the map $\gamma \colon (\mathcal{P}_n, \cdot) \rightarrow (\mathbb{F}_2^{2n}, \oplus)$ defined by $E(a,b) \mapsto [a,b]$ is a homomorphism with kernel $\{ \imath^\kappa I_N,\ \kappa \in \mathbb{Z}_4 \}$.
For details about extending the definition of $E(a,b)$ to $\mathbb{Z}_4$-valued arguments, see~\cite{Rengaswamy-pra19}.

\subsection{Stabilizer Codes and Encoding Unitaries}
\label{sec:stabilizer_codes}

A stabilizer group $S$ is a commutative subgroup of $\mathcal{P}_n$ that does not contain $-I_N$.
If the group has $r \leq n$ independent generators $\varepsilon_i E(a_i,b_i)$, where $\varepsilon_i \in \{ \pm 1 \}$, then $S = \langle \varepsilon_i E(a_i,b_i); \ i = 1,\ldots,r \rangle$ has size $|S| = 2^r$.
Since the generators are Hermitian and unitary, they have eigenvalues $\pm 1$. 
Recollect that commuting matrices can be diagonalized simultaneously.
The stabilizer code defined by $S$ is the common $+1$ eigenspace of all generators, i.e., it is the $2^k$-dimensional subspace, $k = n-r$, fixed by all elements of $S$:
\begin{align}
\mathcal{Q}(S) \coloneqq \{ \dket{\psi} \in \mathbb{C}^N \colon g \dket{\psi} = \dket{\psi}\ \forall\ g \in S \}.
\end{align}

Using the homomorphism $\gamma$, we can write a $r \times (2n+1)$ generator matrix $G_S$ for the stabilizer group: the $i^{\text{th}}$ row of $G_S$ is $[a_i,b_i,\ \varepsilon_i] \in \mathbb{F}_2^{2n} \times \{ \pm 1 \}$.
Since $S$ must be a commutative group, the symplectic inner product between any pair of rows must be zero.
Hence, the subspace of binary mappings of all elements of $S$, denoted $\gamma(S)$, is given by the rowspace of $G_S$.

A CSS (Calderbank-Shor-Steane) code is a special type of stabilizer code for which there exists a set of generators where either $b_i = 0$ or $a_i = 0$ in each generator, i.e., the generators are purely $X$-type and purely $Z$-type operators.
Clearly, for such a code, $G_S$ has a block diagonal form where we can express the $X$-type (resp. $Z$-type) operators as the rowspace of a matrix $[H_X, 0]$ (resp. $[0, H_Z]$), and $0$ represents the all-zeros matrix (of appropriate size).
In this case, the commutativity condition for stabilizers is equivalent to the condition $H_X H_Z^T = 0$.
Therefore, $H_X$ and $H_Z$ can be thought of as generating two classical linear codes $\MCC_X$ and $\MCC_Z$.

The projector onto the $+1$ eigenspace of a Pauli matrix $E(a,b)$ is $\frac{I_N + E(a,b)}{2}$.
Therefore, since $\mathcal{Q}(S)$ is the simultaneous $+1$ eigenspace of $r$ commuting matrices $\varepsilon_i E(a_i,b_i)$, the projector onto the code subspace $\mathcal{Q}(S)$ is
\begin{align}
\label{eq:stabilizer_projector}
\Pi_S = \prod_{i=1}^r \frac{(I_N + \varepsilon_i E(a_i,b_i))}{2} = \frac{1}{2^r} \sum_{m = [m_1,\ldots,m_r] \in \mathbb{F}_2^r} \prod_{i=1}^r \left( \varepsilon_i E(a_i,b_i) \right)^{m_i} = \frac{1}{2^r} \sum_{\varepsilon E(a,b) \in S} \varepsilon E(a,b).
\end{align}

While the stabilizer group $S$ defines the code space, an encoding unitary $\mathcal{U}_{\text{Enc}}(S)$ fully specifies the mapping from logical $k$-qubit states to physical $n$-qubit code states in $\mathcal{Q}(S)$.
The $n$ input qubits to $\mathcal{U}_{\text{Enc}}(S)$ can be split into $k$ logical qubits, whose joint state is arbitrary, and $r = n-k$ ancillary qubits, each of which is initialized in some specific state such as $\dket{0}$.
If ancillas are initialized in the $\dket{0}$ state, then the stabilizer group for these $n$ input qubits is generated by $\{ Z_i \, ; \, i = k+1,k+2,\ldots,n \}$, since $Z \dket{0} = \dket{0}$.
If we conjugate each of these $r$ generators by the encoding unitary $\mathcal{U}_{\text{Enc}}(S)$, then we will obtain $r$ generators $\mathcal{U}_{\text{Enc}}(S) \, Z_i \, \mathcal{U}_{\text{Enc}}(S)^\dagger$ of $S$.
Similarly, if we conjugate the $X_i$ and $Z_i$ operations on the $k$ logical qubits --- which can be used to express arbitrary operations on them since Pauli operators form a basis --- by $\mathcal{U}_{\text{Enc}}(S)$, then we will obtain the generators of logical $X$ and $Z$ operators compatible with the chosen $\mathcal{U}_{\text{Enc}}(S)$.
Therefore, an alternative method to specify $\mathcal{U}_{\text{Enc}}(S)$ is to specify the generators of $S$ as well as the generators of logical $X$ and $Z$ operators.
Since we are requiring $\mathcal{U}_{\text{Enc}}(S)$ to map Paulis to Paulis, it is always Clifford~\cite{Gottesman-icgtmp98}.
Note that $\mathcal{U}_{\text{Enc}}(S)$ is still not unique since we are not specifying how $X_i$ on the ancillas must be mapped, but we do not care about these additional mappings.

There are at least two algorithms provided in the literature for generating the logical Pauli operators of stabilizer codes.
One is by Gottesman~\cite{Gottesman-phd97,Nielsen-2010}, where the idea is to construct the normalizer of the stabilizer group inside the Pauli group, and then perform suitable row operations on the generators of the normalizer.
The other is by Wilde~\cite{Wilde-physreva09}, where he performs a symplectic Gram-Schmidt orthogonalization procedure to arrive at the generators of logical $X$ and logical $Z$ operators.
In this work, as part of our GHZ distillation protocol, we provide a new algorithm to generate logical $X$ and $Z$ operators for any stabilizer code (see Algorithm~\ref{algo:logical_paulis_ghz_msmt}).
The output of the algorithm is compatible with the way logical Paulis must be defined for our analysis of the protocol.
Additionally, the logical $Z$ operators from our algorithm are always guaranteed to be purely $Z$-type operators for any stabilizer code.
If the code is CSS, then the logical $X$ operators are always purely $X$-type.

\subsection{Stabilizer Formalism}
\label{sec:stabilizer_formalism}

When a stabilizer group on $n$ qubits has $n$ independent generators, $\mathcal{Q}(S)$ is a $1$-dimensional subspace that corresponds to a unique quantum state $\dket{\psi(S)}$ (up to an irrelevant global phase), commonly referred to as a stabilizer state~\cite{Gottesman-phd97}.
The actions of unitary operations and measurements on $\dket{\psi(S)}$ can be tracked by updating these $n$ generators accordingly~\cite{Gottesman-icgtmp98,Aaronson-pra04}.
For any element $g$ of $S$, and an arbitrary unitary operation $U$ on $\dket{\psi(S)}$, we observe that
\begin{align}
U \dket{\psi(S)} = U \cdot g \cdot\dket{\psi(S)} = (UgU^\dagger) \cdot U \dket{\psi(S)},
\end{align}
so the stabilizer element $g$ has evolved into the element $g' = UgU^\dagger$ after the action of $U$.
Of course, only if $U$ is a Clifford operation we have that $g'$ is also a Pauli matrix.
Thus, in this case the evolution of the state can be tracked efficiently by simply transforming $G_S$ (and the associated signs) through binary operations (see ``CHP'' algorithm~\cite{Aaronson-pra04}).

The stabilizer formalism also provides a method to systematically update the stabilizers under Pauli measurements of the state $\dket{\psi(S)}$.
Assume that we have $n$ generators for the stabilizer group, namely $\varepsilon_i E(a_i,b_i), i = 1,\ldots,n$, and that we are measuring the Pauli operator $\mu E(u,v)$ to obtain the measurement $(-1)^m, m \in \{0,1\}$. Then, we have the following cases.

% \bane{Shell we define syminn ? }
% \narayanan{I had defined it earlier in Lemma 1 but I have made it more prominent as an equation now.}
% % $\syminn{.}{.}$
% \nithin{Should we say subscript $_s$ means symplectic inner product? I wondered what $_s$ was meaning for a second.}
% \narayanan{Done}

\begin{enumerate}
\item If $\syminn{[u,v]}{[a_i,b_i]} = 0$ for all $i$, then either $E(u,v)$ or $-E(u,v)$ already belongs to $S$, so there is nothing to update.

\item If $\syminn{[u,v]}{[a_j,b_j]} = 1$ for exactly one $j \in \{1,\ldots,n\}$, then we replace $\varepsilon_j E(a_j,b_j)$ by $(-1)^m \mu E(u,v)$.

\item If $\syminn{[u,v]}{[a_i,b_i]} = 1$ for $i \in \mathcal{I} \subseteq \{1,\ldots,n\}$, then we replace $\varepsilon_j E(a_j,b_j)$ by $(-1)^m \mu E(u,v)$ for any one $j \in \mathcal{I}$, and update $\varepsilon_i E(a_i,b_i) \mapsto \varepsilon_i E(a_i,b_i) \cdot \varepsilon_j E(a_j,b_j)$ for all $i \in \mathcal{I} \setminus \{j\}$ (using Lemma~\ref{lem:Eab}(b)).

\end{enumerate}

\begin{example}
\label{eg:measurement}
\normalfont
Consider the standard Bell state $\dket{\Phi^+} = \frac{\dket{00} + \dket{11}}{\sqrt{2}}$, whose stabilizer group is $S = \langle X \otimes X, Z \otimes Z \rangle= \langle E(11,00), E(00,11) \rangle$.
If we measure $Z \otimes I = E(00,10)$, and obtain the result $-1$, then the new stabilizers are $S = \langle -E(00,10), E(00,11) \rangle \equiv \langle -E(00,10), -E(00,01) \rangle$.
This group perfectly stabilizes the post-measurement state $\dket{11}$.

If we instead measure $Y \otimes I = E(10,10)$, and obtain the result $+1$, then the new stabilizers are $S = \langle E(10,10), -E(11,11) \rangle \equiv \langle E(10,10), -E(01,01) \rangle$.
This group perfectly stabilizes the post-measurement state $\frac{(\dket{0} + \imath \dket{1})}{\sqrt{2}} \otimes \frac{(\dket{0} - \imath \dket{1})}{\sqrt{2}}$. \hfill \IEEEQEDhere
\end{example}

\subsection{Bell State Matrix Identity}
\label{sec:bell_state_identity}

Let $n$ standard Bell pairs be shared between Alice and Bob. 
% We can write the joint state as 
% \begin{align}
%     \ket{\psi}_{\text{AB}} &= \ket{\Phi^{+}}^{\otimes n}_{\text{AB}},
% \end{align}
% where $\ket{\Phi^{+}}$ is the Bell state of two qubits:
% \begin{align}
%     \ket{\Phi^{+}} &= \frac{\ket{00}+\ket{11}}{\sqrt{2}}.
% \end{align}
We rearrange the $2n$ qubits to keep Alice's qubits together and Bob's qubits together. 
We can write the joint state as 
\begin{align}    
\label{eq:bell_state_rearranged}
\dket{\Phi^{+}_n}_{\text{AB}} & = \left(\frac{\dket{00}_{\text{AB}} + \dket{11}_{\text{AB}}}{\sqrt{2}}\right)^{\otimes n} = \frac{1}{\sqrt{2^n}} \sum_{x \in \mathbb{F}_2^n} \dket{x}_{\text{A}} \dket{x}_{\text{B}}. 
\end{align}
Let $M = \sum_{x,y \in \mathbb{F}_2^n} M_{xy} \dketbra{x}{y} \in \mathbb{C}^{2^n \times 2^n}$ be any matrix acting on Alice's qubits.
Then, it has been observed that \cite{Bennett-pra96,Wilde-isit10}
\begin{align}
(M \otimes I)\ket{\Phi^{+}_n} & = \frac{1}{\sqrt{2^n}} \sum_{x,y \in \mathbb{F}_2^n} M_{xy} \dket{x}_{\text{A}} \dket{y}_{\text{B}} \\
  & = \frac{1}{\sqrt{2^n}} \sum_{x,y \in \mathbb{F}_2^n} \dket{x}_{\text{A}} (M^T)_{yx} \dket{y}_{\text{B}} \\
  & = (I \otimes M^T) \ket{\Phi^{+}_n}. 
\end{align}

\begin{example}
\label{eg:Y_measurement}
\normalfont
As in Example~\ref{eg:measurement}, consider the standard Bell pair and measure $Y \otimes I$. 
If the measurement result is $+1$, then the projector $P_Y = \frac{I_2 + Y}{2}$ gets applied to the first qubit.
Then, according to the above identity, this is equivalent to applying $P_Y^T = \frac{I_2 - Y}{2}$ on the second qubit.
This exactly agrees with the post-measurement state $\frac{(\dket{0} + \imath \dket{1})}{\sqrt{2}} \otimes \frac{(\dket{0} - \imath \dket{1})}{\sqrt{2}}$. \hfill \IEEEQEDhere
\end{example}

If Alice measures the generators of a stabilizer group $S = \langle \varepsilon_i E(a_i,b_i); \ i = 1,\ldots,r \rangle$, and obtains results $(-1)^{m_i}, m_i \in \{0,1\},$ then $M = \Pi_{S'}$ is the projector onto the subspace $\mathcal{Q}(S')$ of the stabilizer code defined by $S' = \langle (-1)^{m_i} \varepsilon_i E(a_i,b_i); \ i = 1,\ldots,r \rangle$. 
According to the above identity, this is equivalent to projecting Bob's qubits onto the stabilizer code defined by 
\begin{align}
S'' = \langle (-1)^{m_i} \varepsilon_i E(a_i,b_i)^T; \ i = 1,\ldots,r \rangle = \langle (-1)^{m_i + a_i b_i^T} \varepsilon_i E(a_i,b_i); \ i = 1,\ldots,r \rangle,
\end{align}
where we have applied Lemma~\ref{lem:Eab}(a).
Note that, in such cases where $M$ is a projector, we can write
\begin{align}
\label{eq:projector_bell_pair}
(M \otimes I)\ket{\Phi^{+}_n} = (M^2 \otimes I)\ket{\Phi^{+}_n} = (M \otimes M^T)\ket{\Phi^{+}_n},
\end{align}
so that the action of Alice can be interpreted as \emph{both} Alice and Bob projecting their own qubits simultaneously.

\section{Revisiting the Bell Pair Distillation Protocol}
\label{sec:bell_distillation}

\begin{figure}
 \centering
\includegraphics[scale=0.8,keepaspectratio]{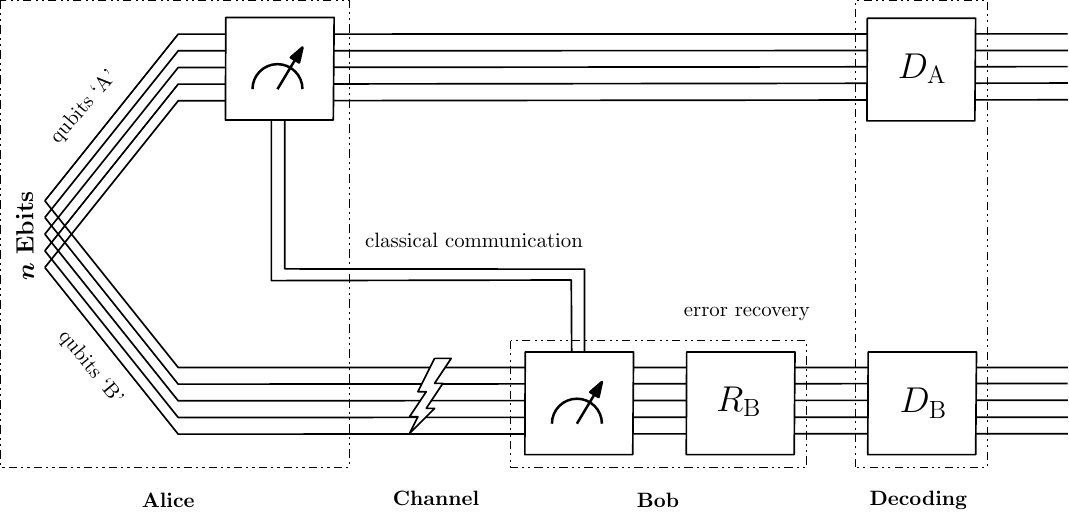}
  \caption{The QEC-based entanglement distillation protocol of Wilde \etal \cite{Wilde-isit10}. Figure adapted from~\cite{Wilde-isit10}. 
%   \bane{Shell we clarify what ``ebit'' stands for? Say it is a nonlocal bipartite Bell state $\dket{\Phi^+}$ as Wilde said in his paper? In the second paragraph of this section where we introduce $\dket{\Phi^+}$? Shell we explain in the text what the output's of Alice and Bob's decoders are.} \narayanan{I think I have addressed this now. Please check.} 
   }
   \label{fig:bell-pair-entanglement}
\end{figure}

\begin{algorithm}[b] % The number tells where the line numbering should start
\DontPrintSemicolon
%\SetAlgoNoLine
\SetAlgoLined
% \KwResult{Logical qubits of the stabilizer codes on Alice and Bob form a Bell state \;}
\KwResult{Alice and Bob share $k$ perfect Bell pairs or at least one of the $k$ pairs has an unknown Pauli error}
\SetKwInOut{Input}{Input}
\SetKwInOut{Output}{Output}
\Input{$n$ Bell pairs $\ket{\Phi^{+}}^{\otimes n}$ at Alice, % one qubit of each pair will be sent to Bob over a noisy channel; 
$\llbr n,k,d \rrbr$ stabilizer code $\mathcal{Q}(S)$ defined by a stabilizer group $S$}
%\Output{Logical qubits of the QEC $\mathcal{C}$ form a Bell pair}
\Output{$k$ Bell pairs of higher quality shared between Alice and Bob if channel introduces a correctable error}
 Initialization: Rearrange the $2n$ qubits in $\ket{\Phi^{+}}^{\otimes n}$ to obtain $\ket{\Phi^{+}_n}$~\eqref{eq:bell_state_rearranged} for processing by Alice and Bob respectively\;
  \;
 Alice \;
 (a) measures all the stabilizer generators $\{ \varepsilon_i E(a_i,b_i) \, ; \, i = 1,2,\ldots,r=n-k \}$ on her $n$ qubits, obtains syndrome, \; 
 %$\{ (-1)^{m_i}, m_i \in \mathbb{F}_2 \}$, \;
%  (a) considers the code $\mathcal{C}$ with stabilizer generators $S_1. S_2, \cdots, S_{n-k}$ .\;
%  (b) performs a suitable Pauli operation to her qubits to bring them to the code space of $\mathcal{Q}(S)$, \;
%  (b) measures all the stabilizer generators $S_1, S_2, \cdots, S_{n-k}$ on $n$ qubits.\;
%  (c) qubits are in the code $\mathcal{C}$.\;
 %(d) performs Clifford operations on Charlie's qubits \;
 (b) sends the remaining $n$ qubits to Bob over a noisy quantum channel, \;
 %\While{  }  
 %{  Alice considers an $\llbr n,k,d \rrbr$ QEC $\mathcal{C}$ with stabilizer generators $S_1. S_2, \cdots, S_{n-k}$  \;
 %\For{$i=1$ \KwTo $n-k$}{
%    Alice measures $S_i$ \;
  %  \uIf{All measurement outcomes are +1 }{
  %   Alice's qubits are in the code $\mathcal{C}$ \;
  %}
  %\uElseIf{Some measurement outcomes are -1 %}{
   % Alice performs local Pauli corrections \;
    %Alice's qubits are in the code $\mathcal{C}$ \;
    %      }
%}
%}
%Alice sends the remaining $n$ qubits to Bob over a noisy quantum channel\;
 (c) sends the stabilizers and syndrome (which together define $\mathcal{Q}(S)$) %and Pauli operation information
 to Bob over a noiseless classical channel. \;
% (e) sends the syndrome information of the stabilizer measurements to Bob over a perfect classical channel \;
 \;
{  Bob \;
 (a) measures all the stabilizer generators $\{ \varepsilon_i E(a_i,b_i) \, ; \, i = 1,2,\ldots,r=n-k \}$ on his $n$ qubits, \;
% (a) considers the code $\mathcal{C}$ with %same $\llbr n,k,d \rrbr$ QEC $\mathcal{C}$  
% stabilizer generators $S_1. S_2, \cdots, S_{n-k}$  \;
% (b) measures all the stabilizer generators $S_1, S_2, \cdots, S_{n-k}$ on the $n$ qubits \;
(b) combines the syndrome information from Alice as well as his measurements and interprets using Section~\ref{sec:bell_state_identity}, \;
(c) performs necessary Pauli corrections on his qubits to bring them to the code space of $\mathcal{Q}(S)$. \;
 %  \uIf{All measurement outcomes are +1 }{
 %    Bob's qubits are in the code $\mathcal{C}$ \;
 % }
 % \uElseIf{Some measurement outcomes are -1 }{
 %   Bob performs local Pauli corrections \;
 %   Bob's qubits are in the code $\mathcal{C}$ \;
 %         }
% \For{$i=1$ \KwTo $n-k$}{
%    Bob measures $S_i$ \;
%    \uIf{Measurement outcome=+1 }{
%     $i=i+1$ \;
%  }
%  \uElseIf{Measurement outcome =-1 }{
 %   Perform local Pauli correction \;
 %    $i=i+1$ \;
 %         }
 %}
  \; 
 // If the channel error was correctable, pairs of logical qubits of Alice's and Bob's codes form $k$ Bell states \;
 // If channel error was NOT correctable, some pair of logical qubits form a Bell state with an unknown Pauli error \;
 Alice and Bob respectively apply the inverse of the encoding unitary for their code on their $n$ qubits \;
 // The encoding unitary is determined by the logical Pauli operators for $\mathcal{Q}(S)$ obtained from Algorithm~\ref{algo:logical_paulis_ghz_msmt}
%  Alice and Bob share $k$ Bell pairs and $n-k$ ancillary qubits \;
}
%   \caption{Algorithm to create $k$ Bell pairs of higher quality from $n$ Bell pairs, using an $\llbr n,k,d \rrbr$ stabilizer code}
  \caption{Algorithm to convert $n$ Bell pairs into $k$ Bell pairs of higher quality, using an $\llbr n,k,d \rrbr$ stabilizer code}
 \label{algo:algo_bell}
\end{algorithm}

In Ref.~\cite{Wilde-isit10}, Wilde \etal described a protocol to distill Bell pairs using an arbitrary quantum stabilizer code.
We reiterate this protocol here and provide more clarity on the reasons behind its working.
Then, in the next section, we will generalize this protocol to distill GHZ states, i.e., the $3$-qubit entangled state $\dket{\text{GHZ}} = \frac{1}{\sqrt{2}} \left( \dket{000} + \dket{111} \right)$.

Initially, Alice generates $n$ copies of the Bell state $\dket{\Phi^+}$ ($n$ ``ebits''), rearranges the qubits as described above, and sends Bob's set of $n$ qubits to him over a noisy channel. 
It is not necessary that Alice must prepare Bell pairs locally and then transmit half the qubits to Bob. 
Indeed, the protocol is applicable as long as Alice and Bob share some initial (noisy) Bell pairs.
Then, Alice measures the stabilizers of a quantum stabilizer code defined by $S = \langle \varepsilon_i E(a_i,b_i); \ i = 1,\ldots,r \rangle$ on her qubits, with $\varepsilon_i = \pm 1$.
Let her measurement results be $(-1)^{m_i}, m_i \in \{0,1\}$.
This projects her qubits onto the codespace fixed by the stabilizers $S' = \langle (-1)^{m_i} \varepsilon_i E(a_i,b_i); \ i = 1,\ldots,r \rangle$. 
Alice applies some suitable Pauli ``correction'' to bring her qubits back to the code subspace $\mathcal{Q}(S)$ (rather than $\mathcal{Q}(S')$), if that is the code she desires to use.
She classically communicates the chosen stabilizers, $S$, the measurements $\{ m_i \}_{i=1}^r$, and the Pauli correction to Bob.

Although we use the term ``correction'', there is really no error on Alice's qubits.
Instead, the terminology is used to indicate that Alice brings the qubits to her desired code space.
Furthermore, even if there is some error on Alice's qubits, one can map it to an equivalent error on Bob's qubits using the Bell state matrix identity (Section~\ref{sec:bell_state_identity}).

Note that the authors of Ref.~\cite{Wilde-isit10} do not explicitly mention that the Pauli correction needs to be communicated, but it could be necessary in situations where Alice's and Bob's decoders are not identical or have some randomness embedded in them.
In a similar vein, Alice might also have to classically communicate the generators of the logical Pauli operators of the code, so that Bob exactly knows which encoding circuit to invert at the end of the protocol.
%These logical Pauli generators need to be ``compatible'' with the protocol, in the sense that the logical Bell stabilizers $\overline{X}_{\text{A}} \overline{X}_{\text{B}}$ and $\overline{Z}_{\text{A}} \overline{Z}_{\text{B}}$ on the $k$ pairs of logical qubits of A and B must be converted into physical Bell stabilizers corresponding to physical Bell pairs.
Though any appropriate definition of logical Pauli generators likely works with the protocol, we employ Algorithm~\ref{algo:logical_paulis_ghz_msmt} to obtain generators that are ``compatible'' with our way of analyzing the protocol (using the stabilizer formalism).
This phenomenon will become more clear after the $\llbr 5,1,3 \rrbr$ code example in this section.
While the algorithm simulates measurements on GHZ states to define logical Paulis, an equivalent algorithm can be constructed that only simulates Bell measurements.

\begin{remark}
\normalfont
In this protocol, whenever the syndrome of Alice is non-trivial, i.e., at least one $m_i$ equals $1$, she can either perform a Pauli correction or just define her code to be $\mathcal{Q}(S')$ and not perform any correction.
If the protocol is defined so that she always does the latter, as depicted in Fig.~\ref{fig:bell-pair-entanglement} where there is no `Recovery' block on Alice's qubits, then Bob can adjust his processing accordingly based on the syndrome information from Alice.
\end{remark}

% \bane{Let's suppose that during this correction operation, Alice's decoder tries hard but makes a misscorrection and introduces a logical error, i.e.,  $E(c,d)$ is a logical operator. Alice's qubits are still in the right code space, but how would this affect Bob's qubits? In the Wilde protocol, does Alice communicate the information about what correction was made to Bob? Does it matter to Bob what correction Alice has made? If Alice send's Bob's qubits only after she makes the correction, it might not. }
% \narayanan{Note that there has not been any ``error'' on Alice's qubits. It is just that her qubits got projected to an equivalent code that is different from her desired one by a Pauli operation. So, she can apply any appropriate Pauli operation to fix the signs, and there is no such thing as a logical error here. However, her Pauli operation might change the signs of other Bell stabilizers we started with. So, if either she communicates the operation to Bob, or if Bob uses the exact same decoder (and it does not have any randomness), then I think Bob can take care of this just using Alice's syndrome information. To keep things clear, I specified above that Alice also communicates the Pauli correction.}

% Due to the aforemetioned matrix identity, once Alice measures her qubits, Bob's qubits would have automatically been projected according to $S'' = \langle (-1)^{m_i} \varepsilon_i E(a_i,b_i)^T; \ i = 1,\ldots,r \rangle$.
Without loss of generality, we can assume that Alice sends Bob's qubits to him only after performing her measurements and any Pauli correction. 
So, the channel applies a Pauli error only after Bob's qubits got projected according to $S'' = \langle (-1)^{m_i + a_i b_i^T} \varepsilon_i E(a_i,b_i); \ i = 1,\ldots,r \rangle$. % and evolved through $E(c,d)^T$.
Now, Bob measures the stabilizers $\varepsilon_i E(a_i,b_i)$ and applies corrections on his qubits using his syndromes as well as Alice's syndromes (and the Bell matrix identity, which in particular involves the transpose). 
This projects his qubits to the same codespace as Alice.
Finally, Alice and Bob locally apply the inverse of the encoding unitary for their code, $\mathcal{U}_{\text{Enc}}(S)^\dagger$. 
If Bob's correction was successful, this converts the $k$ logical Bell pairs into $k$ physical Bell pairs that are of higher quality than the $n$ Bell pairs initially shared between them.
% Wilde \etal showed that logical qubits of Alice and Bob are entangled to form a Bell state  \cite{Wilde-isit10}. 
This protocol is shown in Figure~\ref{fig:bell-pair-entanglement} and summarized in Algorithm~\ref{algo:algo_bell}. 
% \bane{In introduction, we may want to give a sentence or two explaining a general concept of Wilde's distillation to ``convert the logical Bell pairs into higher quality physical Bell pairs.''}

While the steps of the protocol are clear, it is not obvious why the logical qubits of Alice and Bob have to be $k$ copies of the Bell pair, assuming all errors were corrected successfully.
To get some intuition, let us quickly consider the example of the $3$-qubit bit-flip code defined by $S = \langle ZZI, IZZ \rangle$.
According to~\eqref{eq:stabilizer_projector}, the projector onto $\mathcal{Q}(S)$ is $\Pi_S = \frac{(I_8 + ZZI)}{2} \frac{(I_8 + IZZ)}{2}$.
% \bane{Perhaps we can introduce the encoding unitary $\mathcal{U}_{\text{Enc}}$ in general in the background section or at the beginning of this section. It will become handy later for Eq. (16)} \narayanan{Done}
The encoding unitary, as described in Section~\ref{sec:stabilizer_codes}, is $\mathcal{U}_{\text{Enc}} = \text{CNOT}_{1 \rightarrow 2}\, \text{CNOT}_{1 \rightarrow 3}$.
Since $Z^T = Z$, Alice's measurements will project Bob's qubits onto the same code subspace as her's.
For convenience, assume that Alice obtains the trivial syndrome $(+1,+1)$ and that the channel does not introduce any error.
Then, according to~\eqref{eq:projector_bell_pair}, the resulting (unnormalized) state after Alice's measurements is $(\Pi_S \otimes \Pi_S) \dket{\Phi_3^+}$.

Consider the action of $(I_8 + ZZI)$ on a computational basis state $\dket{x}, x = [x_1,x_2,x_3] \in \mathbb{F}_2^3$: 
\begin{align}
(I_8 + ZZI) \dket{x} = \dket{x} + E(000,110) \dket{x} = \dket{x} + (-1)^{[1,1,0] x^T} \dket{x} = 
\begin{cases}
2 \dket{x} & \ \text{if}\ x_1 \oplus x_2 = 0, \\
0          & \ \text{otherwise}.
\end{cases}
\end{align}
Hence, after the action of $(\Pi_S \otimes \Pi_S)$ and inversion of the encoding unitary by Alice and Bob, we obtain
\begin{align}
(\Pi_S \otimes \Pi_S) \dket{\Phi_3^+} & = \frac{1}{16 \cdot \sqrt{2^3}} \sum_{x \in \{000,111\}} 4\dket{x}_{\text{A}} \otimes 4\dket{x}_{\text{B}} \\
  & \propto \dket{000}_{\text{A}} \dket{000}_{\text{B}} + \dket{111}_{\text{A}} \dket{111}_{\text{B}} \\
  & \xrightarrow{(\mathcal{U}_{\text{Enc}}^{\dagger})_{\text{A}} \otimes (\mathcal{U}_{\text{Enc}}^{\dagger})_{\text{B}}} \dket{000}_{\text{A}} \dket{000}_{\text{B}} + \dket{100}_{\text{A}} \dket{100}_{\text{B}} \\
  & = \dket{00}_{\text{AB}} \otimes \dket{00}_{\text{AB}} \otimes \dket{00}_{\text{AB}} + \dket{11}_{\text{AB}} \otimes \dket{00}_{\text{AB}} \otimes \dket{00}_{\text{AB}} \\
  & = \left( \dket{00}_{\text{AB}} + \dket{11}_{\text{AB}} \right) \otimes \dket{00}_{\text{AB}} \otimes \dket{00}_{\text{AB}}.
\end{align}
Thus, the output is a single Bell pair and ancillary qubits on Alice and Bob.
In Appendix~\ref{sec:logical_bell_CSS}, we show this phenomenon for arbitrary CSS codes by generalizing the state vector approach used above.

\subsection{Bell Pair Distillation using the 5-Qubit Code}

In the remainder of this section, with the $\llbr 5,1,3 \rrbr$ code~\cite{Bennett-pra96,Laflamme-prl96} as an example, we use the stabilizer formalism to show that the above phenomenon is true for any stabilizer code.
Recall that this code is defined by
\begin{align}
S = \langle XZZXI, \ IXZZX, \ XIXZZ, \ ZXIXZ \rangle.
\end{align}
As described in Section~\ref{sec:stabilizer_codes}, the corresponding binary stabilizer matrix is given by
\begin{align}
G_S = 
\left[
\begin{array}{ccccc|ccccc|c}
1 & 0 & 0 & 1 & 0   &   0 & 1 & 1 & 0 & 0  &  +1 \\
0 & 1 & 0 & 0 & 1   &   0 & 0 & 1 & 1 & 0  &  +1 \\
1 & 0 & 1 & 0 & 0   &   0 & 0 & 0 & 1 & 1  &  +1 \\
0 & 1 & 0 & 1 & 0   &   1 & 0 & 0 & 0 & 1  &  +1
\end{array}
\right].
\end{align}
Initially, Alice starts with $5$ copies of the standard Bell state $\dket{\Phi^+}^{\otimes 5}$, and marks one qubit of each copy as Bob's.
She does not yet send Bob's qubits to him.
The stabilizer group for this joint state of 5 ``ebits'' (or ``EPR pairs'') is
% \bane{The letter $\mathcal{B}$ may be associated to Bob. Consider using other letter for joint stabilizers, say  $\mathcal{J}$.}
% \narayanan{How about $\mathcal{E}$ for ``ebits'' or ``EPR''? I would like the notation to be specific to Bell pairs so that I can pick another notation later for GHZ. Since joint stabilizers apply to both Bell and GHZ, I am hesitating to use that.}
\begin{align}
\mathcal{E}_5 & = \langle X_{\text{A}_1} X_{\text{B}_1}, \, Z_{\text{A}_1} Z_{\text{B}_1}, \ X_{\text{A}_2} X_{\text{B}_2}, \, Z_{\text{A}_2} Z_{\text{B}_2}, \ X_{\text{A}_3} X_{\text{B}_3}, \, Z_{\text{A}_3} Z_{\text{B}_3}, \ X_{\text{A}_4} X_{\text{B}_4}, \, Z_{\text{A}_4} Z_{\text{B}_4}, \ X_{\text{A}_5} X_{\text{B}_5}, \, Z_{\text{A}_5} Z_{\text{B}_5} \rangle \\
  & = \langle X_{\text{A}_i} X_{\text{B}_i} =  E([e_i^{\text{A}}, e_i^{\text{B}}], [0^{\text{A}}, 0^{\text{B}}]), \ Z_{\text{A}_i} Z_{\text{B}_i} = E([0^{\text{A}}, 0^{\text{B}}],[e_i^{\text{A}}, e_i^{\text{B}}]) \, ; \, i = 1,2,\ldots,5  \rangle,
\end{align}
where $e_i \in \mathbb{F}_2^5$ is the standard basis vector with a $1$ in position $i$ and zeros elsewhere, $0 \in \mathbb{F}_2^5$ is the all-zeros vector, and the $X$- and $Z$- components in the $E(a,b)$ notation have been split into Alice's qubits and Bob's qubits.
Observe that this is a maximal stabilizer group on $10$ qubits and hence, there are no non-trivial logical operators associated with this group, i.e., the normalizer of $\mathcal{E}_5$ in $\mathcal{P}_{10}$ is itself.

It will be convenient to adopt a tabular format for these generators, where the first column of each row gives the sign of the generator, the next two columns give the $X$-components of Alice and Bob in that generator, the subsequent two columns give the $Z$-components of Alice and Bob in that generator, and the last column gives the Pauli representation of that generator for clarity.
Hence, the above generators are written as follows.
% \bane{How about using +1 and -1 for sign in first columns of tables? The next one and Table \ref{tab:bell_protocol}.}

\begin{center}
\begin{tabularx}{.8\linewidth}{c || *{2}{C} | *{2}{C} || c}
\toprule
Sign & \multicolumn{2}{c|}{$X$-Components} & \multicolumn{2}{c||}{$Z$-Components} & Pauli Representation \\
     &   A & B   &   A & B   &   \\
\midrule
\midrule
$+1$ &    $e_i$ & $e_i$   &   $0$ & $0$    & $X_{\text{A}_i} X_{\text{B}_i} =  E([e_i^{\text{A}}, e_i^{\text{B}}], [0^{\text{A}}, 0^{\text{B}}])$ \\
\midrule
$+1$ &    $0$ & $0$   & $e_i$ & $e_i$   &   $Z_{\text{A}_i} Z_{\text{B}_i} =  E([0^{\text{A}}, 0^{\text{B}}], [e_i^{\text{A}}, e_i^{\text{B}}])$ \\
\bottomrule
\bottomrule
\end{tabularx}
\end{center}

\begin{table}
\caption{\label{tab:bell_protocol}
Steps of the Bell-pair distillation protocol based on the $\llbr 5,1,3 \rrbr$ code. Any `$0$' that is not part of a string represents $00000$, and $e_i \in \mathbb{F}_2^5$ is the standard basis vector with a $1$ in the $i$-th position and zeros elsewhere. Code stabilizers are typeset in boldface. An additional left arrow indicates which row is being replaced with a code stabilizer, i.e., the first row that anticommutes with the stabilizer. Other updated rows are highlighted in gray. Classical communications: A $\rightarrow$ B.}
\centering
\begin{tabularx}{.8\linewidth}{c c || *{2}{C} | *{2}{C} || c}
\toprule
Step & Sign & \multicolumn{2}{c|}{$X$-Components} & \multicolumn{2}{c||}{$Z$-Components} & Pauli Representation \\
     &      &   A & B   &   A & B   &   \\
\midrule
\midrule
     &     &        &       &         &          &                   \\
$(0)$ & $+1$ &    $e_1$ & $e_1$   &   $0$ & $0$    & $X_{\text{A}_1} X_{\text{B}_1}$ \\
     & $+1$ &    $e_2$ & $e_2$   &   $0$ & $0$    & $X_{\text{A}_2} X_{\text{B}_2}$ \\
     & $+1$ &    $e_3$ & $e_3$   &   $0$ & $0$    & $X_{\text{A}_3} X_{\text{B}_3}$ \\
     & $+1$ &    $e_4$ & $e_4$   &   $0$ & $0$    & $X_{\text{A}_4} X_{\text{B}_4}$ \\
     & $+1$ &    $e_5$ & $e_5$   &   $0$ & $0$    & $X_{\text{A}_5} X_{\text{B}_5}$ \\
\cmidrule(lr){2-7}
     & $+1$ &    $0$ & $0$   &   $e_1$ & $e_1$    & $Z_{\text{A}_1} Z_{\text{B}_1}$ \\
     & $+1$ &    $0$ & $0$   &   $e_2$ & $e_2$    & $Z_{\text{A}_2} Z_{\text{B}_2}$ \\
     & $+1$ &    $0$ & $0$   &   $e_3$ & $e_3$    & $Z_{\text{A}_3} Z_{\text{B}_3}$ \\
     & $+1$ &    $0$ & $0$   &   $e_4$ & $e_4$    & $Z_{\text{A}_4} Z_{\text{B}_4}$ \\
     & $+1$ &    $0$ & $0$   &   $e_5$ & $e_5$    & $Z_{\text{A}_5} Z_{\text{B}_5}$ \\
     &     &        &       &         &          &                   \\
\midrule
     &     &        &       &         &          &                   \\
$(1)$ & $+1$ &    $e_1$ & $e_1$   &   $0$ & $0$    & $X_{\text{A}_1} X_{\text{B}_1}$ \\
\rowcolor{lightgray}
     & $+1$ &    $e_2$ & $e_2$   &   $e_4$ & $e_4$    & $X_{\text{A}_2} X_{\text{B}_2} Z_{\text{A}_4} Z_{\text{B}_4}$ \\
\rowcolor{lightgray}
     & $+1$ &    $e_3$ & $e_3$   &   $e_4$ & $e_4$    & $X_{\text{A}_3} X_{\text{B}_3} Z_{\text{A}_4} Z_{\text{B}_4}$ \\
     & $+1$ &    $e_4$ & $e_4$   &   $0$ & $0$    & $X_{\text{A}_4} X_{\text{B}_4}$ \\
     & $+1$ &    $e_5$ & $e_5$   &   $0$ & $0$    & $X_{\text{A}_5} X_{\text{B}_5}$ \\
\cmidrule(lr){2-7}
\rowcolor{lightgray}
     & $+1$ &    $0$ & $0$   &   $e_1+e_4$ & $e_1+e_4$    & $Z_{\text{A}_1} Z_{\text{B}_1} Z_{\text{A}_4} Z_{\text{B}_4}$ \\
     & $+1$ &    $0$ & $0$   &   $e_2$ & $e_2$    & $Z_{\text{A}_2} Z_{\text{B}_2}$ \\
     & $+1$ &    $0$ & $0$   &   $e_3$ & $e_3$    & $Z_{\text{A}_3} Z_{\text{B}_3}$ \\
% \rowcolor{lightgray}
     & $\boldsymbol{ \varepsilon_1 }$ &    $\boldsymbol{ 10010 }$ & $\boldsymbol{ 00000 }$   &   $\boldsymbol{ 01100 }$ & $\boldsymbol{ 00000 }$    & \qquad \ \  $\boldsymbol{ \varepsilon_1 \, X_{\text{A}_1} Z_{\text{A}_2} Z_{\text{A}_3} X_{\text{A}_4} } \ \ \boldsymbol{\longleftarrow} $ \\
     & $+1$ &    $0$ & $0$   &   $e_5$ & $e_5$    & $Z_{\text{A}_5} Z_{\text{B}_5}$ \\
     &     &        &       &         &          &                   \\
\midrule
     &     &        &       &         &          &                   \\
$(2)$ & $+1$ &    $e_1$ & $e_1$   &   $0$ & $0$    & $X_{\text{A}_1} X_{\text{B}_1}$ \\
     & $+1$ &    $e_2$ & $e_2$   &   $e_4$ & $e_4$    & $X_{\text{A}_2} X_{\text{B}_2} Z_{\text{A}_4} Z_{\text{B}_4}$ \\
\rowcolor{lightgray}
     & $+1$ &    $e_3$ & $e_3$   &   $e_4+e_5$ & $e_4+e_5$    & $X_{\text{A}_3} X_{\text{B}_3} Z_{\text{A}_4} Z_{\text{B}_4} Z_{\text{A}_5} Z_{\text{B}_5}$ \\
     & $+1$ &    $e_4$ & $e_4$   &   $e_5$ & $e_5$    & $X_{\text{A}_4} X_{\text{B}_4} Z_{\text{A}_5} Z_{\text{B}_5}$ \\
     & $+1$ &    $e_5$ & $e_5$   &   $0$ & $0$    & $X_{\text{A}_5} X_{\text{B}_5}$ \\
\cmidrule(lr){2-7}
     & $+1$ &    $0$ & $0$   &   $e_1+e_4$ & $e_1+e_4$    & $Z_{\text{A}_1} Z_{\text{B}_1} Z_{\text{A}_4} Z_{\text{B}_4}$ \\
\rowcolor{lightgray}
     & $+1$ &    $0$ & $0$   &   $e_2+e_5$ & $e_2+e_5$    & $Z_{\text{A}_2} Z_{\text{B}_2} Z_{\text{A}_5} Z_{\text{B}_5}$ \\
     & $+1$ &    $0$ & $0$   &   $e_3$ & $e_3$    & $Z_{\text{A}_3} Z_{\text{B}_3}$ \\
     & $\boldsymbol{ \varepsilon_1 }$ &    $\boldsymbol{ 10010 }$ & $\boldsymbol{ 00000 }$   &   $\boldsymbol{ 01100 }$ & $\boldsymbol{ 00000 }$    & $\boldsymbol{ \varepsilon_1 \, X_{\text{A}_1} Z_{\text{A}_2} Z_{\text{A}_3} X_{\text{A}_4} }$ \\
% \rowcolor{lightgray}
     & $\boldsymbol{ \varepsilon_2 }$ &    $\boldsymbol{ 01001 }$ & $\boldsymbol{ 00000 }$   &   $\boldsymbol{ 00110 }$ & $\boldsymbol{ 00000 }$    & \qquad \ \  $\boldsymbol{ \varepsilon_2 \, X_{\text{A}_2} Z_{\text{A}_3} Z_{\text{A}_4} X_{\text{A}_5} } \ \ \boldsymbol{\longleftarrow}$ \\
     &     &        &       &         &          &                   \\
\midrule
     &     &        &       &         &          &                   \\
$(3)$ & $+1$ &    $e_1$ & $e_1$   &   $0$ & $0$    & $X_{\text{A}_1} X_{\text{B}_1}$ \\
     & $+1$ &    $e_2$ & $e_2$   &   $e_4$ & $e_4$    & $X_{\text{A}_2} X_{\text{B}_2} Z_{\text{A}_4} Z_{\text{B}_4}$ \\
     & $+1$ &    $e_3$ & $e_3$   &   $e_4+e_5$ & $e_4+e_5$    & $X_{\text{A}_3} X_{\text{B}_3} Z_{\text{A}_4} Z_{\text{B}_4} Z_{\text{A}_5} Z_{\text{B}_5}$ \\
\rowcolor{lightgray}
     & $-1$ &    $e_4+e_5$ & $e_4+e_5$   &   $e_5$ & $e_5$    & $- \, X_{\text{A}_4} X_{\text{B}_4} Z_{\text{A}_5} Z_{\text{B}_5} X_{\text{A}_5} X_{\text{B}_5}$ \\
% \rowcolor{lightgray}
     & $\boldsymbol{ \varepsilon_3 }$ &    $\boldsymbol{ 10100 }$ & $\boldsymbol{ 00000 }$   &   $\boldsymbol{ 00011 }$ & $\boldsymbol{ 00000 }$    & \qquad \ \  $\boldsymbol{ \varepsilon_3 \, X_{\text{A}_1} X_{\text{A}_3} Z_{\text{A}_4} Z_{\text{A}_5} } \ \ \boldsymbol{\longleftarrow}$ \\
\cmidrule(lr){2-7}
\rowcolor{lightgray}
     & $+1$ &    $e_5$ & $e_5$   &   $e_1+e_4$ & $e_1+e_4$    & $Z_{\text{A}_1} Z_{\text{B}_1} Z_{\text{A}_4} Z_{\text{B}_4} X_{\text{A}_5} X_{\text{B}_5}$ \\
     & $+1$ &    $0$ & $0$   &   $e_2+e_5$ & $e_2+e_5$    & $Z_{\text{A}_2} Z_{\text{B}_2} Z_{\text{A}_5} Z_{\text{B}_5}$ \\
\rowcolor{lightgray}
     & $+1$ &    $e_5$ & $e_5$   &   $e_3$ & $e_3$    & $Z_{\text{A}_3} Z_{\text{B}_3} X_{\text{A}_5} X_{\text{B}_5}$ \\
     & $\boldsymbol{ \varepsilon_1 }$ &    $\boldsymbol{ 10010 }$ & $\boldsymbol{ 00000 }$   &   $\boldsymbol{ 01100 }$ & $\boldsymbol{ 00000 }$    & $\boldsymbol{ \varepsilon_1 \, X_{\text{A}_1} Z_{\text{A}_2} Z_{\text{A}_3} X_{\text{A}_4} }$ \\
     & $\boldsymbol{ \varepsilon_2 }$ &    $\boldsymbol{ 01001 }$ & $\boldsymbol{ 00000 }$   &   $\boldsymbol{ 00110 }$ & $\boldsymbol{ 00000 }$    & $\boldsymbol{ \varepsilon_2 \, X_{\text{A}_2} Z_{\text{A}_3} Z_{\text{A}_4} X_{\text{A}_5} }$ \\
     &     &        &       &         &          &                   \\
\midrule
     &     &        &       &         &          &                   \\
%
% \rowcolor{lightgray}
$(4)$ & $\boldsymbol{ \varepsilon_4 }$ &    $\boldsymbol{ 01010 }$ & $\boldsymbol{ 00000 }$   &   $\boldsymbol{ 10001 }$ & $\boldsymbol{ 00000 }$    & \qquad \ \  $\boldsymbol{ \varepsilon_4 \, Z_{\text{A}_1} X_{\text{A}_2} X_{\text{A}_4} Z_{\text{A}_5} } \ \ \boldsymbol{\longleftarrow}$ \\
\rowcolor{lightgray}
     & $+1$ &    $e_1+e_2$ & $e_1+e_2$   &   $e_4$ & $e_4$    & $X_{\text{A}_2} X_{\text{B}_2} Z_{\text{A}_4} Z_{\text{B}_4} X_{\text{A}_1} X_{\text{B}_1}$ \\
\rowcolor{lightgray}
     & $+1$ &    $e_1+e_3$ & $e_1+e_3$   &   $e_4+e_5$ & $e_4+e_5$    & $X_{\text{A}_3} X_{\text{B}_3} Z_{\text{A}_4} Z_{\text{B}_4} Z_{\text{A}_5} Z_{\text{B}_5} X_{\text{A}_1} X_{\text{B}_1}$ \\
\rowcolor{lightgray}
     & $-1$ &    $e_1+e_4+e_5$ & $e_1+e_4+e_5$   &   $e_5$ & $e_5$    & $- \, X_{\text{A}_4} X_{\text{B}_4} Z_{\text{A}_5} Z_{\text{B}_5} X_{\text{A}_5} X_{\text{B}_5} X_{\text{A}_1} X_{\text{B}_1}$ \\
     & $\boldsymbol{ \varepsilon_3 }$ &    $\boldsymbol{ 10100 }$ & $\boldsymbol{ 00000 }$   &   $\boldsymbol{ 00011 }$ & $\boldsymbol{ 00000 }$    & $\boldsymbol{ \varepsilon_3 \, X_{\text{A}_1} X_{\text{A}_3} Z_{\text{A}_4} Z_{\text{A}_5} }$ \\
\cmidrule(lr){2-7}
     & $+1$ &    $e_5$ & $e_5$   &   $e_1+e_4$ & $e_1+e_4$    & $Z_{\text{A}_1} Z_{\text{B}_1} Z_{\text{A}_4} Z_{\text{B}_4} X_{\text{A}_5} X_{\text{B}_5}$ \\
\rowcolor{lightgray}
     & $+1$ &    $e_1$ & $e_1$   &   $e_2+e_5$ & $e_2+e_5$    & $Z_{\text{A}_2} Z_{\text{B}_2} Z_{\text{A}_5} Z_{\text{B}_5} X_{\text{A}_1} X_{\text{B}_1}$ \\
\rowcolor{lightgray}
     & $+1$ &    $e_1+e_5$ & $e_1+e_5$   &   $e_3$ & $e_3$    & $Z_{\text{A}_3} Z_{\text{B}_3} X_{\text{A}_5} X_{\text{B}_5} X_{\text{A}_1} X_{\text{B}_1}$ \\
     & $\boldsymbol{ \varepsilon_1 }$ &    $\boldsymbol{ 10010 }$ & $\boldsymbol{ 00000 }$   &   $\boldsymbol{ 01100 }$ & $\boldsymbol{ 00000 }$    & $\boldsymbol{ \varepsilon_1 \, X_{\text{A}_1} Z_{\text{A}_2} Z_{\text{A}_3} X_{\text{A}_4} }$ \\
     & $\boldsymbol{ \varepsilon_2 }$ &    $\boldsymbol{ 01001 }$ & $\boldsymbol{ 00000 }$   &   $\boldsymbol{ 00110 }$ & $\boldsymbol{ 00000 }$    & $\boldsymbol{ \varepsilon_2 \, X_{\text{A}_2} Z_{\text{A}_3} Z_{\text{A}_4} X_{\text{A}_5} }$ \\
     &     &        &       &         &          &                   \\
\bottomrule
\bottomrule
\end{tabularx}
\end{table}

Given this ``initialization'', let us track these $10$ stabilizers through each step of the protocol, as shown in Table~\ref{tab:bell_protocol}. \\

\begin{enumerate}

\item[(1)] Alice measures the first stabilizer generator $X_{\text{A}_1} Z_{\text{A}_2} Z_{\text{A}_3} X_{\text{A}_4}$, and %for convenience 
assume that the measurement result is $\varepsilon_1 \in \{ \pm 1 \}$. %$+1$.
% If the measurement yields $\varepsilon_1 = -1$, Alice can simply perform a suitable Pauli correction at the end to fix the sign of this generator as well as subsequent ones.
We apply the stabilizer formalism for measurements from Section~\ref{sec:stabilizer_formalism} to update $\mathcal{E}_5$.
Since there are several elements of $\mathcal{E}_5$ that anticommute with this generator, we choose to remove%
\footnote{Later, in the GHZ protocol, we restrict this choice to be the first element in the table that anticommutes with the measured stabilizer.}
$Z_{\text{A}_4} Z_{\text{B}_4} = E([0^{\text{A}},0^{\text{B}}], [e_4^{\text{A}},e_4^{\text{B}}])$ and replace all other anticommuting elements by their product with $Z_{\text{A}_4} Z_{\text{B}_4}$.
Let this updated group in Step (1) of Table~\ref{tab:bell_protocol} be denoted as $\mathcal{E}_5^{(1)}$.
For visual clarity, code stabilizer rows are boldfaced and binary vectors are written out in full. \\

% For visual clarity, we enclose changed elements within curly braces and color the non-code ones in blue. 
% To distinguish the code's stabilizer generators, we color them in red and also avoid the $E(a,b)$ notation.
% Hence,
% \begin{align}
% \mathcal{E}_5^{(1)} & = \langle \ \ E([e_1^{\text{A}}, e_1^{\text{B}}], [0^{\text{A}}, 0^{\text{B}}]), \, {\color{blue} \{ E([0^{\text{A}}, 0^{\text{B}}],[(e_1+e_4)^{\text{A}}, (e_1+e_4)^{\text{B}}]) \} }, \nonumber \\ 
% %
%   & \hspace{1cm} {\color{blue} \{ E([e_2^{\text{A}}, e_2^{\text{B}}], [e_4^{\text{A}}, e_4^{\text{B}}]) \} }, \, E([0^{\text{A}}, 0^{\text{B}}],[e_2^{\text{A}}, e_2^{\text{B}}]), \ {\color{blue} \{ E([e_3^{\text{A}}, e_3^{\text{B}}], [e_4^{\text{A}}, e_4^{\text{B}}]) \} }, \, E([0^{\text{A}}, 0^{\text{B}}],[e_3^{\text{A}}, e_3^{\text{B}}]), \nonumber \\
% %
%   & \hspace{1cm} E([e_4^{\text{A}}, e_4^{\text{B}}], [0^{\text{A}}, 0^{\text{B}}]), \, {\color{red} \{ X_{\text{A}_1} Z_{\text{A}_2} Z_{\text{A}_3} X_{\text{A}_4} \} }, \ E([e_5^{\text{A}}, e_5^{\text{B}}], [0^{\text{A}}, 0^{\text{B}}]), \, E([0^{\text{A}}, 0^{\text{B}}],[e_5^{\text{A}}, e_5^{\text{B}}]) \ \  \rangle.
% \end{align}
Now, we observe that if Bob measures the same generator on his qubits, i.e., $X_{\text{B}_1} Z_{\text{B}_2} Z_{\text{B}_3} X_{\text{B}_4}$, then it is trivial because it commutes with all elements in $\mathcal{E}_5^{(1)}$ and 
% \bane{make it precise - in ${B}_5^{(1)}$}. \narayanan{Done}
% Since $\mathcal{E}_5^{(1)}$ is a maximal sized stabilizer group, there are no non-trivial logical Pauli operators, hence $X_{\text{B}_1} Z_{\text{B}_2} Z_{\text{B}_3} X_{\text{B}_4}$ 
hence is already contained in $\mathcal{E}_5^{(1)}$.
This is a manifestation of the Bell state matrix identity discussed in Section~\ref{sec:bell_state_identity}.
% Indeed, Bob's generator can be obtained by multiplying the first four uncolored elements above (which do not have curly braces) with the red colored $X_{\text{A}_1} Z_{\text{A}_2} Z_{\text{A}_3} X_{\text{A}_4}$.
Indeed, Bob's generator can be obtained by multiplying $X_{\text{A}_1} X_{\text{B}_1}, X_{\text{A}_4} X_{\text{B}_4}, Z_{\text{A}_2} Z_{\text{B}_2}$, $Z_{\text{A}_3} Z_{\text{B}_3},$ and $X_{\text{A}_1} Z_{\text{A}_2} Z_{\text{A}_3} X_{\text{A}_4}$ in Step (1) of Table~\ref{tab:bell_protocol}. \\

\item[(2)] Alice measures the second stabilizer generator $X_{\text{A}_2} Z_{\text{A}_3} Z_{\text{A}_4} X_{\text{A}_5}$, and %for convenience 
assume that the measurement result is $\varepsilon_2 \in \{ \pm 1 \}$. % $+1$.
Then, the new joint stabilizer group, $\mathcal{E}_5^{(2)}$, is given in Step (2) of Table~\ref{tab:bell_protocol}.
% \begin{align}
% \mathcal{E}_5^{(2)} & = \langle \ \ E([e_1^{\text{A}}, e_1^{\text{B}}], [0^{\text{A}}, 0^{\text{B}}]), \, E([0^{\text{A}}, 0^{\text{B}}],[(e_1+e_4)^{\text{A}}, (e_1+e_4)^{\text{B}}]), \ E([e_2^{\text{A}}, e_2^{\text{B}}], [e_4^{\text{A}}, e_4^{\text{B}}]),  \nonumber \\ 
% %
%   & \hspace{1cm} {\color{blue} \{ E([0^{\text{A}}, 0^{\text{B}}],[(e_2+e_5)^{\text{A}}, (e_2+e_5)^{\text{B}}]) \} }, \ {\color{blue} \{ E([e_3^{\text{A}}, e_3^{\text{B}}], [(e_4+e_5)^{\text{A}}, (e_4+e_5)^{\text{B}}]) \} }, \, E([0^{\text{A}}, 0^{\text{B}}],[e_3^{\text{A}}, e_3^{\text{B}}]), \nonumber \\
% %
%   & \hspace{1cm} {\color{blue} \{ E([e_4^{\text{A}}, e_4^{\text{B}}], [e_5^{\text{A}}, e_5^{\text{B}}]) \} }, \, {\color{red} X_{\text{A}_1} Z_{\text{A}_2} Z_{\text{A}_3} X_{\text{A}_4} }, \ E([e_5^{\text{A}}, e_5^{\text{B}}], [0^{\text{A}}, 0^{\text{B}}]), \, {\color{red} \{ X_{\text{A}_2} Z_{\text{A}_3} Z_{\text{A}_4} X_{\text{A}_5} \} } \ \  \rangle.
% \end{align}
This stabilizer generator anticommutes with the third row of the top block and the second and fifth rows of the bottom block.
We have replaced $Z_{\text{A}_5} Z_{\text{B}_5}$ (fifth row of the bottom block) with this generator and multiplied the other anticommuting elements with $Z_{\text{A}_5} Z_{\text{B}_5}$.
It can be verified that the second stabilizer generator of Bob is already in $\mathcal{E}_5^{(2)}$. \\

\item[(3)] Alice measures the third stabilizer generator $X_{\text{A}_1} X_{\text{A}_3} Z_{\text{A}_4} Z_{\text{A}_5}$, and %for convenience 
assume that the measurement result is $\varepsilon_3 \in \{ \pm 1 \}$. % $+1$.
Then, the new joint stabilizer group, $\mathcal{E}_5^{(3)}$, is given in Step (3) of Table~\ref{tab:bell_protocol}.
% \begin{align}
% \mathcal{E}_5^{(3)} & = \langle \ \ E([e_1^{\text{A}}, e_1^{\text{B}}], [0^{\text{A}}, 0^{\text{B}}]), \, {\color{blue} \{ E([e_5^{\text{A}}, e_5^{\text{B}}],[(e_1+e_4)^{\text{A}}, (e_1+e_4)^{\text{B}}]) \} }, \ E([e_2^{\text{A}}, e_2^{\text{B}}], [e_4^{\text{A}}, e_4^{\text{B}}]),  \nonumber \\ 
% %
%   & \hspace{1cm} E([0^{\text{A}}, 0^{\text{B}}],[(e_2+e_5)^{\text{A}}, (e_2+e_5)^{\text{B}}]), \ E([e_3^{\text{A}}, e_3^{\text{B}}], [(e_4+e_5)^{\text{A}}, (e_4+e_5)^{\text{B}}]), \, {\color{blue} \{ E([e_5^{\text{A}}, e_5^{\text{B}}],[e_3^{\text{A}}, e_3^{\text{B}}]) \} }, \nonumber \\
% %
%   & \hspace{1cm} {\color{blue} \{ - E([(e_4+e_5)^{\text{A}}, (e_4+e_5)^{\text{B}}], [e_5^{\text{A}}, e_5^{\text{B}}]) \} }, \, {\color{red} X_{\text{A}_1} Z_{\text{A}_2} Z_{\text{A}_3} X_{\text{A}_4} }, \ {\color{red} \{ X_{\text{A}_1} X_{\text{A}_3} Z_{\text{A}_4} Z_{\text{A}_5} \} }, \, {\color{red} X_{\text{A}_2} Z_{\text{A}_3} Z_{\text{A}_4} X_{\text{A}_5} } \ \  \rangle.
% \end{align}
Once again, it can be verified that the third stabilizer generator of Bob is already in $\mathcal{E}_5^{(3)}$.
The minus sign in the fourth row of the top block gets introduced when we apply the multiplication rule for $E(a,b)$ from Lemma~\ref{lem:Eab}(b). \\

\item[(4)] Alice measures the final stabilizer generator $Z_{\text{A}_1} X_{\text{A}_2} X_{\text{A}_4} Z_{\text{A}_5}$, and %for convenience 
assume that the measurement result is $\varepsilon_4 \in \{ \pm 1 \}$. % $+1$.
Then, the new joint stabilizer group, $\mathcal{E}_5^{(4)}$, is given in Step (4) of Table~\ref{tab:bell_protocol}.
% \begin{align}
% \mathcal{E}_5^{(4)} & = \langle \ \ {\color{red} \{ Z_{\text{A}_1} X_{\text{A}_2} X_{\text{A}_4} Z_{\text{A}_5} \} }, \, E([e_5^{\text{A}}, e_5^{\text{B}}],[(e_1+e_4)^{\text{A}}, (e_1+e_4)^{\text{B}}]), \ {\color{blue} \{ E([(e_1+e_2)^{\text{A}}, (e_1+e_2)^{\text{B}}], [e_4^{\text{A}}, e_4^{\text{B}}]) \} },  \nonumber \\ 
% %
%   & \hspace{1cm} {\color{blue} \{ E([e_1^{\text{A}}, e_1^{\text{B}}],[(e_2+e_5)^{\text{A}}, (e_2+e_5)^{\text{B}}]) \} }, \ {\color{blue} \{ E([(e_1+e_3)^{\text{A}}, (e_1+e_3)^{\text{B}}], [(e_4+e_5)^{\text{A}}, (e_4+e_5)^{\text{B}}]) \} }, \nonumber \\
% %
%   & \hspace{1cm} {\color{blue} \{ E([(e_1+e_5)^{\text{A}}, (e_1+e_5)^{\text{B}}],[e_3^{\text{A}}, e_3^{\text{B}}]) \} }, \ {\color{blue} \{ -E([(e_1+e_4+e_5)^{\text{A}}, (e_1+e_4+e_5)^{\text{B}}], [e_5^{\text{A}}, e_5^{\text{B}}]) \} }, \nonumber \\ 
% %
%   & \hspace{1cm} {\color{red} X_{\text{A}_1} Z_{\text{A}_2} Z_{\text{A}_3} X_{\text{A}_4} }, \ {\color{red} X_{\text{A}_1} X_{\text{A}_3} Z_{\text{A}_4} Z_{\text{A}_5} }, \, {\color{red} X_{\text{A}_2} Z_{\text{A}_3} Z_{\text{A}_4} X_{\text{A}_5} } \ \  \rangle.
% \end{align}
As before, it can be verified that the final stabilizer generator of Bob is already in $\mathcal{E}_5^{(4)}$. 
This completes all measurements of Alice, and she now sends Bob's qubits over the channel.
To understand the working of the protocol in the ideal scenario, assume that no errors occur. \\

\end{enumerate}

Since we know that all stabilizer generators of Bob are in $\mathcal{E}_5^{(4)}$, we conveniently perform the following replacements:
\begin{align}
E([e_1^{\text{A}}, e_1^{\text{B}}],[(e_2+e_5)^{\text{A}}, (e_2+e_5)^{\text{B}}]) & \mapsto X_{\text{B}_1} Z_{\text{B}_2} Z_{\text{B}_3} X_{\text{B}_4}, \nonumber \\
E([(e_1+e_2)^{\text{A}}, (e_1+e_2)^{\text{B}}], [e_4^{\text{A}}, e_4^{\text{B}}]) & \mapsto X_{\text{B}_2} Z_{\text{B}_3} Z_{\text{B}_4} X_{\text{B}_5}, \nonumber \\
E([(e_1+e_3)^{\text{A}}, (e_1+e_3)^{\text{B}}], [(e_4+e_5)^{\text{A}}, (e_4+e_5)^{\text{B}}]) & \mapsto X_{\text{B}_1} X_{\text{B}_3} Z_{\text{B}_4} Z_{\text{B}_5}, \nonumber \\
E([e_5^{\text{A}}, e_5^{\text{B}}],[(e_1+e_4)^{\text{A}}, (e_1+e_4)^{\text{B}}]) & \mapsto Z_{\text{B}_1} X_{\text{B}_2} X_{\text{B}_4} Z_{\text{B}_5}.
\end{align}
% Then, by coloring Alice's stabilizers in red and Bob's stabilizers in blue, the group $\mathcal{E}_5^{(4)}$ can be rewritten as
% \begin{align}
% \mathcal{E}_5^{(4)} & = \langle \ \ {\color{red} {\color{red} X_{\text{A}_1} Z_{\text{A}_2} Z_{\text{A}_3} X_{\text{A}_4} }, \ {\color{red} X_{\text{A}_2} Z_{\text{A}_3} Z_{\text{A}_4} X_{\text{A}_5} }, \ {\color{red} X_{\text{A}_1} X_{\text{A}_3} Z_{\text{A}_4} Z_{\text{A}_5} }, \ Z_{\text{A}_1} X_{\text{A}_2} X_{\text{A}_4} Z_{\text{A}_5} }, \nonumber \\
% %
%  & \hspace{1cm} {\color{dkgreen} E([(e_1+e_5)^{\text{A}}, (e_1+e_5)^{\text{B}}],[e_3^{\text{A}}, e_3^{\text{B}}]) }, \ {\color{dkgreen} -E([(e_1+e_4+e_5)^{\text{A}}, (e_1+e_4+e_5)^{\text{B}}], [e_5^{\text{A}}, e_5^{\text{B}}]) }, \nonumber \\
% %
%   & \hspace{1cm} {\color{blue} X_{\text{B}_1} Z_{\text{B}_2} Z_{\text{B}_3} X_{\text{B}_4} }, \ {\color{blue} X_{\text{B}_2} Z_{\text{B}_3} Z_{\text{B}_4} X_{\text{B}_5} }, \ {\color{blue} X_{\text{B}_1} X_{\text{B}_3} Z_{\text{B}_4} Z_{\text{B}_5}, \ {\color{blue} Z_{\text{B}_1} X_{\text{B}_2} X_{\text{B}_4} Z_{\text{B}_5} } } \ \  \rangle.
% \end{align}
Recollect that for the $\llbr 5,1,3 \rrbr$ code, the logical Pauli operators are $\overline{X} = X_1 X_2 X_3 X_4 X_5 = E([11111,00000])$ and $\overline{Z} = Z_1 Z_2 Z_3 Z_4 Z_5 = E([00000,11111])$.
If we used Algorithm~\ref{algo:logical_paulis_ghz_msmt}, we would obtain the same $\overline{Z}$ and $\overline{X} = - Y_1 Z_3 Z_4$.
Then, by grouping Alice's code stabilizers and Bob's code stabilizers, the group $\mathcal{E}_5^{(4)}$ can be rewritten as
\begin{align}
\mathcal{E}_5^{(4)} & = \langle \ \ \varepsilon_1 \, X_{\text{A}_1} Z_{\text{A}_2} Z_{\text{A}_3} X_{\text{A}_4} , \ \varepsilon_2 \,  X_{\text{A}_2} Z_{\text{A}_3} Z_{\text{A}_4} X_{\text{A}_5} , \ \varepsilon_3 \, X_{\text{A}_1} X_{\text{A}_3} Z_{\text{A}_4} Z_{\text{A}_5} , \ \varepsilon_4 \, Z_{\text{A}_1} X_{\text{A}_2} X_{\text{A}_4} Z_{\text{A}_5} , \nonumber \\
 & \hspace{1cm} E([(e_1+e_5)^{\text{A}}, (e_1+e_5)^{\text{B}}],[e_3^{\text{A}}, e_3^{\text{B}}]) , \ -E([(e_1+e_4+e_5)^{\text{A}}, (e_1+e_4+e_5)^{\text{B}}], [e_5^{\text{A}}, e_5^{\text{B}}]) , \nonumber \\
  & \hspace{1cm} \varepsilon_1 \, X_{\text{B}_1} Z_{\text{B}_2} Z_{\text{B}_3} X_{\text{B}_4} , \ \varepsilon_2 \, X_{\text{B}_2} Z_{\text{B}_3} Z_{\text{B}_4} X_{\text{B}_5} , \ \varepsilon_3 \, X_{\text{B}_1} X_{\text{B}_3} Z_{\text{B}_4} Z_{\text{B}_5}, \ \varepsilon_4 \, Z_{\text{B}_1} X_{\text{B}_2} X_{\text{B}_4} Z_{\text{B}_5} \ \  \rangle.
\end{align}
% The $4$ red and $4$ blue operators correspond to Alice's and Bob's stabilizer generators, respectively.
% Recollect that for the $\llbr 5,1,3 \rrbr$ code, the logical Pauli operators are $\overline{X} = X_1 X_2 X_3 X_4 X_5 = E([11111,00000])$ and $\overline{Z} = Z_1 Z_2 Z_3 Z_4 Z_5 = E([00000,11111])$.
Using some manipulations, we recognize that the two %green 
operators on the second line in $\mathcal{E}_5^{(4)}$ are
\begin{align}
E([(e_1+e_5)^{\text{A}}, (e_1+e_5)^{\text{B}}],[e_3^{\text{A}}, e_3^{\text{B}}]) = (X_{\text{A}_1} Z_{\text{A}_3} X_{\text{A}_5}) (X_{\text{B}_1} Z_{\text{B}_3} X_{\text{B}_5}) & \equiv \overline{Z}_{\text{A}} \overline{Z}_{\text{B}}, \nonumber \\
-E([(e_1+e_4+e_5)^{\text{A}}, (e_1+e_4+e_5)^{\text{B}}], [e_5^{\text{A}}, e_5^{\text{B}}]) = (\imath X_{\text{A}_1} X_{\text{A}_4} Y_{\text{A}_5}) (\imath X_{\text{B}_1} X_{\text{B}_4} Y_{\text{B}_5}) & \equiv \overline{X}_{\text{A}} \overline{X}_{\text{B}}.
\end{align}
Thus, $\mathcal{E}_5^{(4)}$ can be interpreted as having $8$ stabilizer generators (Alice and Bob combined) and a pair of logical $X_{\text{A}} X_{\text{B}}$ and logical $Z_{\text{A}} Z_{\text{B}}$ operators, which implies that the pair of logical qubits shared between Alice and Bob forms a Bell pair. 
This can be converted into a physical Bell pair by performing the inverse of the encoding unitary on both Alice's and Bob's qubits locally. 
Note that this encoding unitary must be compatible with the above definition of the logical Paulis for the $\llbr 5,1,3 \rrbr$ code, i.e., when the physical $X$ and $Z$ on the input (logical) qubit to the encoder is conjugated by the chosen encoding unitary, the result must be the above logical Paulis $\overline{X}$ and $\overline{Z}$, respectively, potentially multiplied by some stabilizer element.

\begin{remark} \label{rem:Bell_local_msmt}
\normalfont
In this example, we have assumed that Bob's qubits do not suffer any error, so that we can clearly show the existence of the correct logical Bell stabilizers.
If, however, the channel introduced an error, then Alice and Bob can \emph{jointly} deduce the error by measuring the signs of all generators of $\mathcal{E}_5^{(1)}$ and applying the necessary Pauli correction.
Since there are no non-trivial logical Pauli operators, any syndrome-matched correction can differ from the true error only by a stabilizer, so any error is correctable by the joint action of Alice and Bob.
But, since we prohibit \emph{non-local} measurements between Alice and Bob, our error correction capability is limited to that of the code (on Bob's side).
If the channel introduces a correctable Pauli error for the chosen code and Bob's decoder, then the protocol will output $k$ perfect Bell pairs.
However, if the Pauli error is miscorrected by Bob's decoder, then there will be a logical error on the code, and hence at least one of the $k$ output Bell pairs will suffer from an unknown Pauli error.
% \bane{this statement about local measurements only shell eb also mentioned in introduction.}
% This discussion applies to the subsequent measurements of Alice as well.
\end{remark}

We can arrive at the above conclusion without knowing the specific logical operators for the code.
After Alice measures all her stabilizer generators, we know that Bob's stabilizer generators will also be present in the group, simply based on the Bell state matrix identity from Section~\ref{sec:bell_state_identity}.
For this example, the transpose in that identity did not make a difference, but for other codes this can only introduce an additional minus sign since $Y^T = -Y$.
For an $\llbr n,k,d \rrbr$ code, we now have a $2n$-qubit stabilizer group $\mathcal{E}_n^{(n-k)}$ where $2(n-k)$ generating elements are Alice's and Bob's stabilizer generators.
We are left with $2n - 2(n-k) = 2k$ elements in the generators, each of which \emph{must} jointly involve Alice's \emph{and} Bob's qubits.
These commute with each other and with the $2(n-k)$ stabilizer generators of Alice and Bob, and are independent, so we can rename them as the logical $X_{\text{A}_j} X_{\text{B}_j}$ and logical $Z_{\text{A}_j} Z_{\text{B}_j}$ for $j=1,2,\ldots,k$.
Thus, \emph{by definition}, the $k$ pairs of logical qubits form $k$ logical Bell pairs.
% As long as Alice and Bob pick the encoding unitary matching these logical operators for the code, they can produce physical Bell pairs by simultaneously inverting this unitary locally.
Alice and Bob can produce physical Bell pairs by simultaneously inverting the (same) encoding unitary for the code locally.
This is the key idea behind the working of the Bell pair distillation protocol employed by Wilde \etal in~\cite{Wilde-isit10}.

% \ankur{To be added:EXAMPLES} 

\section{Distillation of Greenberger-Horne-Zeilinger (GHZ) States}
\label{sec:ghz_distillation}

% We extend this idea to the GHZ states in this section.
In this section, we extend the above Bell pair distillation protocol to distill GHZ states, $\dket{\text{GHZ}} = \frac{(\dket{000} + \dket{111})}{\sqrt{2}}$.
%  \ankur{To be added: EXAMPLES} 
Let $n$ GHZ states be shared between Alice, Bob, and Charlie. 
% We can write the joint state as 
% \begin{align}
%     \ket{\psi}_{\text{ABC}} &= \ket{\mathrm{GHZ}}^{\otimes n}_{\text{ABC}},
% \end{align}
% where $\ket{\mathrm{GHZ}}$ is the GHZ state of three qubits:
% \begin{align}
%     \ket{\mathrm{GHZ}} &= \frac{\ket{000}+\ket{111}}{\sqrt{2}}.
% \end{align}
We rearrange all the qubits to keep Alice's, Bob's and Charlie's qubits together respectively. 
Hence, this joint state can be expressed as
%\begin{align}    
%\ket{\psi}_{\text{AB}}    &= \left(\frac{\ket{000}_{\text{AB}}+\ket{111}_{\text{ABC}}}{\sqrt{2}}\right)^{\otimes n} = \frac{1}{\sqrt{2^n}} \displaystyle \sum_{x \in \mathbb{F}^n_2} \ket{x}_{\text{A}} \ket{x}_{\text{B}} \ket{x}_{\text{C}} \coloneqq \ket{\Phi^{+}_n}. 
%\end{align}
%
\begin{align}    
\label{eq:ghz_state_rearranged}
\dket{\text{GHZ}_n}_{\text{ABC}} & = \left(\frac{\dket{000}_{\text{ABC}} + \dket{111}_{\text{ABC}}}{\sqrt{2}}\right)^{\otimes n} = \frac{1}{\sqrt{2^n}} \sum_{x \in \mathbb{F}_2^n} \dket{x}_{\text{A}} \dket{x}_{\text{B}} \dket{x}_{\text{C}}. 
\end{align}
Since the GHZ state has stabilizers $S_{\text{GHZ}} = \langle Z_{\text{A}} Z_{\text{B}} I_{\text{C}}, \ I_{\text{A}} Z_{\text{B}} Z_{\text{C}}, \ X_{\text{A}} X_{\text{B}} X_{\text{C}} \rangle$, the stabilizers for $\dket{\text{GHZ}_n}_{\text{ABC}}$ are
\begin{align}
\label{eq:ghz_n_stabilizers}
S_{\text{GHZ}}^{\otimes n} = \langle \ Z_{\text{A}_i} Z_{\text{B}_i} I_{\text{C}_i}, \ I_{\text{A}_i} Z_{\text{B}_i} Z_{\text{C}_i}, \ X_{\text{A}_i} X_{\text{B}_i} X_{\text{C}_i} \ ; \ i = 1,2,\ldots,n \ \rangle.
\end{align}
Thus, we have identified the GHZ version of the basic properties of Bell states that was needed in the Bell pair distillation protocol.
However, the critical part of the Wilde \etal protocol was the transpose trick that formed the Bell matrix identity in Section~\ref{sec:bell_state_identity}.
When applied to stabilizer codes, this implied that each stabilizer generator $\varepsilon E(a,b)$ of Alice is transformed into the generator $\varepsilon E(a,b)^T = \varepsilon (-1)^{ab^T} E(a,b)$ (using Lemma~\ref{lem:Eab}(a)) for Bob.
Naturally, we need to determine the equivalent phenomenon for GHZ states before we can proceed to constructing a distillation protocol.

\subsection{GHZ State Matrix Identity}
\label{sec:ghz_state_identity}

In the following lemma, we generalize the Bell state matrix identity in Section~\ref{sec:bell_state_identity} to the GHZ state.

\begin{lemma}
\label{lem:ghz_state_identity}
Let $M = \sum_{x,y \in \mathbb{F}_2^n} M_{xy} \dketbra{x}{y} \in \mathbb{C}^{2^n \times 2^n}$ be any matrix acting on Alice's qubits.
Then, 
% \begin{align}
% (M_{\text{A}} \otimes I_{\text{BC}}) \dket{\text{GHZ}_n} &= \left( I_{\text{A}} \otimes \left(\ghzmap{M^T}\right)_{\text{BC}} \right) \dket{\text{GHZ}_n},
%     % \ghzmap{M}_{k-1}^T & = \ghzmap{M_{k-1}^T}
% \end{align}
% where the `GHZ-map' is defined as $\ghzmap{M} \coloneqq \sum_{x,y \in \mathbb{F}_2^n} M_{xy} \dketbra{x}{y} \otimes \dketbra{x}{y}$.
\begin{align*}
(M_{\text{A}} \otimes I_{\text{BC}}) \dket{\text{GHZ}_n}_{\text{ABC}} &= \left( I_{\text{A}} \otimes \left(\ghzmap{M^T}\right)_{\text{BC}} \right) \dket{\text{GHZ}_n}_{\text{ABC}} \ ; \ \ 
\text{`GHZ-map'} \colon M \mapsto \ghzmap{M} \coloneqq \sum_{x,y \in \mathbb{F}_2^n} M_{xy} \dketbra{x,x}{y,y} \in \mathbb{C}^{2^{2n} \times 2^{2n}}.
\end{align*}
\end{lemma}
\begin{IEEEproof}
Similar to the Bell case, we calculate
\begin{align}
(M_{\text{A}} \otimes I_{\text{BC}}) \dket{\text{GHZ}_n}_{\text{ABC}} & = \frac{1}{\sqrt{2^n}} \sum_{x,y \in \mathbb{F}_2^n} M_{xy} \dket{x}_{\text{A}} \dket{y}_{\text{B}} \dket{y}_{\text{C}} \\
  & = \frac{1}{\sqrt{2^n}} \sum_{x,y \in \mathbb{F}_2^n} \dket{x}_{\text{A}} (M^T)_{yx} \dket{y}_{\text{B}} \dket{y}_{\text{C}} \\
  & = \left( I_{\text{A}} \otimes \left(\ghzmap{M^T}\right)_{\text{BC}} \right) \dket{\text{GHZ}_n}_{\text{ABC}}.
\end{align}
This completes the proof and establishes the identity.
\end{IEEEproof}

As our next result, we prove some properties of the GHZ-map defined in the above lemma.

\begin{lemma}
The GHZ-map $M \in \mathbb{C}^{2^n \times 2^n} \mapsto \ghzmap{M} \in \mathbb{C}^{2^{2n} \times 2^{2n}}$ in Lemma~\ref{lem:ghz_state_identity} is an algebra homomorphism~\cite{Dummit-2004}: %satisfies the following properties:
\begin{enumerate}

\item[(a)] Linear: If $M = \alpha A + \beta B$, where $\alpha, \beta \in \mathbb{C}$, then $\ghzmap{M} = \alpha \ghzmap{A} + \beta \ghzmap{B}$.

\item[(b)] Multiplicative: If $M = A B$, then $\ghzmap{M} = \ghzmap{A} \ghzmap{B}$.

\item[(c)] Projector-preserving: If $M$ is a projector, then $\ghzmap{M}$ is also a projector.

\end{enumerate}
\end{lemma}
\begin{IEEEproof}
We prove these properties via the definition of the mapping.
\begin{enumerate}
    
\item[(a)] Since $M_{xy} = \dbra{x} M \dket{y} = \alpha \dbra{x} A \dket{y} + \beta \dbra{x} B \dket{y} = \alpha A_{xy} + \beta B_{xy}$, the property follows.

\item[(b)] We observe that
\begin{align}
\ghzmap{A} \ghzmap{B} & = \sum_{x,y \in \mathbb{F}_2^n} A_{xy} \dketbra{x,x}{y,y} \cdot \sum_{x',y' \in \mathbb{F}_2^n} B_{x'y'} \dketbra{x',x'}{y',y'} \\
  & = \sum_{x,y' \in \mathbb{F}_2^n} \left[ \sum_{y \in \mathbb{F}_2^n} A_{xy} B_{yy'} \right] \dketbra{x,x}{y',y'} \\
  & = \sum_{x,y' \in \mathbb{F}_2^n} (AB)_{xy'} \dketbra{x,x}{y',y'} \\
  & = \ghzmap{AB} = \ghzmap{M}.
\end{align}
% So, the map is multiplicative.

\item[(c)] This simply follows from the multiplicative property via the special case $A = B = M$.
    
\end{enumerate}
This completes the proof and establishes the said properties of the GHZ-map.
\end{IEEEproof}

We are interested in performing stabilizer measurements at Alice and deducing the effect on Bob's and Charlie's qubits.
The above properties greatly simplify the analysis, given that the code projector for a stabilizer code~\eqref{eq:stabilizer_projector} is a product of sums.
Due to the multiplicativity of the GHZ-map $M \mapsto \ghzmap{M}$, we only have to analyze the case where Alice's code has a single stabilizer generator $\varepsilon E(a,b)$, i.e., her code projector is simply $M = \frac{I_N + \varepsilon E(a,b)}{2}$, where $N = 2^n$.
Now, using linearity, we just need to determine $\ghzmap{I_N}$ and $\ghzmap{E(a,b)}$.
Then, due to Lemma~\ref{lem:Eab}(a), we have $\ghzmap{M^T} = \frac{1}{2} \left( \ghzmap{I_N} + (-1)^{ab^T} \ghzmap{E(a,b)} \right)$.

% \begin{theorem}
% \label{thm:ghz_stabilizer_measurement}
% Given $n$ copies of the GHZ state shared between Alice, Bob and Charlie, measuring $\varepsilon E(a,b)$ on Alice's $n$ qubits is equivalent to measuring $\varepsilon E(a,b)_{\text{B}}^T \otimes E(a,0)_{\text{C}} = \varepsilon (-1)^{ab^T} E(a,b)_{\text{B}} \otimes E(a,0)_{\text{C}}$ and $\{ Z_{\text{B}_i} Z_{\text{C}_i} = E([0^{\text{B}},0^{\text{C}}],[e_i^{\text{B}},e_i^{\text{C}}]) \ ; \ i = 1,2,\ldots,n \}$ on the $2n$ qubits of Bob and Charlie.
% \end{theorem}
% \begin{IEEEproof}
% See Appendix~\ref{sec:proof_ghz_stabilizer_measurement}.
% \end{IEEEproof}

\begin{theorem}
\label{thm:ghz_stabilizer_measurement}
Given $n$ copies of the GHZ state shared between Alice, Bob and Charlie, measuring $E(a,b)$ on Alice's $n$ qubits and obtaining the result $\varepsilon \in \{ \pm 1 \}$ is equivalent to measuring the following with results $+1$ on the qubits of Bob and Charlie: 
\[ \varepsilon E(a,b)_{\text{B}}^T \otimes E(a,0)_{\text{C}} = \varepsilon (-1)^{ab^T} E(a,b)_{\text{B}} \otimes E(a,0)_{\text{C}} \ \ \text{and} \ \ \{ Z_{\text{B}_i} Z_{\text{C}_i} = E(0,e_i)_{\text{B}} \otimes E(0,e_i)_{\text{C}} \ ; \ i = 1,2,\ldots,n \}, \]
where $Z_{\text{B}_i}$ (resp. $Z_{\text{C}_i}$) refers to $Z$ on $i$-th qubit of Bob (resp. Charlie), and $e_i$ is the zero vector with a $1$ in $i$-th position.
\end{theorem}
\begin{IEEEproof}
See Appendix~\ref{sec:proof_ghz_stabilizer_measurement}.
\end{IEEEproof}

\begin{example}
\normalfont
Consider $n=1$ and the case when Alice applies $M = \frac{I+Z}{2} = \frac{I+E(0,1)}{2}$, with $a = 0, b = 1$.
Then $\ghzmap{I} = \frac{I \otimes I + Z \otimes Z}{2}$ and $\ghzmap{E(0,1)^T} = (E(0,1)^T \otimes E(0,0)) \cdot \ghzmap{I} = (Z \otimes I) \cdot \ghzmap{I}$.
Therefore, the stabilizers for BC are $\langle Z \otimes I, Z \otimes Z \rangle$.

If we had an $X$-measurement for Alice, where $a=1, b=0$, then $E(a,b)^T \otimes E(a,0) = X \otimes X$.
Combined with the $Z \otimes Z$ from $\ghzmap{I}$, the qubits on BC are projected to the Bell state.

More interestingly, if we consider a $Y$-measurement for Alice, where $a=b=1$, then $E(a,b)^T \otimes E(a,0) = Y^T \otimes X = - Y \otimes X$.
Thus, assuming the measurement result is $+1$, the new BC stabilizers  are $\langle - Y \otimes X, Z \otimes Z \rangle$.
It can be verified that the post-measurement state for this case will be $\frac{(\dket{0} + \imath \dket{1})}{\sqrt{2}} \otimes \frac{(\dket{00} - \imath \dket{11})}{\sqrt{2}}$, which is stabilized by the above stabilizer. \hfill \IEEEQEDhere
\end{example}

\begin{theorem}
Given $n$ copies of the GHZ state, assume that Alice has applied some matrix to her qubits. 
Then, Bob applying another matrix $W$ to his qubits is equivalent to Charlie applying $W$ to his qubits and then exchanging his qubits with Bob.
\end{theorem}
\begin{IEEEproof}
Consider two arbitrary matrices $M = \sum_{x,y \in \mathbb{F}_2^n} M_{xy} \dketbra{x}{y}$ and $W = \sum_{u,v \in \mathbb{F}_2^n} W_{uv} \dketbra{u}{v}$.
Then, we have
\begin{align}
W_{\text{B}} M_{\text{A}} \dket{\text{GHZ}_n} & = W_{\text{B}} \sum_{x,y \in \mathbb{F}_2^n} M_{xy} \dketbra{x}{y}_{\text{A}} \cdot \frac{1}{\sqrt{2^n}} \sum_{z \in \mathbb{F}_2^n} \dket{z}_{\text{A}} \dket{z}_{\text{B}} \dket{z}_{\text{C}} \\
  & = \frac{1}{\sqrt{2^n}} \sum_{u,v \in \mathbb{F}_2^n} W_{uv} \dketbra{u}{v}_{\text{B}} \cdot \sum_{x,z \in \mathbb{F}_2^n} M_{xz} \dket{x}_{\text{A}} \dket{z}_{\text{B}} \dket{z}_{\text{C}} \\
  & = \frac{1}{\sqrt{2^n}} \sum_{u,x,z \in \mathbb{F}_2^n} M_{xz} \dket{x}_{\text{A}} W_{uz} \dket{u}_{\text{B}} \dket{z}_{\text{C}}.
\end{align}
Similarly, we have
\begin{align}
\text{Swap}_{\text{BC}} \, W_{\text{C}} M_{\text{A}} \dket{\text{GHZ}_n} & = \text{Swap}_{\text{BC}} \, \sum_{u,v \in \mathbb{F}_2^n} W_{uv} \dketbra{u}{v}_{\text{C}} \cdot \frac{1}{\sqrt{2^n}} \sum_{x,z \in \mathbb{F}_2^n} M_{xz} \dket{x}_{\text{A}} \dket{z}_{\text{B}} \dket{z}_{\text{C}} \\
  & = \text{Swap}_{\text{BC}} \, \frac{1}{\sqrt{2^n}} \sum_{u,x,z \in \mathbb{F}_2^n} M_{xz} \dket{x}_{\text{A}} \dket{z}_{\text{B}} W_{uz} \dket{u}_{\text{C}} \\
  & = \frac{1}{\sqrt{2^n}} \sum_{u,x,z \in \mathbb{F}_2^n} M_{xz} \dket{x}_{\text{A}} W_{uz} \dket{u}_{\text{B}} \dket{z}_{\text{C}} \\
  & = W_{\text{B}} M_{\text{A}} \dket{\text{GHZ}_n}.
\end{align}
This proves the identity.
\end{IEEEproof}

Note that this is result is non-trivial --- while Charlie can obviously obtain Bob's qubits initially, apply $W$, and then return them to Bob, the identity shows that Charlie can apply $W$ to his qubits and then exchange them with Bob.
While our GHZ distillation protocol does not utilize this property, we state it here since it might be useful in other settings.

\subsection{GHZ Distillation using a 3-Qubit Code}
\label{sec:ghz_3-qubit_code}

We now have all the tools to investigate a stabilizer code based GHZ distillation protocol that generalizes the Bell pair distillation protocol discussed in Section~\ref{sec:bell_distillation}.
Let us consider the $3$-qubit code with stabilizers $S = \langle YYI, IYY \rangle$ to understand the subtleties in the steps of the protocol.
First, similar to the Bell pair scenario, we have the following stabilizer group for $3$ copies of the GHZ state:
\begin{align}
\label{eq:ghz_stabilizers_std}
\mathcal{G}_3 & = \langle \ Z_{\text{A}_i} Z_{\text{B}_i} I_{\text{C}_i}, \ I_{\text{A}_i} Z_{\text{B}_i} Z_{\text{C}_i}, \ X_{\text{A}_i} X_{\text{B}_i} X_{\text{C}_i} \ ; \ i = 1,2,3 \ \rangle \\
  & = \langle \ E([0^{\text{A}},0^{\text{B}},0^{\text{C}}],[e_i^{\text{A}},e_i^{\text{B}},0^{\text{C}}]), \ E([0^{\text{A}},0^{\text{B}},0^{\text{C}}],[0^{\text{A}},e_i^{\text{B}},e_i^{\text{C}}]), \ E([e_i^{\text{A}},e_i^{\text{B}},e_i^{\text{C}}],[0^{\text{A}},0^{\text{B}},0^{\text{C}}]) \ ; \ i = 1,2,3 \ \rangle \\
  & = \langle \ E([0^{\text{A}},0^{\text{B}},0^{\text{C}}],[e_i^{\text{A}},e_i^{\text{B}},0^{\text{C}}]), \ E([0^{\text{A}},0^{\text{B}},0^{\text{C}}],[0^{\text{A}},e_i^{\text{B}},e_i^{\text{C}}]), \ - E([e_i^{\text{A}},e_i^{\text{B}},e_i^{\text{C}}],[e_i^{\text{A}},e_i^{\text{B}},0^{\text{C}}]) \ ; \ i = 1,2,3 \ \rangle \\
  & = \langle \ Z_{\text{A}_i} Z_{\text{B}_i} I_{\text{C}_i}, \ I_{\text{A}_i} Z_{\text{B}_i} Z_{\text{C}_i}, \ - Y_{\text{A}_i} Y_{\text{B}_i} X_{\text{C}_i} \ ; \ i = 1,2,3 \ \rangle.
\end{align}
For some codes, it is convenient to work with the stabilizer $-YYX$ (or other combinations of the generators in~\eqref{eq:ghz_stabilizers_std}) rather than $XXX$, so we have shown this in the last equality.

\begin{table}
\caption{\label{tab:ghz_protocol} Steps of the GHZ distillation protocol based on the $\llbr 3,1,1 \rrbr$ code defined by $S = \langle YYI, IYY \rangle$. Each `$0$' below represents $000$, and $e_i \in \mathbb{F}_2^3$ is the standard basis vector with a $1$ in the $i$-th position and zeros elsewhere. Code stabilizers are typeset in boldface. An additional left arrow indicates which row is being replaced with a code stabilizer, i.e., the first row that anticommutes with the stabilizer. Other updated rows are highlighted in gray. Classical communications: A $\rightarrow$ B, B $\rightarrow$ C.}
\centering
\begin{tabularx}{\linewidth}{c c || *{3}{C} | *{3}{C} || c}
\toprule
Step & Sign ($\pm 1$) & \multicolumn{3}{c|}{$X$-Components} & \multicolumn{3}{c||}{$Z$-Components} & Pauli Representation \\
     &     &   $A$ & $B$ & $C$   &   $A$ & $B$ & $C$   &   \\
\midrule
\midrule
     &     \multicolumn{8}{c}{Alice creates $3$ copies of the ideal GHZ state and writes down their $9$ stabilizer generators; she replaces $XXX$ with $-YYX$ for convenience}   \\
\midrule
%
%     &     &       &     &       &       &     &       &   \\
%
$(0)$ & $+1$ &    $0$ & $0$ & $0$   &   $e_1$ & $e_1$ & $0$    & $Z_{\text{A}_1} Z_{\text{B}_1}$ \\
     & $+1$ &    $0$ & $0$ & $0$   &   $e_2$ & $e_2$ & $0$    & $Z_{\text{A}_2} Z_{\text{B}_2}$ \\
     & $+1$ &    $0$ & $0$ & $0$   &   $e_3$ & $e_3$ & $0$    & $Z_{\text{A}_3} Z_{\text{B}_3}$ \\
\cmidrule(lr){2-9}
     & $+1$ &    $0$ & $0$ & $0$   &   $0$ & $e_1$ & $e_1$    & $Z_{\text{B}_1} Z_{\text{C}_1}$ \\
     & $+1$ &    $0$ & $0$ & $0$   &   $0$ & $e_2$ & $e_2$    & $Z_{\text{B}_2} Z_{\text{C}_2}$ \\
     & $+1$ &    $0$ & $0$ & $0$   &   $0$ & $e_3$ & $e_3$    & $Z_{\text{B}_3} Z_{\text{C}_3}$ \\
\cmidrule(lr){2-9}
     & $-1$ &   $e_1$ & $e_1$ & $e_1$   &   $e_1$ & $e_1$ & $0$    & $- Y_{\text{A}_1} Y_{\text{B}_1} X_{\text{C}_1}$ \\
     & $-1$ &   $e_2$ & $e_2$ & $e_2$   &   $e_2$ & $e_2$ & $0$    & $- Y_{\text{A}_2} Y_{\text{B}_2} X_{\text{C}_2}$ \\
     & $-1$ &   $e_3$ & $e_3$ & $e_3$   &   $e_3$ & $e_3$ & $0$    & $- Y_{\text{A}_3} Y_{\text{B}_3} X_{\text{C}_3}$ \\
     &     &       &     &       &       &     &       &   \\
\midrule
     &     \multicolumn{8}{c}{Alice measures code stabilizers on her qubits, updates first $3$ rows, and rewrites $-YYX$ entries to introduce joint BC stabilizers by Theorem~\ref{thm:ghz_stabilizer_measurement}}   \\
\midrule
%
%     &     &       &     &       &       &     &       &   \\
%
$(1)$ & $\boldsymbol{\varepsilon_1^{\text{A}}}$ &    $\boldsymbol{110}$ & $\boldsymbol{000}$ & $\boldsymbol{000}$   &   $\boldsymbol{110}$ & $\boldsymbol{000}$ & $\boldsymbol{000}$    & \qquad \ \ $\boldsymbol{\varepsilon_1^{\text{A}} \, Y_{\text{A}_1} Y_{\text{A}_2} I_{\text{A}_3}} \ \ \boldsymbol{\longleftarrow}$ \\
\rowcolor{lightgray}
     & $+1$ &    $0$ & $0$ & $0$   &   $e_1+e_2$ & $e_1+e_2$ & $0$    & $Z_{\text{A}_2} Z_{\text{B}_2} Z_{\text{A}_1} Z_{\text{B}_1}$ \\
     & $+1$ &    $0$ & $0$ & $0$   &   $e_3$ & $e_3$ & $0$    & $Z_{\text{A}_3} Z_{\text{B}_3}$ \\
\cmidrule(lr){2-9}
     & $+1$ &    $0$ & $0$ & $0$   &   $0$ & $e_1$ & $e_1$    & $Z_{\text{B}_1} Z_{\text{C}_1}$ \\
     & $+1$ &    $0$ & $0$ & $0$   &   $0$ & $e_2$ & $e_2$    & $Z_{\text{B}_2} Z_{\text{C}_2}$ \\
     & $+1$ &    $0$ & $0$ & $0$   &   $0$ & $e_3$ & $e_3$    & $Z_{\text{B}_3} Z_{\text{C}_3}$ \\
\cmidrule(lr){2-9}
\rowcolor{lightgray}
     & $\varepsilon_1^{\text{A}}$ &   $0$ & $e_1+e_2$ & $e_1+e_2$   &   $0$ & $e_1+e_2$ & $0$    & $\varepsilon_1^{\text{A}} \, (Y_{\text{B}_1} Y_{\text{B}_2} I_{\text{B}_3}) (X_{\text{C}_1} X_{\text{C}_2} I_{\text{C}_3})$ \\
     & $-1$ &   $e_2$ & $e_2$ & $e_2$   &   $e_2$ & $e_2$ & $0$    & $- Y_{\text{A}_2} Y_{\text{B}_2} X_{\text{C}_2}$ \\
     & $-1$ &   $e_3$ & $e_3$ & $e_3$   &   $e_3$ & $e_3$ & $0$    & $- Y_{\text{A}_3} Y_{\text{B}_3} X_{\text{C}_3}$ \\
     &     &       &     &       &       &     &       &   \\
\midrule
%
%     &     &       &     &       &       &     &       &   \\
%
$(2)$ & $\boldsymbol{\varepsilon_1^{\text{A}}}$ &    $\boldsymbol{110}$ & $\boldsymbol{000}$ & $\boldsymbol{000}$   &   $\boldsymbol{110}$ & $\boldsymbol{000}$ & $\boldsymbol{000}$    & $\boldsymbol{\varepsilon_1^{\text{A}} \, Y_{\text{A}_1} Y_{\text{A}_2} I_{\text{A}_3}}$ \\
     & $\boldsymbol{\varepsilon_2^{\text{A}}}$ &    $\boldsymbol{011}$ & $\boldsymbol{000}$ & $\boldsymbol{000}$   &   $\boldsymbol{011}$ & $\boldsymbol{000}$ & $\boldsymbol{000}$    & \qquad \ \ $\boldsymbol{\varepsilon_2^{\text{A}} \, I_{\text{A}_1} Y_{\text{A}_2} Y_{\text{A}_3}} \ \ \boldsymbol{\longleftarrow}$ \\
\rowcolor{lightgray}
     & $+1$ &    $0$ & $0$ & $0$   &   $e_1+e_2+e_3$ & $e_1+e_2+e_3$ & $0$    & $Z_{\text{A}_3} Z_{\text{B}_3} Z_{\text{A}_2} Z_{\text{B}_2} Z_{\text{A}_1} Z_{\text{B}_1}$ \\
\cmidrule(lr){2-9}
     & $+1$ &    $0$ & $0$ & $0$   &   $0$ & $e_1$ & $e_1$    & $Z_{\text{B}_1} Z_{\text{C}_1}$ \\
     & $+1$ &    $0$ & $0$ & $0$   &   $0$ & $e_2$ & $e_2$    & $Z_{\text{B}_2} Z_{\text{C}_2}$ \\
     & $+1$ &    $0$ & $0$ & $0$   &   $0$ & $e_3$ & $e_3$    & $Z_{\text{B}_3} Z_{\text{C}_3}$ \\
\cmidrule(lr){2-9}
     & $\varepsilon_1^{\text{A}}$ &   $0$ & $e_1+e_2$ & $e_1+e_2$   &   $0$ & $e_1+e_2$ & $0$    & $\varepsilon_1^{\text{A}} \, (Y_{\text{B}_1} Y_{\text{B}_2} I_{\text{B}_3}) (X_{\text{C}_1} X_{\text{C}_2} I_{\text{C}_3})$ \\
\rowcolor{lightgray}
     & $\varepsilon_2^{\text{A}}$ &   $0$ & $e_2+e_3$ & $e_2+e_3$   &   $0$ & $e_2+e_3$ & $0$    & $\varepsilon_2^{\text{A}} \, (I_{\text{B}_1} Y_{\text{B}_2} Y_{\text{B}_3}) (I_{\text{C}_1} X_{\text{C}_2} X_{\text{C}_3})$ \\
     & $-1$ &   $e_3$ & $e_3$ & $e_3$   &   $e_3$ & $e_3$ & $0$    & $- Y_{\text{A}_3} Y_{\text{B}_3} X_{\text{C}_3}$ \\
     &     &       &     &       &       &     &       &   \\
\midrule
     &     \multicolumn{8}{c}{Alice applies $P^\dagger = \begin{bsmallmatrix} 1 & 0 \\ 0 & -\imath \end{bsmallmatrix}$ to all $3$ qubits of C system to change $X$'s to $Y$'s, then sends B's and C's qubits to Bob over noisy Pauli channel}   \\
\midrule
%
%     &     &       &     &       &       &     &       &   \\
%
$(3)$ & $\boldsymbol{\varepsilon_1^{\text{A}}}$ &    $\boldsymbol{110}$ & $\boldsymbol{000}$ & $\boldsymbol{000}$   &   $\boldsymbol{110}$ & $\boldsymbol{000}$ & $\boldsymbol{000}$    & $\boldsymbol{\varepsilon_1^{\text{A}} \, Y_{\text{A}_1} Y_{\text{A}_2} I_{\text{A}_3}}$ \\
     & $\boldsymbol{\varepsilon_2^{\text{A}}}$ &    $\boldsymbol{011}$ & $\boldsymbol{000}$ & $\boldsymbol{000}$   &   $\boldsymbol{011}$ & $\boldsymbol{000}$ & $\boldsymbol{000}$    & $\boldsymbol{\varepsilon_2^{\text{A}} \, I_{\text{A}_1} Y_{\text{A}_2} Y_{\text{A}_3}}$ \\
\rowcolor{lightgray}
     & $\eta$ &    $0$ & $0$ & $0$   &   $e_1+e_2+e_3$ & $e_1+e_2+e_3$ & $0$    & $\eta \, Z_{\text{A}_3} Z_{\text{B}_3} Z_{\text{A}_2} Z_{\text{B}_2} Z_{\text{A}_1} Z_{\text{B}_1}$ \\
\cmidrule(lr){2-9}
\rowcolor{lightgray}
     & $\nu_1$ &    $0$ & $0$ & $0$   &   $0$ & $e_1$ & $e_1$    & $\nu_1 \, Z_{\text{B}_1} Z_{\text{C}_1}$ \\
\rowcolor{lightgray}
     & $\nu_2$ &    $0$ & $0$ & $0$   &   $0$ & $e_2$ & $e_2$    & $\nu_2 \, Z_{\text{B}_2} Z_{\text{C}_2}$ \\
\rowcolor{lightgray}
     & $\nu_3$ &    $0$ & $0$ & $0$   &   $0$ & $e_3$ & $e_3$    & $\nu_3 \, Z_{\text{B}_3} Z_{\text{C}_3}$ \\
\cmidrule(lr){2-9}
\rowcolor{lightgray}
     & $\mu_1 \varepsilon_1^{\text{A}}$ &   $\boldsymbol{000}$ & $\boldsymbol{110}$ & $\boldsymbol{110}$   &   $\boldsymbol{000}$ & $\boldsymbol{110}$ & $\boldsymbol{110}$    & \tiny $\boldsymbol{\mu_1 \varepsilon_1^{\text{A}} \, (Y_{\text{B}_1} Y_{\text{B}_2} I_{\text{B}_3}) \, (Y_{\text{C}_1} Y_{\text{C}_2} I_{\text{C}_3})}$ \\
\rowcolor{lightgray}
     & $\mu_2 \varepsilon_2^{\text{A}}$ &   $\boldsymbol{000}$ & $\boldsymbol{011}$ & $\boldsymbol{011}$   &   $\boldsymbol{000}$ & $\boldsymbol{011}$ & $\boldsymbol{011}$    & \tiny $\boldsymbol{\mu_2 \varepsilon_2^{\text{A}} \, (I_{\text{B}_1} Y_{\text{B}_2} Y_{\text{B}_3}) \, (I_{\text{C}_1} Y_{\text{C}_2} Y_{\text{C}_3})}$ \\
\rowcolor{lightgray}
     & $\mu_3$ &   $e_3$ & $e_3$ & $e_3$   &   $e_3$ & $e_3$ & $e_3$    & $\mu_3 \, Y_{\text{A}_3} Y_{\text{B}_3} Y_{\text{C}_3}$ \\
     &     &       &     &       &       &     &       &   \\
\midrule
     &     \multicolumn{8}{c}{Bob measures the $5$ joint BC stabilizers, corrects errors (signs), measures B stabilizers, and sends C's qubits to Charlie over noisy Pauli channel}   \\
\midrule
%
%     &     &       &     &       &       &     &       &   \\
%
% \rowcolor{lightgray}
$(4)$ & $\boldsymbol{\varepsilon_1^{\text{A}}}$ &    $\boldsymbol{110}$ & $\boldsymbol{000}$ & $\boldsymbol{000}$   &   $\boldsymbol{110}$ & $\boldsymbol{000}$ & $\boldsymbol{000}$    & $\boldsymbol{\varepsilon_1^{\text{A}} \, Y_{\text{A}_1} Y_{\text{A}_2} I_{\text{A}_3}}$ \\
     & $\boldsymbol{\varepsilon_2^{\text{A}}}$ &    $\boldsymbol{011}$ & $\boldsymbol{000}$ & $\boldsymbol{000}$   &   $\boldsymbol{011}$ & $\boldsymbol{000}$ & $\boldsymbol{000}$    & $\boldsymbol{\varepsilon_2^{\text{A}} \, I_{\text{A}_1} Y_{\text{A}_2} Y_{\text{A}_3}}$ \\
     & $+1$ &    $0$ & $0$ & $0$   &   $e_1+e_2+e_3$ & $e_1+e_2+e_3$ & $0$    & \qquad \quad $\overline{Z}_{\text{A}} \overline{Z}_{\text{B}} \overline{I}_{\text{C}}$ (logical) \\
\cmidrule(lr){2-9}
     & $\boldsymbol{\varepsilon_1^{\text{B}}}$ &    $\boldsymbol{000}$ & $\boldsymbol{110}$ & $\boldsymbol{000}$   &   $\boldsymbol{000}$ & $\boldsymbol{110}$ & $\boldsymbol{000}$    & \qquad \ \ $\boldsymbol{\varepsilon_1^{\text{B}} \, Y_{\text{B}_1} Y_{\text{B}_2} I_{\text{B}_3}} \ \ \boldsymbol{\longleftarrow}$ \\
     & $\boldsymbol{\varepsilon_2^{\text{B}}}$ &    $\boldsymbol{000}$ & $\boldsymbol{011}$ & $\boldsymbol{000}$   &   $\boldsymbol{000}$ & $\boldsymbol{011}$ & $\boldsymbol{000}$    & \qquad \ \ $\boldsymbol{\varepsilon_2^{\text{B}} \, I_{\text{B}_1} Y_{\text{B}_2} Y_{\text{B}_3}} \ \ \boldsymbol{\longleftarrow}$ \\
\rowcolor{lightgray}
     & $\beta$ &    $0$ & $0$ & $0$   &   $0$ & $e_1+e_2+e_3$ & $e_1+e_2+e_3$    & \qquad \quad $\beta \, \overline{I}_{\text{A}} \overline{Z}_{\text{B}} \overline{Z}_{\text{C}}$ (logical) \\
\cmidrule(lr){2-9}
\rowcolor{lightgray}
     & $\boldsymbol{\alpha_1 \varepsilon_1^{\text{A}} \varepsilon_1^{\text{B}}}$ &   $\boldsymbol{000}$ & $\boldsymbol{000}$ & $\boldsymbol{110}$   &   $\boldsymbol{000}$ & $\boldsymbol{000}$ & $\boldsymbol{110}$    & $\boldsymbol{\alpha_1 \varepsilon_1^{\text{A}} \varepsilon_1^{\text{B}} \, Y_{\text{C}_1} Y_{\text{C}_2} I_{\text{C}_3}}$ \\
\rowcolor{lightgray}
     & $\boldsymbol{\alpha_2 \varepsilon_2^{\text{A}} \varepsilon_2^{\text{B}}}$ &   $\boldsymbol{000}$ & $\boldsymbol{000}$ & $\boldsymbol{011}$   &   $\boldsymbol{000}$ & $\boldsymbol{000}$ & $\boldsymbol{011}$    & $\boldsymbol{\alpha_2 \varepsilon_2^{\text{A}} \varepsilon_2^{\text{B}} \, I_{\text{C}_1} Y_{\text{C}_2} Y_{\text{C}_3}}$ \\
\rowcolor{lightgray}
     & $\alpha_3$ &   $e_3$ & $e_3$ & $e_3$   &   $e_3$ & $e_3$ & $e_3$    & \qquad \quad $\alpha_3 \, \overline{X}_{\text{A}} \overline{X}_{\text{B}} \overline{X}_{\text{C}}$ (logical) \\
     &     &       &     &       &       &     &       &   \\
\midrule
     &     \multicolumn{8}{c}{Charlie measures his code stabilizers, corrects errors, inverts the encoding unitary; Alice and Bob invert their encoding after sending qubits}   \\
\bottomrule
\bottomrule
\end{tabularx}
\end{table}

\begin{figure}
 \centering
  \includegraphics[scale=0.81,keepaspectratio]{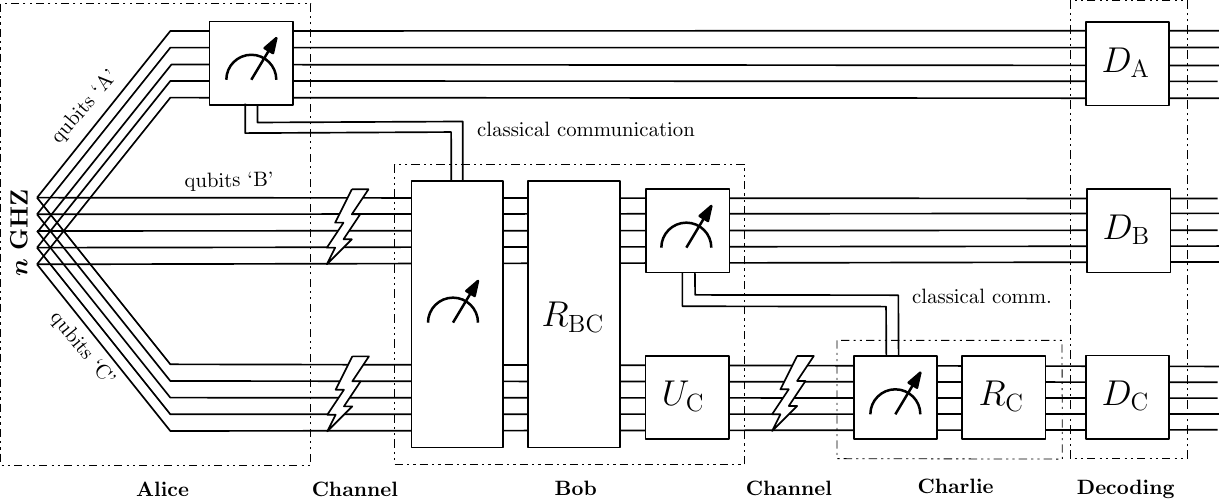}
  \caption{A QEC based protocol to distill GHZ states shared between three parties. Each of the $n$ GHZ states initially prepared by Alice has a qubit marked `A', a qubit marked `B', and a qubit marked `C'. The channel is assumed to introduce Pauli errors. Both classical communications are noiseless and involve the communication of stabilizers, syndrome, and logical Pauli operators. The block $U_{\text{C}}$ represents Bob's application of an appropriate diagonal Clifford unitary on qubits `C', as prescribed in Appendix~\ref{sec:ghz_implementation}. The recovery blocks $R_{\text{BC}}$ and $R_{\text{C}}$ represent Pauli corrections applied by Bob on qubits `BC' (B and C) and by Charlie on qubits `C' respectively. The decoding blocks represent (locally) inverting the respective encoding unitaries of the codes of Alice, Bob, and Charlie, and this step can be asynchronous (unlike what the picture may suggest).}
%   \caption{QEC based GHZ-distillation protocol. Alice holds $n$ GHZ qubits and sends $2n$ qubits  to Bob over noisy quantum channel after measuring QEC stabilizers on her qubits. Alice also sends the syndrome information to Bob. Bob performs measurement on $2n$ qubits, corrects for the quantum errors on the $2n$ qubit channel; Bob then measures stabilizers of the same QEC on the same $n$ qubits  and forwards $n$ qubits to Charlie who combines this information and the syndrome measurement information on his qubits, correcting the errors on the $n$ qubit channel. In this way we get entangled logical qubits in the GHZ state.}
   \label{fig:ghz-entanglement}
\end{figure}

\begin{algorithm} % The number tells where the line numbering should start
\DontPrintSemicolon
%\SetAlgoNoLine
\SetAlgoLined
% \KwResult{Logical qubits of the stabilizer codes on Alice and Bob form a Bell state \;}
%\KwResult{Alice and Bob either share $k$ perfect GHZ states or at least one of the $k$ states has an unknown Pauli error}
\SetKwInOut{Input}{Input}
\SetKwInOut{Output}{Output}
\Input{$n$ GHZ states $\ket{\text{GHZ}}^{\otimes n}$ at Alice, % one qubit of each pair will be sent to Bob over a noisy channel; 
$\llbr n,k,d \rrbr$ stabilizer code $\mathcal{Q}(S)$ defined by a stabilizer group $S$}
%\Output{Logical qubits of the QEC $\mathcal{C}$ form a Bell pair}
\Output{$k$ GHZ states of higher quality shared between Alice and Bob if channel introduces a correctable error}
Initialization: Rearrange the $3n$ qubits in $\ket{\text{GHZ}}^{\otimes n}$ to obtain $\ket{\text{GHZ}_n}$~\eqref{eq:ghz_state_rearranged} for processing by Alice and Bob, respectively\;
  \;
 Alice \;
 (a) measures the stabilizer generators $\{ E(a_i,b_i) \, ; \, i = 1,2,\ldots,r=n-k \}$ on her $n$ qubits and obtains syndrome $\{ \varepsilon_i^{\text{A}} \}$, \;
%  (a) considers the code $\mathcal{C}$ with stabilizer generators $S_1. S_2, \cdots, S_{n-k}$ .\;
%  (b) performs a suitable Pauli correction to her qubits to bring them to the code space of $\mathcal{Q}(S)$, \;
%  (b) measures all the stabilizer generators $S_1, S_2, \cdots, S_{n-k}$ on $n$ qubits.\;
%  (c) qubits are in the code $\mathcal{C}$.\;
 %(d) performs Clifford operations on Charlie's qubits \;
 (b) sends the remaining $2n$ qubits to Bob over a noisy quantum channel, \;
 %\While{  }  
 %{  Alice considers an $\llbr n,k,d \rrbr$ QEC $\mathcal{C}$ with stabilizer generators $S_1. S_2, \cdots, S_{n-k}$  \;
 %\For{$i=1$ \KwTo $n-k$}{
%    Alice measures $S_i$ \;
  %  \uIf{All measurement outcomes are +1 }{
  %   Alice's qubits are in the code $\mathcal{C}$ \;
  %}
  %\uElseIf{Some measurement outcomes are -1 %}{
   % Alice performs local Pauli corrections \;
    %Alice's qubits are in the code $\mathcal{C}$ \;
    %      }
%}
%}
%Alice sends the remaining $n$ qubits to Bob over a noisy quantum channel\;
 (c) sends the stabilizers, syndrome and logical Pauli operators to Bob over a perfect classical channel. \;
% (e) sends the syndrome information of the stabilizer measurements to Bob over a perfect classical channel \;
 \;
{  Bob \;
 (a) uses Theorem~\ref{thm:ghz_stabilizer_measurement} to define the $2n$-qubit joint BC code and measures all the $(2n-k)$ stabilizer generators 
 $$\{ \varepsilon_i^{\text{A}} E(a_i,b_i)_{\text{B}}^T \otimes E(a_i,0)_{\text{C}} \, , \, Z_{\text{B}_j} Z_{\text{C}_j} = E(0,e_j)_{\text{B}} \otimes E(0,e_j)_{\text{C}} \, ; \, i = 1,2,\ldots,r=n-k \, , \, j = 1,2,\ldots,n \}$$ 
 on the received $2n$ qubits, \;
 (b) performs necessary Pauli corrections on all qubits to bring them to the code space of the joint BC code, \;
 (c) measures the stabilizer generators $\{ E(a_i,b_i) \, ; \, i = 1,2,\ldots,r=n-k \}$ on the $n$ qubits of subsystem B and obtains syndrome $\{ \varepsilon_i^{\text{B}} \}$; for purely $Z$-type stabilizers $E(0,b_i)$ the sign is $\varepsilon_i^{\text{A}}$, so we set $\varepsilon_i^{\text{B}} \coloneqq +1$ for them, \;
% (a) considers the code $\mathcal{C}$ with %same $\llbr n,k,d \rrbr$ QEC $\mathcal{C}$  
% stabilizer generators $S_1. S_2, \cdots, S_{n-k}$  \;
% (b) measures all the stabilizer generators $S_1, S_2, \cdots, S_{n-k}$ on the $n$ qubits \;
(d) sends the stabilizers, syndrome and logical Pauli operators to Charlie over a perfect classical channel, \;
(e) performs appropriate (see Appendix~\ref{sec:diagonal_clifford}) local diagonal Clifford on qubits C, \;
(f) sends qubits C to Charlie over a noisy Pauli channel. \;
 %  \uIf{All measurement outcomes are +1 }{
 %    Bob's qubits are in the code $\mathcal{C}$ \;
 % }
 % \uElseIf{Some measurement outcomes are -1 }{
 %   Bob performs local Pauli corrections \;
 %   Bob's qubits are in the code $\mathcal{C}$ \;
 %         }
% \For{$i=1$ \KwTo $n-k$}{
%    Bob measures $S_i$ \;
%    \uIf{Measurement outcome=+1 }{
%     $i=i+1$ \;
%  }
%  \uElseIf{Measurement outcome =-1 }{
 %   Perform local Pauli correction \;
 %    $i=i+1$ \;
 %         }
 %}
  \;
 Charlie \; 
 (a) uses $\mathcal{Q}(S)$, Theorem~\ref{thm:ghz_stabilizer_measurement}, and Bob's syndrome to determine the signs $\varepsilon_i^{\text{A}} \varepsilon_i^{\text{B}} (-1)^{a_i b_i^T}$ of his stabilizers, and then measures the generators $\{ \varepsilon_i^{\text{A}} \varepsilon_i^{\text{B}} (-1)^{a_i b_i^T} E(a_i,b_i) \, ; \, i = 1,2,\ldots,r=n-k \}$ on his $n$ qubits, \;
 (b) performs the necessary Pauli corrections on all qubits to bring them to the code space of his code. \;
  \;
 // If the channel error was correctable, triples of logical qubits of Alice's, Bob's and Charlie's codes form $k$ GHZ states \;
 // If channel error was NOT correctable, some triple of logical qubits form a GHZ state with an unknown Pauli error \;
 Alice, Bob, and Charlie respectively apply the inverse of the encoding unitary for their code on their $n$ qubits\;
 // The encoding unitary is determined by the logical Pauli operators obtained from Algorithm~\ref{algo:logical_paulis_ghz_msmt} \;
%  Alice and Bob share $k$ Bell pairs and $n-k$ ancillary qubits \;
}
%   \caption{Algorithm to create $k$ Bell pairs of higher quality from $n$ Bell pairs, using an $\llbr n,k,d \rrbr$ stabilizer code}
  \caption{Algorithm to convert $n$ GHZ states into $k$ GHZ states of higher quality, using an $\llbr n,k,d \rrbr$ stabilizer code}
 \label{algo:algo_ghz}
\end{algorithm}

\begin{algorithm}

\DontPrintSemicolon
\SetAlgoLined

%\KwResult{Generators of logical $X$ and logical $Z$ generators for the given stabilizer code}

\SetKwInOut{Input}{Input}
\SetKwInOut{Output}{Output}

\Input{An $\llbr n,k,d \rrbr$ stabilizer code defined by its stabilizer generators $\{ \varepsilon_i E(a_i,b_i) \, ; \, i=1,2,\ldots,r=n-k \}$}

\Output{The logical $X$ generators $\{ \nu_j E(c_j,d_j) \, ; \, j=1,\ldots,k \}$ and logical $Z$ generators $\{ E(0,f_j) \, ; \, j=1,\ldots,k \}$}

Initialization: Form a $r \times (2n+1)$ binary parity-check matrix $H$ for the code, whose rows are $[a_i,b_i, \ \varepsilon_i]$. 
Preprocess the matrix so that its first $2n$ columns take the form $H_{1 : 2n} = \begin{bsmallmatrix} 0 & H_Z \\ H_1 & H_2 \end{bsmallmatrix}$, where $H_Z$ is a $r_Z \times n$ matrix of full rank, and $H_1$ is a $r_X \times n$ matrix of full rank ($r_X + r_Z = r = n-k$). 
The rows of $H_Z$ provide the generators for all purely $Z$-type stabilizers of the code. 
While performing row operations on $H_{1 : 2n}$, care must be taken to adhere to Pauli multiplication arithmetic (Lemma~\ref{lem:Eab}(b)). \;

\;

Simulate the creation of $n$ copies of the GHZ state as follows.
Create a $2n \times (6n+1)$ GHZ stabilizer matrix $S_{\text{GHZ}}$ whose first $n$ rows take the form $[0, 0, 0,\ e_i, e_i, 0,\ +1]$ and the second $n$ rows take the form $[e_i, e_i, e_i,\ 0, 0, 0,\ +1]$, where $i=1,2,\ldots,n$.
This matrix is almost the same as Step (0) in Table~\ref{tab:ghz_protocol}, but we have omitted the middle section. \;

\;

\For{$p=1$ to $r$}
{
  Simulate the measurement of $H^{(p)}$, the $p$-th row of $H$, on subsystem A of the GHZ states, using Section~\ref{sec:stabilizer_formalism}: \;
  Replace the first anticommuting row of $S_{\text{GHZ}}$ with $H^{(p)}$ and multiply subsequent anticommuting rows by $H^{(p)}$, using Lemma~\ref{lem:Eab}(b) \;
}

\;

\For{$q$ in the set of non-replaced rows of $S_{\text{GHZ}}$ with (row) index at most $n$}
{
  \eIf{$S_{\text{GHZ}}^{(q)}$ (only the $2n$ columns of subsystem A) is linearly independent from all rows of $H$}{
    Define a new logical $Z$ generator $E(0,f_j)$ from the A-columns of $S_{\text{GHZ}}^{(q)}$, with sign $+1$ \;
    Append $[0,\ f_j,\ +1]$ as a new row to $H$ \;
  }{
    continue \;
  }
}

// Now, $H$ has $n$ rows where the last $k$ rows correspond to the logical $Z$ generators $E(0,f_j)$ \;

\;

\For{$q'$ in the set of non-replaced rows of $S_{\text{GHZ}}$ with (row) index at least $(n+1)$}
{
  \eIf{$S_{\text{GHZ}}^{(q')}$ (only the $2n$ columns of subsystem A) is linearly independent from all rows of $H$}{
    Define a new logical $X$ generator $\nu_j E(c_j,d_j)$ from the A-columns of $S_{\text{GHZ}}^{(q')}$, with sign $\nu_j$ (last column of $S_{\text{GHZ}}^{(q')}$) \;
    Append $[c_j,\ d_j,\ \nu_j]$ as a new row to $H$ \;
  }{
    continue \;
  }
}

// Now, we have $k$ logical $Z$ and logical $X$ generators, but they might not pair up appropriately \;

\;

Compute the $k \times k$ symplectic inner product matrix $T$ with entries $T_{ij} = \syminn{[0,f_i]}{[c_j,d_j]}$ for $i,j \in \{1,\ldots,k\}$ \;

\eIf{$T$ is not the $k \times k$ identity matrix}{
  Compute the binary inverse $T^{-1}$ of $T$ \;
  Form a $k \times n$ matrix $F$ whose rows are $f_j$ \;
  Define the new $f_j$'s as the rows of $T^{-1} F$ \;
}{
  Retain the definitions of logical $Z$ generators $E(0,f_j)$ and logical $X$ generators $\nu_j E(c_j,d_j)$ \;
}

\;

return $\{ \overline{Z}_j = E(0,f_j) \, , \, \overline{X}_j = \nu_j E(c_j,d_j) \, ; \, j = 1,2,\ldots,k \}$ \;

\caption{Algorithm to generate logical Paulis of a stabilizer code through GHZ measurements (see Appendix~\ref{sec:logical_paulis_ghz_msmt})}
\label{algo:logical_paulis_ghz_msmt}

\end{algorithm}

The steps of the protocol, with this particular $\llbr 3,1,1 \rrbr$ code as an example, are shown in Table~\ref{tab:ghz_protocol}.
Again, we use the stabilizer formalism for measurements from Section~\ref{sec:stabilizer_formalism}.
We will explain each step below and discuss the potential subtleties that can arise.
It could be useful to imagine the three parties as being three nodes A --- B --- C on a linear network chain.
For other network topologies, the protocol can be modified appropriately, as we discuss later in Section~\ref{sec:ghz_variations}. \\ 
% \narayanan{Need to future-reference to the discussion of the protocol where Alice can measure both her qubits as well as Bob's qubits before transmitting the qubits.} \\
% As before, Alice measures her two stabilizer generators followed by Bob.
\begin{enumerate}

\item[(0)] Alice locally prepares $3$ copies of the perfect GHZ state and groups her qubits together for further processing. She keeps aside the grouped qubits of Bob's and Charlie's but does not send those to them yet. She also writes down the parity check matrix for the $9$ qubits, based on only GHZ stabilizers, along with signs, as shown in Step (0) of Table~\ref{tab:ghz_protocol}. \\

\item[(1)] Alice measures the stabilizer $Y_{\text{A}_1} Y_{\text{A}_2} I_{\text{A}_3} = E([(e_1+e_2)^{\text{A}},0^{\text{B}},0^{\text{C}}],[(e_1+e_2)^{\text{A}},0^{\text{B}},0^{\text{C}}])$ and the group $\mathcal{G}_3$ gets updated as shown in Step (1) of Table~\ref{tab:ghz_protocol}, assuming that the measurement result is $\varepsilon_1^{\text{A}} \in \{ \pm 1 \}$.
%follows, assuming a measurement result $+1$: 
% \begin{align}
% \mathcal{G}_3^{(1)} & = \langle \ {\color{red} \{ Y_{\text{A}_1} Y_{\text{A}_2} I_{\text{A}_3} \} }, \ {\color{blue} \{ E([0^{\text{A}},0^{\text{B}},0^{\text{C}}],[(e_1+e_2)^{\text{A}},(e_1+e_2)^{\text{B}},0^{\text{C}}]) \} }, \ E([0^{\text{A}},0^{\text{B}},0^{\text{C}}],[e_3^{\text{A}},e_3^{\text{B}},0^{\text{C}}]), \nonumber \\
% %
%   & \hspace{1cm} E([0^{\text{A}},0^{\text{B}},0^{\text{C}}],[0^{\text{A}},e_i^{\text{B}},e_i^{\text{C}}]), \ - E([e_i^{\text{A}},e_i^{\text{B}},e_i^{\text{C}}],[e_i^{\text{A}},e_i^{\text{B}},0^{\text{C}}]) \ ; \ i = 1,2,3 \ \rangle.
% \end{align}
Based on the stabilizer formalism (Section~\ref{sec:stabilizer_formalism}), the measured stabilizer replaces the first row (as indicated by the left arrow) and the second row is multiplied with the previous first row.
For visual clarity, code stabilizer rows are boldfaced and binary vectors are written out in full.
Furthermore, as per Theorem~\ref{thm:ghz_stabilizer_measurement}, this measurement of $E(e_1+e_2, e_1+e_2)$ by Alice should imply that %the stabilizer 
$$ \varepsilon_1^{\text{A}} E(e_1+e_2, e_1+e_2)_{\text{B}}^T \otimes E(e_1+e_2, 0)_{\text{C}} = \varepsilon_1^{\text{A}} E(e_1+e_2, e_1+e_2)_{\text{B}} \otimes E(e_1+e_2, 0)_{\text{C}} $$ 
automatically belongs to the (new) stabilizer group. %$\mathcal{G}_3^{(1)}$.
Indeed, this element can be produced by multiplying the elements $- E([e_i^{\text{A}},e_i^{\text{B}},e_i^{\text{C}}],[e_i^{\text{A}},e_i^{\text{B}},0^{\text{C}}])$ for $i=1,2$ along with $Y_{\text{A}_1} Y_{\text{A}_2} I_{\text{A}_3} = E([(e_1+e_2)^{\text{A}},0^{\text{B}},0^{\text{C}}],[(e_1+e_2)^{\text{A}},0^{\text{B}},0^{\text{C}}])$ using Lemma~\ref{lem:Eab}(b).
This is exactly how the seventh row gets updated. \\

\item[(2)] Alice measures the second stabilizer $I_{\text{A}_1} Y_{\text{A}_2} Y_{\text{A}_3} = E([(e_2+e_3)^{\text{A}},0^{\text{B}},0^{\text{C}}],[(e_2+e_3)^{\text{A}},0^{\text{B}},0^{\text{C}}])$ and the group gets updated as shown in Step (2) of Table~\ref{tab:ghz_protocol}. %follows, assuming a measurement result $+1$:
% \begin{align}
% \mathcal{G}_3^{(2)} & = \langle \ {\color{red} Y_{\text{A}_1} Y_{\text{A}_2} I_{\text{A}_3} }, \ {\color{blue} \{ E([0^{\text{A}},0^{\text{B}},0^{\text{C}}],[(e_1+e_2+e_3)^{\text{A}},(e_1+e_2+e_3)^{\text{B}},0^{\text{C}}]) \} }, \ {\color{red} \{ I_{\text{A}_1} Y_{\text{A}_2} Y_{\text{A}_3} \} }, \nonumber \\
% %
%   & \hspace{1cm} E([0^{\text{A}},0^{\text{B}},0^{\text{C}}],[0^{\text{A}},e_i^{\text{B}},e_i^{\text{C}}]), \ - E([e_i^{\text{A}},e_i^{\text{B}},e_i^{\text{C}}],[e_i^{\text{A}},e_i^{\text{B}},0^{\text{C}}]) \ ; \ i = 1,2,3 \ \rangle.
% \end{align}
The procedure is very similar to that in Step (1). \\

Since Alice has measured all her stabilizer generators, and the stabilizer formalism preserves the commutativity of the elements in the group, the third row in the first block of 3 rows must necessarily commute with Alice's stabilizers.
Thus, the Alice component of the third row must form a logical operator for Alice's code, and we define it to be the logical $Z$ operator, i.e., $\overline{Z}_{\text{A}} = ZZZ = E(0,e_1+e_2+e_3)$.
We will see shortly that Bob's qubits get the same code (possibly with sign changes for the stabilizers), so this third row can be written as the logical GHZ stabilizer $\overline{Z}_{\text{A}} \overline{Z}_{\text{B}} \overline{I}_{\text{C}}$. \\

This phenomenon also generalizes to any $\llbr n,k,d \rrbr$ stabilizer code, with some caveats when the code has some purely $Z$-type stabilizers, and we determine the logical $Z$ operators either after Alice's set of measurements or apriori using some linear algebraic arguments (see Appendix~\ref{sec:logical_paulis_ghz_msmt} for details).
Note that it is convenient to choose the logical $Z$ operators such that they respect the GHZ structure of our analysis, e.g., $\overline{Z} = IIY$ will not be compatible here. \\

\item[(3)] If we consider the parity-check matrix after Step (2), we see that rows 4 through 8 are the stabilizers promised by Theorem~\ref{thm:ghz_stabilizer_measurement} that act only on B and C systems.
However, due to the same result, the C parts of rows 7 and 8 only have $X$-s instead of $Y$-s.
So, to change them back to $Y$-s, Alice applies the inverse of the Phase (i.e., $\sqrt{Z}$) gate to all $3$ qubits of the C system.
She specifically applies the inverse, rather than $\sqrt{Z}$ itself, to get rid of the $-1$ sign for the last row. \\

For a general stabilizer code, the appropriate diagonal Clifford must be chosen as discussed in Appendix~\ref{sec:diagonal_clifford}.
This operation converts the BC stabilizers $\varepsilon_i^{\text{A}} (-1)^{a_ib_i^T} E(a_i,b_i)_{\text{B}} \otimes E(a_i,0)_{\text{C}}$ into $\varepsilon_i^{\text{A}} (-1)^{a_ib_i^T} E(a_i,b_i)_{\text{B}} \otimes E(a_i,b_i)_{\text{C}}$, which ensures that Charlie gets the same code (up to signs of stabilizers) as Alice and Bob.
Since the Clifford is guaranteed to be diagonal, it leaves purely $Z$-type stabilizers unchanged.
Later, in Section~\ref{sec:local_clifford}, we show that it is better for Bob to perform this Clifford on Charlie's qubits, rather than Alice. \\

Though we have used the $-YYX$ GHZ stabilizer here for convenience, for a general code we can simply continue to use $XXX$.
After Alice has measured her stabilizer generators, this last block of 3 rows could have changed but they still commute with the generators.
Since the middle block never gets affected by Alice's measurements, we can guarantee using Theorem~\ref{thm:ghz_stabilizer_measurement} that two of the last 3 rows must be the joint BC stabilizers induced by Alice's two generators.
Hence, the remaining row's Alice component must form a logical operator for Alice's code, and will be distinct from the previously defined logical $Z$ operator.
We define this to be the logical $X$ operator, i.e., $\overline{X}_{\text{A}} = IIY = Y_3 = E(e_3,e_3)$.
As we will see shortly, both Bob and Charlie get the same code, so this last row of the third block can be written as the logical GHZ stabilizer $\overline{X}_{\text{A}} \overline{X}_{\text{B}} \overline{X}_{\text{C}}$.
The generalization to arbitrary stabilizer codes is discussed in Appendix~\ref{sec:logical_paulis_ghz_msmt}. \\

Then, she sends Bob both his qubits as well as Charlie's qubits over a noisy Pauli channel, which introduces the signs $\eta, \nu_i, \mu_i \in \{ \pm 1 \}, i = 1,2,3$.
She also classically communicates the code stabilizers, her syndromes $\{\varepsilon_1^{\text{A}}, \varepsilon_2^{\text{A}}\}$, and the logical $Z$ and $X$ operators to him. \\
%the current parity-check matrix, which is the same as shown in Step 3) of Table~\ref{tab:ghz_protocol} but without these new signs (which she does not know), i.e., the signs are the ones in Step 2) but the last row also has a positive sign due to Alice's operations.
%She indicates her code stabilizer rows to Bob. \\

\item[(4)] Now, based on Alice's classical communication, Bob applies Theorem~\ref{thm:ghz_stabilizer_measurement} to obtain the stabilizer generators
\begin{align*}
\varepsilon_1^{\text{A}} (Y_{\text{B}_1} Y_{\text{B}_2} I_{\text{B}_3}) \, (Y_{\text{C}_1} Y_{\text{C}_2} I_{\text{C}_3}) & = \varepsilon_1^{\text{A}} E(e_1+e_2,e_1+e_2)_{\text{B}}^T \otimes E(e_1+e_2,e_1+e_2)_{\text{C}} , \\ 
\varepsilon_2^{\text{A}} (I_{\text{B}_1} Y_{\text{B}_2} Y_{\text{B}_3}) \, (I_{\text{C}_1} Y_{\text{C}_2} Y_{\text{C}_3}) & = \varepsilon_2^{\text{A}} E(e_2+e_3,e_2+e_3)_{\text{B}}^T \otimes E(e_2+e_3,e_2+e_3)_{\text{C}}
\end{align*}
for the induced joint code on his as well as Charlie's qubits.
As per Theorem~\ref{thm:ghz_stabilizer_measurement}, he also includes the $IZZ$-type GHZ stabilizers, i.e., $\{ Z_{\text{B}_i} Z_{\text{C}_i} = E([0^{\text{A}},0^{\text{B}},0^{\text{C}}],[0^{\text{A}},e_i^{\text{B}},e_i^{\text{C}}]) \ ; \ i = 1,2,3 \}$, as stabilizer generators for the $\llbr 6,1 \rrbr$ joint code on BC systems.
He measures these $5$ stabilizers to deduce and correct the error introduced by the channel on the $6$ qubits sent by Alice.
Assuming perfect error correction, the signs will be back to the ones in Alice's final parity-check matrix. \\

When Bob sends Charlie's qubits to him, the channel might introduce errors on those $3$ qubits.
To deduce and correct these errors, there must have been a code induced on Charlie's qubits \emph{even before the transmission}.
Hence, after correcting errors on the $6$ qubits of BC systems, Bob measures the same stabilizers as Alice's code but on his qubits.
This produces rows $4$ and $5$ in Step (4) of Table~\ref{tab:ghz_protocol}, and row $6$ gets updated as per the stabilizer formalism.
When looking at rows $7$ and $8$ of Step (3), it is evident that one can correspondingly multiply them with these new rows $4$ and $5$ to produce the same code just on Charlie's qubits. \\

This phenomenon extends to general stabilizer codes as well, where the joint stabilizers $\varepsilon_i^{\text{A}} (-1)^{a_ib_i^T} E(a_i,b_i)_{\text{B}} \otimes E(a_i,b_i)_{\text{C}}$ are multiplied with Bob's stabilizers $\varepsilon_i^{\text{B}} E(a_i,b_i)_{\text{B}}$ to obtain Charlie's stabilizers $\varepsilon_i^{\text{A}} \varepsilon_i^{\text{B}} (-1)^{a_ib_i^T} E(a_i,b_i)_{\text{C}}$ (see Algorithm~\ref{algo:algo_ghz}).
Any purely $Z$-type stabilizer directly carries over to Charlie (without the above argument) as follows. 
At the beginning of the protocol, we can rewrite the $E([0^{\text{A}},0^{\text{B}},0^{\text{C}}],[e_i^{\text{A}},e_i^{\text{B}},0^{\text{C}}])$ rows such that a subset of them correspond to $E([0^{\text{A}},0^{\text{B}},0^{\text{C}}],[z^{\text{A}},z^{\text{B}},0^{\text{C}}])$, where $E(0,z)$-s are the purely $Z$-type stabilizer generators of the code.
This subset of rows will never be replaced by stabilizer measurements since they commute with other stabilizers.
After Alice's measurements, there will be rows corresponding to $\varepsilon_z^{\text{A}} E(0,z)_{\text{A}}$, which can be multiplied respectively with the aforesaid subset of rows to obtain $\varepsilon_z^{\text{A}} E(0,z)_{\text{B}}$.
Just like we rewrote a subset of the first $n$ rows, we can rewrite a subset of the second $n$ rows to obtain $E([0^{\text{A}},0^{\text{B}},0^{\text{C}}],[0^{\text{A}},z^{\text{B}},z^{\text{C}}])$, which when multiplied with $\varepsilon_z^{\text{A}} E(0,z)_{\text{B}}$ produces the desired $\varepsilon_z^{\text{A}} E(0,z)_{\text{C}}$ for Charlie's code.
See Appendix~\ref{sec:logical_paulis_ghz_msmt} for some related discussion.
In order to merge this phenomenon for purely $Z$-type operators with the general signs $\varepsilon_i^{\text{A}} \varepsilon_i^{\text{B}} (-1)^{a_ib_i^T}$ for Charlie, we set $\varepsilon_i^{\text{B}} \coloneqq +1$ whenever $a_i=0$, as mentioned in Algorithm~\ref{algo:algo_ghz}. \\
% See Appendix~\ref{sec:ghz_implementation} for the extension to general stabilizer codes. \\

Now, Bob has the parity-check matrix shown in Step (4) but without the new signs $\beta, \alpha_1, \alpha_2, \alpha_3$, which will be introduced by the channel during transmission of Charlie's qubits.
He sends Charlie his qubits (over a noisy Pauli channel), the code stabilizer generators, along with the corresponding signs $\{ \varepsilon_1^{\text{A}} \varepsilon_1^{\text{B}}, \varepsilon_2^{\text{A}} \varepsilon_2^{\text{B}} \}$, and the logical $Z$ and $X$ operators. \\

% \item Now, Bob measures the stabilizer $Y_{\text{B}_1} Y_{\text{B}_2} I_{\text{B}_3} = E([0^{\text{A}},(e_1+e_2)^{\text{B}},0^{\text{C}}],[0^{\text{A}},(e_1+e_2)^{\text{B}},0^{\text{C}}])$ and the group $\mathcal{G}_3^{(2)}$ gets updated as follows, assuming a measurement result $+1$: 
% \begin{align}
% \mathcal{G}_3^{(2)} & = \langle \ {\color{red} Y_{\text{A}_1} Y_{\text{A}_2} I_{\text{A}_3} }, \ E([0^{\text{A}},0^{\text{B}},0^{\text{C}}],[(e_1+e_2+e_3)^{\text{A}},(e_1+e_2+e_3)^{\text{B}},0^{\text{C}}]), \ {\color{red} I_{\text{A}_1} Y_{\text{A}_2} Y_{\text{A}_3} }, \nonumber \\
% %
%   & \hspace{1cm} {\color{red} \{ Y_{\text{B}_1} Y_{\text{B}_2} I_{\text{B}_3} \} }, \  {\color{blue} \{ E([0^{\text{A}},0^{\text{B}},0^{\text{C}}],[0^{\text{A}},(e_1+e_2)^{\text{B}},(e_1+e_2)^{\text{C}}]) \} }, \ E([0^{\text{A}},0^{\text{B}},0^{\text{C}}],[0^{\text{A}},e_3^{\text{B}},e_3^{\text{C}}]), \nonumber \\
% %
%   & \hspace{1cm} - E([e_i^{\text{A}},e_i^{\text{B}},e_i^{\text{C}}],[e_i^{\text{A}},e_i^{\text{B}},0^{\text{C}}]) \ ; \ i = 1,2,3 \ \rangle.
% \end{align}
% Observe that we can produce a stabilizer purely on Charlie by multiplying Alice's and Bob's first stabilizers and $- E([e_i^{\text{A}},e_i^{\text{B}},e_i^{\text{C}}],[e_i^{\text{A}},e_i^{\text{B}},0^{\text{C}}])$ for $i=1,2$.
% The resulting stabilizer for Charlie is $E((e_1+e_2),0) = X_{\text{C}_1} X_{\text{C}_2} I_{\text{C}_3}$, which is not the same as for Alice and Bob.

\end{enumerate}

Finally, Charlie measures these generators and fixes errors based on discrepancies in signs with respect to $\{ \varepsilon_1^{\text{A}} \varepsilon_1^{\text{B}}, \varepsilon_2^{\text{A}} \varepsilon_2^{\text{B}} \}$ (the additional signs $(-1)^{a_i b_i^T}$ do not make a difference for this example).
In the matrix in Step (4) of Table~\ref{tab:ghz_protocol}, after excluding the three sets of code stabilizers, we see that there are $3$ rows left which exactly correspond to the logical GHZ stabilizers, where we have defined the logical operators $\overline{Z} = ZZZ, \overline{X} = IIY = Y_3$ for the code.
Therefore, we have shown that after all steps of the protocol, the logical qubits of A, B, and C are in the GHZ state.
Since the signs of the stabilizer generators can be different for each of the three parties, although their logical $X$ and $Z$ operators are the same, the encoding unitary can be slightly different.
If they each perform the inverse of their respective encoding unitaries on their qubits (called ``Decoding'' in Figure~\ref{fig:ghz-entanglement}), then the logical GHZ state is converted into a physical GHZ state.

It might seem like this last step requires coordination among all three of them, which would require two-way communications between parties.
However, this is not necessary as Alice can perform the unitary on her qubits once she sends the $6$ qubits to Bob, and Bob can perform the unitary on his qubits once he sends the $3$ qubits to Charlie.
Subsequent operations will necessarily commute with these local unitaries as those qubits are not touched by the remaining parties in the protocol.

Hence, we have illustrated a complete GHZ distillation protocol, although much care must be taken while executing the steps for an arbitrary code.
For example, the local Clifford on C must be determined by solving a set of linear equations and finding a binary symmetric matrix that specifies the diagonal Clifford, via the connection to binary symplectic matrices~\cite{Dehaene-physreva03,Rengaswamy-tqe20}.
This is discussed in detail in Appendix~\ref{sec:diagonal_clifford}.
% questions such as which rows in Steps (1) and (2) should be replaced with the $BC$ joint stabilizers must be addressed.
% We discuss this in more detail in Appendix~\ref{sec:ghz_implementation}.
Similarly, the logical operators of the code that are compatible with our analysis of the protocol must be determined by simulating Alice's part of the protocol and applying some linear algebraic arguments. % as discussed in Appendix~\ref{sec:ghz_implementation}. %with a perfect channel and observing the non-code-stabilizer rows at the end.
For a general $\llbr n,k,d \rrbr$ code, there will be $3k$ non-code-stabilizer rows at the end, and one needs to identify $k$ pairs of logical $X$ and $Z$ operators for the code from these rows.
Although any valid definition of logical Paulis would likely suffice, we use Algorithm~\ref{algo:logical_paulis_ghz_msmt} to define them so that they naturally fit our analysis.
The explanations for the steps involved in this algorithm are given in Appendix~\ref{sec:logical_paulis_ghz_msmt}.
% Once the code is fixed, these subtle issues can be handled by sharing the same algorithm for the stabilizer formalism between all three parties, by simulating the protocol under perfect channel conditions, and by setting some conventions between all three parties.
In order to keep the main paper accessible, we have moved the discussion on implementation details to Appendix~\ref{sec:ghz_implementation}.

\subsection{Placement of Local Clifford and Distillation Performance}
\label{sec:local_clifford}

In our description of the protocol above, we mentioned that Alice performs the local Clifford $\left( \sqrt{Z}^\dagger \right)^{\otimes 3}$ on the qubits of system C in order to make Charlie's code the same as Alice's and Bob's.
However, due to this operation, the joint BC code (in Step (4)) cannot distinguish between Bob's qubits and Charlie's qubits.
Indeed, consider the two-qubit operator $X_{\text{B}_1} X_{\text{C}_1}$.
This commutes with all $5$ stabilizer generators of this code, although just $X_{\text{B}_1}$ or just $X_{\text{C}_1}$ would have anticommuted with the first generator $(Y_{\text{B}_1} Y_{\text{B}_2} I_{\text{B}_3}) \, (Y_{\text{C}_1} Y_{\text{C}_2} I_{\text{C}_3})$.
Therefore, if the true error is $X_{\text{C}_1}$, then the maximum likelihood decoder will correct it with $X_{\text{B}_1}$, which results in a logical error.
Of course, the $3$-qubit code only has distance $1$, but even if we consider the same $5$-qubit code as in Section~\ref{sec:bell_distillation}, the above phenomenon will still occur.
In effect, the induced joint BC code has distance dropping to $2$ whenever Alice's code does not have any purely $Z$-type stabilizer.
If we do not perform the diagonal Clifford at all, then in such cases Charlie's code will have only distance $1$.

To mitigate this, we can instead make \emph{Bob} perform the same diagonal Clifford operation on Charlie's qubits.
This ensures that the stabilizers for the BC code induced by Alice's code are of the form $E(a,b)_{\text{B}} \otimes E(a,0)_{\text{C}}$, or just $E(0,b)_{\text{B}}$ whenever the stabilizer is purely $Z$-type.
If Alice's code has good distance properties, then this joint BC code will have at least that much protection for Bob's qubits.
Although the C parts of the stabilizers are purely $X$-type, the additional GHZ stabilizers $Z_{\text{B}_i} Z_{\text{C}_i}$ help in detecting $X$-errors on system C as well.
Alternatively, one could make Alice perform one type of diagonal Clifford and Bob perform another diagonal Clifford, both on system C, to make both the BC code as well as Charlie's code as good as possible.
In future work, we will investigate these interesting degrees of freedom.

\begin{figure}

% \vspace{-20pt}

\begin{center}
    % \scalebox{0.75}{%
    %     \input{figures/ghz_5qubit_code.tex}
    % }
    \includegraphics[scale=0.65,keepaspectratio]{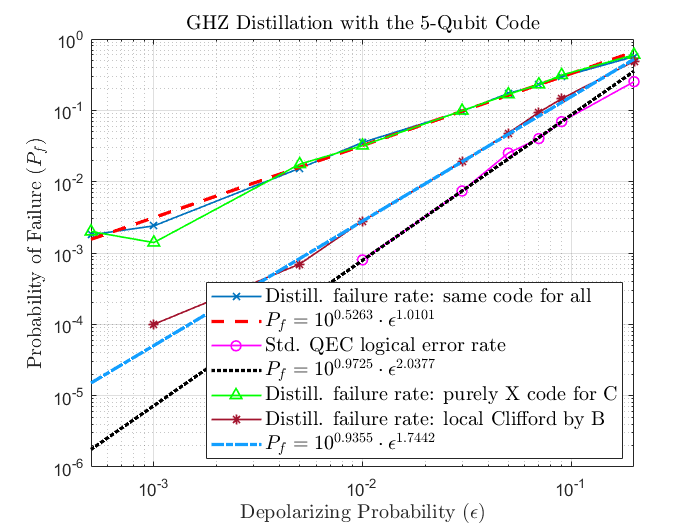}
\end{center}

\caption{\label{fig:GHZ_5qubit_code}Performance of variations of the GHZ distillation protocol using the $\llbr 5,1,3 \rrbr$ perfect code, and comparison with the standard QEC performance of the same code on the depolarizing channel. The decoder employs a maximum likelihood decoding scheme that identifies a minimal weight error pattern matching the syndrome.}

% \vspace{-10pt}

\end{figure}

We developed a MATLAB simulation of this protocol% 
\footnote{Implementation available online: \url{https://github.com/nrenga/ghz_distillation_qec}}
and tested it using the $\llbr 5,1,3 \rrbr$ perfect code defined by $S = \langle \, XZZXI,\, IXZZX,\, XIXZZ,\, ZXIXZ \, \rangle$.
The result is shown as the green curve marked `purely X code for C' in Fig.~\ref{fig:GHZ_5qubit_code}.
When compared to the standard QEC performance of this code on the depolarizing channel, we see that the exponent is worse.
This is because, by the arguments above, all non-purely $Z$-type stabilizers of $\mathcal{Q}(S)$ get converted into a purely $X$-type stabilizer for Charlie's code.
To mitigate this, we make Alice perform a local diagonal Clifford operation on qubits C to transform $E(a_i,0)_{\text{C}}$ into $E(a_i,b_i)_{\text{C}}$ later.
The performance of this is shown as the solid blue curve marked `same code for all', which is equally worse.
This time the reason is that the BC code $\mathcal{Q}(S')$ has stabilizers of the form $E(a_i,b_i)_{\text{B}} \otimes E(a_i,b_i)_{\text{C}}$, which means that the code cannot distinguish the $i$-th qubit of B and the $i$-th qubit of C.
Finally, we make \emph{Bob} perform the aforesaid diagonal Clifford on qubits C, and this produces the solid dark red curve marked `local Clifford by B'.
Clearly, the exponent now approaches the ``fundamental limit'' set by standard QEC on the depolarizing channel.

\subsection{Output Fidelity of the Protocol}

During the protocol, if error correction at Bob and/or Charlie miscorrects and introduces a logical error, then the final effect is a change in the signs of some of the logical GHZ stabilizers.
This in turn means that after the decoding step, some of the $k$ triples will be the standard GHZ state corrupted by an unknown Pauli operation.
Hence, the output of the protocol is correct with probability $(1-P_f)$, and produces at least one Pauli corrupted GHZ state with probability $P_f$, using the notation in Fig.~\ref{fig:GHZ_5qubit_code}.
To make this precise, denote by $\dket{\text{GHZ}_0}, \dket{\text{GHZ}_1}, \ldots, \dket{\text{GHZ}_7}$ the eight possible variants of the GHZ state under Pauli operations, i.e., each variant has the stabilizer group $\langle \alpha_1 \, ZZI, \ \alpha_2 \, IZZ, \ \alpha_3 \, XXX \rangle$ with $\alpha_1, \alpha_2, \alpha_3 \in \{ \pm 1 \}$.
Then, assuming all variants are equally likely conditioned on a failure event, the density matrix representing the output of the protocol is 
\begin{align}
\rho_{\text{out}} = (1 - P_f) \dketbra{\text{GHZ}_{00\cdots 0}} + P_f \sum_{i=1}^{8^k-1} \frac{1}{8^k-1} \dketbra{\text{GHZ}_{i_1 i_2 \cdots i_k}},
\end{align}
where $\dketbra{\text{GHZ}_{00\cdots 0}} = \dketbra{\text{GHZ}}^{\otimes k}$, $(i_1 i_2 \cdots i_k)$ is the base-$8$ expansion of $i$, and $\dketbra{\text{GHZ}_{i_1 i_2 \cdots i_k}} = \dketbra{\text{GHZ}_{i_1}} \otimes \dketbra{\text{GHZ}_{i_2}} \otimes \cdots \otimes \dketbra{\text{GHZ}_{i_k}}$.

Similar to the case of triorthogonal codes in magic state distillation~\cite{Bravyi-pra12}, it is likely useful to consider the reduced density matrix for one of the $k$ output triples, and relate its fidelity (with respect to $\dketbra{\text{GHZ}}$) to properties of the code and decoder.
In Ref.~\cite{Bravyi-pra12}, the authors adopted exactly such a strategy for distillation of $T$-states, under a purely $Z$-error model and relying on post-selection where non-trivial syndromes are discarded.
In recent work~\cite{Krishna-arxiv18}, it has been shown that performing error correction rather than just detection (and post-selection) leads to better performance of triorthogonal codes.
For our scenario of GHZ distillation, it is an interesting problem to construct codes and decoders for this protocol where we can relate the output fidelity to code properties and arrive at analytical scaling arguments with increasing code size.
This would be useful for comparing with fundamental limits of entanglement distillation~\cite{Fang-it19} and assessing the optimality of this protocol.
In particular, by employing the recent exciting classes of quantum low-density parity-check (QLDPC) codes that have linearly scaling rate and almost linearly scaling distance with code length~\cite{Hastings-stoc21,Panteleev-arxiv20,Breuckmann-it21,Breuckmann-arxiv21}, and using efficient iterative decoders such as belief-propagation (BP)~\cite{Kschischang-it01,Panteleev-arxiv19}, one can hope to achieve high-rate high-output-fidelity GHZ distillation schemes.

\subsection{Variations of the GHZ Protocol}
\label{sec:ghz_variations}

In the GHZ protocol we have discussed so far, Alice starts by preparing $n$ GHZ states and measuring the $n$-qubit stabilizers of her code on qubits A.
Then, using Theorem~\ref{thm:ghz_stabilizer_measurement}, Bob proceeds by measuring the $2n$-qubit stabilizers of the code induced on qubits B and C by Alice's choice of code on qubits A.
Subsequently, he also measures the same $n$-qubit stabilizers as Alice but on qubits B, so that there is a code induced just on qubits C and Charlie can use that code to correct errors from the channel.
If we imagine the three parties being on a linear network topology, then this protocol seems reasonable since each party retains his/her qubits and passes on all remaining qubits to the next hop in the chain.
However, there is an asymmetry in the operations since Bob needs to perform two rounds of measurements and one involves twice the number of qubits.

To address this, the protocol can be modified so that Alice starts by measuring qubits A and qubits B separately.
Though this does not circumvent the issue of performing twice the number of measurements at one of the nodes, this avoids the need of $2n$-qubit measurements.
Of course, when Alice measures $\{ \varepsilon_i E(a_i,b_i) \}$ qubits A, Theorem~\ref{thm:ghz_stabilizer_measurement} still dictates that there is a $2n$-qubit code $\{ \varepsilon_i E(a_i,b_i)_{\text{B}}^T \otimes E(a_i,0)_{\text{C}} \}$ jointly on qubits B and C.
But, when she measures the same stabilizers $\{ \varepsilon_i E(a_i,b_i) \}$ on qubits B, one can multiply with the corresponding $2n$-qubit stabilizer to see that the joint stabilizers can be broken into purely B and purely C stabilizers.
Therefore, once Alice performs the two rounds of measurements, she can send qubits B to Bob and qubits C to Charlie, along with the necessary classical information.
As individual codes have been induced separately on qubits B and qubits C, Bob and Charlie can still perform local $n$-qubit measurements to fix errors during qubit transmission.
Finally, this scheme suits other network topologies such as an isosceles one where the distance from A to B and A to C are roughly the same.
Since this variation relies on the same intuitions from Theorem~\ref{thm:ghz_stabilizer_measurement}, we do not elaborate further.
We also note that there can be further variations based on other practical considerations.

% \section{Using noisy GHZ states}
% \label{sec:noisy-GHZ}

% In this section, we consider noisy GHZ states to begin with and apply the same principles. For instance, consider the Werner state of two qubits:
% \begin{align}
% \rho_{W} &= \frac{\lambda}{4} \, \ket{\Phi^{+}}\bra{\Phi^{+}} + \frac{1-\lambda}{4} \, \mathbb{I}_4. 
% \end{align}

\section{Conclusion and Future Work}
\label{sec:conclusion}

In this work, we began by describing the Bell pair distillation protocol introduced in Ref.~\cite{Wilde-isit10}, and used the stabilizer formalism to understand its working.
We identified that the Bell state matrix identity (Section~\ref{sec:bell_state_identity}) plays a critical role in that protocol.
As our first result, we proved the equivalent matrix identity for GHZ states, where we introduced the GHZ-map and showed that it is an algebra homomorphism.
Using the GHZ-map, we proved our main result (Theorem~\ref{thm:ghz_stabilizer_measurement}) that describes the effect of Alice's stabilizer measurements (on qubits A) on qubits B and qubits C.
Then, we constructed a complete GHZ distillation protocol whose steps were guided by the aforementioned main result.
We demonstrated that the placement of a certain local Clifford on qubits C in the protocol has an immense effect on the performance of the protocol.
We described the relation between the probability of failure of the protocol and the output fidelity of the GHZ states.
As part of our protocol, we also developed a new algorithm to generate logical Pauli operators for an arbitrary stabilizer code.

It will be interesting to investigate codes and decoders where the output fidelity can be analytically related to code properties.
The presence of the local Clifford in the protocol introduces a non-trivial degree of freedom where there is scope for optimization.
In future work, we will combine this degree of freedom with other optimization methods that consider noisy local operations at each node~\cite{Krastanov-quantum19}.
We will also extend our methods to $t$-qubit GHZ for $t > 3$ and investigate if the size and number of rounds of measurements at each node can be bounded independent of $t$, since that will be practically attractive.

\section*{Acknowledgements}

The authors would like to thank Kaushik Seshadreesan and Saikat Guha for helpful discussions, and Mark Wilde for describing his protocol in our quantum error correction meeting.
This work is funded by the NSF Center for Quantum Networks (CQN), under the grant NSF-ERC 1941583, and also by the NSF grants CIF-1855879, CIF-2106189,
CCF-2100013 and ECCS/CCSS-2027844. 
This research was carried out in part at the Jet Propulsion Laboratory, California Institute of Technology, under a contract with the National Aeronautics and Space Administration and funded through JPL’s Strategic University Research Partnerships (SURP) program.
Bane Vasi\'c has disclosed an outside interest in his startup company Codelucida to the University of Arizona.  
Conflicts of interest resulting from this interest are being managed by The University of Arizona in accordance with its policies.

%\bibliographystyle{IEEEtran}
%\bibliography{./bib/bibliography}

\begin{thebibliography}{10}
\providecommand{\url}[1]{#1}
\csname url@samestyle\endcsname
\providecommand{\newblock}{\relax}
\providecommand{\bibinfo}[2]{#2}
\providecommand{\BIBentrySTDinterwordspacing}{\spaceskip=0pt\relax}
\providecommand{\BIBentryALTinterwordstretchfactor}{4}
\providecommand{\BIBentryALTinterwordspacing}{\spaceskip=\fontdimen2\font plus
\BIBentryALTinterwordstretchfactor\fontdimen3\font minus
  \fontdimen4\font\relax}
\providecommand{\BIBforeignlanguage}[2]{{%
\expandafter\ifx\csname l@#1\endcsname\relax
\typeout{** WARNING: IEEEtran.bst: No hyphenation pattern has been}%
\typeout{** loaded for the language `#1'. Using the pattern for}%
\typeout{** the default language instead.}%
\else
\language=\csname l@#1\endcsname
\fi
#2}}
\providecommand{\BIBdecl}{\relax}
\BIBdecl

\bibitem{Bennett-prl96}
\BIBentryALTinterwordspacing
C.~H. Bennett, G.~Brassard, S.~Popescu, B.~Schumacher, J.~A. Smolin, and W.~K.
  Wootters, ``{Purification of Noisy Entanglement and Faithful Teleportation
  via Noisy Channels},'' \emph{Phys. Rev. Lett.}, vol.~76, no.~5, p. 722, Jan
  1996. [Online]. Available: \url{https://arxiv.org/abs/quant-ph/9511027}
\BIBentrySTDinterwordspacing

\bibitem{Bennett-pra96}
\BIBentryALTinterwordspacing
C.~H. Bennett, D.~P. DiVincenzo, J.~A. Smolin, and W.~K. Wootters,
  ``{Mixed-state entanglement and quantum error correction},'' \emph{Phys. Rev.
  A}, vol.~54, no.~5, pp. 3824--3851, 1996. [Online]. Available:
  \url{https://arxiv.org/abs/quant-ph/9604024}
\BIBentrySTDinterwordspacing

\bibitem{Miyake-prl05}
A.~Miyake and H.~J. Briegel, ``Distillation of multipartite entanglement by
  complementary stabilizer measurements,'' \emph{Phys. Rev. Lett.}, vol.~95, p.
  220501, November 2005.

\bibitem{Dur-rpp07}
W.~D{\"u}r and H.~J. Briegel, ``Entanglement purification and quantum error
  correction,'' \emph{Rep. Prog. Phys.}, vol.~70, no.~8, p. 1381, November
  2007.

\bibitem{Leditzky-it17}
\BIBentryALTinterwordspacing
F.~Leditzky, N.~Datta, and G.~Smith, ``Useful states and entanglement
  distillation,'' \emph{IEEE Transactions on Information Theory}, vol.~64,
  no.~7, pp. 4689--4708, 2017. [Online]. Available:
  \url{https://arxiv.org/abs/1701.03081}
\BIBentrySTDinterwordspacing

\bibitem{Fang-it19}
K.~Fang, X.~Wang, M.~Tomamichel, and R.~Duan, ``Non-asymptotic entanglement
  distillation,'' \emph{IEEE Trans. on Inf. Theory}, vol.~65, pp. 6454--6465,
  November 2019.

\bibitem{Wilde-isit10}
\BIBentryALTinterwordspacing
M.~M. Wilde, H.~Krovi, and T.~A. Brun, ``Convolutional entanglement
  distillation,'' \emph{Proc. IEEE Intl. Symp. Inf. Theory}, pp. 2657--2661,
  June 2010. [Online]. Available: \url{https://arxiv.org/abs/0708.3699}
\BIBentrySTDinterwordspacing

\bibitem{Rozpedek-pra18}
\BIBentryALTinterwordspacing
F.~Rozp{\c e}dek, T.~Schiet, D.~Elkouss, A.~C. Doherty, S.~Wehner
  \emph{et~al.}, ``Optimizing practical entanglement distillation,''
  \emph{Physical Review A}, vol.~97, no.~6, p. 062333, 2018. [Online].
  Available: \url{https://arxiv.org/abs/1803.10111}
\BIBentrySTDinterwordspacing

\bibitem{Krastanov-quantum19}
\BIBentryALTinterwordspacing
S.~Krastanov, V.~V. Albert, and L.~Jiang, ``Optimized entanglement
  purification,'' \emph{Quantum}, vol.~3, p. 123, 2019. [Online]. Available:
  \url{https://arxiv.org/abs/1712.09762}
\BIBentrySTDinterwordspacing

\bibitem{Dur-pra99}
\BIBentryALTinterwordspacing
W.~D{\"u}r, H.-J. Briegel, J.~I. Cirac, and P.~Zoller, ``Quantum repeaters
  based on entanglement purification,'' \emph{Physical Review A}, vol.~59,
  no.~1, p. 169, 1999. [Online]. Available:
  \url{https://arxiv.org/abs/quant-ph/9808065}
\BIBentrySTDinterwordspacing

\bibitem{Murao-pra98}
\BIBentryALTinterwordspacing
M.~Murao, M.~B. Plenio, S.~Popescu, V.~Vedral, and P.~L. Knight,
  ``{Multiparticle entanglement purification protocols},'' \emph{Phys. Rev. A},
  vol.~57, no.~6, p. R4075, Jun 1998. [Online]. Available:
  \url{https://arxiv.org/abs/quant-ph/9712045}
\BIBentrySTDinterwordspacing

\bibitem{Hastings-stoc21}
M.~B. Hastings, J.~Haah, and R.~O'Donnell, ``Fiber bundle codes: breaking the
  $n^{1/2}$ polylog ($n$) barrier for quantum {LDPC} codes,'' in
  \emph{Proceedings of the 53rd Annual ACM SIGACT Symposium on Theory of
  Computing}, 2021, pp. 1276--1288.

\bibitem{Panteleev-arxiv20}
\BIBentryALTinterwordspacing
P.~Panteleev and G.~Kalachev, ``Quantum {LDPC} codes with almost linear minimum
  distance,'' \emph{arXiv preprint arXiv:2012.04068}, 2020. [Online].
  Available: \url{https://arxiv.org/abs/2012.04068}
\BIBentrySTDinterwordspacing

\bibitem{Breuckmann-it21}
N.~P. Breuckmann and J.~N. Eberhardt, ``Balanced product quantum codes,''
  \emph{IEEE Transactions on Information Theory}, 2021.

\bibitem{Breuckmann-arxiv21}
\BIBentryALTinterwordspacing
------, ``{LDPC} quantum codes,'' \emph{arXiv preprint arXiv:2103.06309}, 2021.
  [Online]. Available: \url{https://arxiv.org/abs/2103.06309}
\BIBentrySTDinterwordspacing

\bibitem{Aschauer-phd05}
\BIBentryALTinterwordspacing
H.~Aschauer, ``Quantum communication in noisy environments,'' Ph.D.
  dissertation, Ludwig-Maximilians-Universit{\"a}t M{\"u}nchen, April 2005.
  [Online]. Available: \url{http://nbn-resolving.de/urn:nbn:de:bvb:19-35882}
\BIBentrySTDinterwordspacing

\bibitem{Calderbank-physreva96}
A.~R. Calderbank and P.~W. Shor, ``Good quantum error-correcting codes exist,''
  \emph{Phys. Rev. A}, vol.~54, pp. 1098--1105, Aug 1996.

\bibitem{Steane-physreva96}
A.~M. Steane, ``{Simple quantum error-correcting codes},'' \emph{Phys. Rev. A},
  vol.~54, no.~6, pp. 4741--4751, 1996.

\bibitem{Gottesman-icgtmp98}
\BIBentryALTinterwordspacing
D.~Gottesman, ``The {H}eisenberg representation of quantum computers,'' in
  \emph{Intl. Conf. on Group Theor. Meth. Phys.}\hskip 1em plus 0.5em minus
  0.4em\relax International Press, Cambridge, MA, 1998, pp. 32--43. [Online].
  Available: \url{https://arxiv.org/abs/quant-ph/9807006}
\BIBentrySTDinterwordspacing

\bibitem{Gottesman-phd97}
\BIBentryALTinterwordspacing
------, ``Stabilizer codes and quantum error correction,'' Ph.D. dissertation,
  California Institute of Technology, 1997. [Online]. Available:
  \url{https://arxiv.org/abs/quant-ph/9705052}
\BIBentrySTDinterwordspacing

\bibitem{Calderbank-it98}
\BIBentryALTinterwordspacing
R.~Calderbank, E.~Rains, P.~Shor, and N.~Sloane, ``Quantum error correction via
  codes over {GF}(4),'' \emph{IEEE Trans. Inf. Theory}, vol.~44, no.~4, pp.
  1369--1387, Jul 1998. [Online]. Available:
  \url{https://arxiv.org/abs/quant-ph/9608006}
\BIBentrySTDinterwordspacing

\bibitem{Laflamme-prl96}
\BIBentryALTinterwordspacing
R.~Laflamme, C.~Miquel, J.~P. Paz, and W.~H. Zurek, ``{Perfect Quantum Error
  Correcting Code},'' \emph{Phys. Rev. Lett.}, vol.~77, no.~1, pp. 198--201,
  1996. [Online]. Available:
  \url{https://link.aps.org/doi/10.1103/PhysRevLett.77.198}
\BIBentrySTDinterwordspacing

\bibitem{Pan-nature01}
\BIBentryALTinterwordspacing
J.~W. Pan, C.~Simon, {\v{C}}.~Brukner, and A.~Zeilinger, ``{Entanglement
  purification for quantum communication},'' \emph{Nature}, vol. 410, no. 6832,
  pp. 1067--1070, Apr 2001. [Online]. Available:
  \url{https://arxiv.org/abs/quant-ph/0012026}
\BIBentrySTDinterwordspacing

\bibitem{Nickerson-ncomms13}
\BIBentryALTinterwordspacing
N.~H. Nickerson, Y.~Li, and S.~C. Benjamin, ``{Topological quantum computing
  with a very noisy network and local error rates approaching one percent},''
  \emph{Nat. Commun.}, vol.~4, no.~1, pp. 1--5, Apr 2013. [Online]. Available:
  \url{https://arxiv.org/abs/1211.2217}
\BIBentrySTDinterwordspacing

\bibitem{Dummit-2004}
D.~S. Dummit and R.~M. Foote, \emph{Abstract algebra}.\hskip 1em plus 0.5em
  minus 0.4em\relax Wiley Hoboken, 2004, vol.~3.

\bibitem{Wang-oc17}
X.-W. Wang, S.-Q. Tang, J.-B. Yuan, and D.-Y. Zhang, ``Distilling perfect ghz
  states from two copies of non-ghz-diagonal mixed states,'' \emph{Optics
  Communications}, vol. 392, pp. 185--189, 2017.

\bibitem{Wilde-2013}
M.~M. Wilde, \emph{Quantum {I}nformation {T}heory}.\hskip 1em plus 0.5em minus
  0.4em\relax Cambridge University Press, 2013.

\bibitem{Rengaswamy-pra19}
\BIBentryALTinterwordspacing
N.~Rengaswamy, R.~Calderbank, and H.~D. Pfister, ``Unifying the {C}lifford
  hierarchy via symmetric matrices over rings,'' \emph{Phys. Rev. A}, vol. 100,
  no.~2, p. 022304, 2019. [Online]. Available:
  \url{http://arxiv.org/abs/1902.04022}
\BIBentrySTDinterwordspacing

\bibitem{Nielsen-2010}
M.~A. Nielsen and I.~L. Chuang, \emph{Quantum {C}omputation and {Q}uantum
  {I}nformation}.\hskip 1em plus 0.5em minus 0.4em\relax Cambridge University
  Press, 2010.

\bibitem{Wilde-physreva09}
\BIBentryALTinterwordspacing
M.~M. Wilde, ``{Logical operators of quantum codes},'' \emph{Phys. Rev. A},
  vol.~79, no.~6, p. 062322, 2009. [Online]. Available:
  \url{https://arxiv.org/abs/0903.5256}
\BIBentrySTDinterwordspacing

\bibitem{Aaronson-pra04}
\BIBentryALTinterwordspacing
S.~Aaronson and D.~Gottesman, ``{Improved simulation of stabilizer circuits},''
  \emph{Phys. Rev. A}, vol.~70, no.~5, p. 052328, 2004. [Online]. Available:
  \url{https://journals.aps.org/pra/pdf/10.1103/PhysRevA.70.052328}
\BIBentrySTDinterwordspacing

\bibitem{Dehaene-physreva03}
J.~Dehaene and B.~{De Moor}, ``{Clifford group, stabilizer states, and linear
  and quadratic operations over GF(2)},'' \emph{Phys. Rev. A}, vol.~68, no.~4,
  p. 042318, Oct 2003.

\bibitem{Rengaswamy-tqe20}
\BIBentryALTinterwordspacing
N.~Rengaswamy, R.~Calderbank, S.~Kadhe, and H.~D. Pfister, ``Logical {C}lifford
  synthesis for stabilizer codes,'' \emph{IEEE Trans. Quantum Engg.}, vol.~1,
  2020. [Online]. Available: \url{http://arxiv.org/abs/1907.00310}
\BIBentrySTDinterwordspacing

\bibitem{Bravyi-pra12}
\BIBentryALTinterwordspacing
S.~Bravyi and J.~Haah, ``{Magic-state distillation with low overhead},''
  \emph{Phys. Rev. A}, vol.~86, no.~5, p. 052329, 2012. [Online]. Available:
  \url{http://arxiv.org/abs/1209.2426}
\BIBentrySTDinterwordspacing

\bibitem{Krishna-arxiv18}
\BIBentryALTinterwordspacing
A.~Krishna and J.-P. Tillich, ``Magic state distillation with punctured polar
  codes,'' \emph{arXiv preprint arXiv:1811.03112}, 2018. [Online]. Available:
  \url{http://arxiv.org/abs/1811.03112}
\BIBentrySTDinterwordspacing

\bibitem{Kschischang-it01}
\BIBentryALTinterwordspacing
F.~Kschischang, B.~Frey, and H.-A. Loeliger, ``Factor graphs and the
  sum-product algorithm,'' \emph{IEEE Trans. Inf. Theory}, vol.~47, no.~2, pp.
  498--519, 2001. [Online]. Available:
  \url{http://ieeexplore.ieee.org/document/910572/}
\BIBentrySTDinterwordspacing

\bibitem{Panteleev-arxiv19}
\BIBentryALTinterwordspacing
P.~Panteleev and G.~Kalachev, ``Degenerate quantum {LDPC} codes with good
  finite length performance,'' \emph{arXiv preprint arXiv:1904.02703}, 2019.
  [Online]. Available: \url{https://arxiv.org/abs/1904.02703}
\BIBentrySTDinterwordspacing

\end{thebibliography}


\appendices

\section{Logical Bell Pairs for Arbitrary CSS Codes}
\label{sec:logical_bell_CSS}

In this appendix, we show that when $n$ raw Bell pairs are projected onto the subspace of a CSS code through stabilizer measurements, the induced logical state is that of $k$ Bell pairs.
We take a meet-in-the-middle approach where we first consider $k$ Bell pairs and show how their encoded state looks like, and then we project $n$ Bell pairs to prove that the resulting state is the same as the aforesaid encoded state.

Let $\MCC_1, \MCC_2$ be two binary linear codes such that $\MCC_2 \subset \MCC_1$.
For the $\llbr n,k \rrbr$ CSS code defined by these codes, $\MCC_2$ produces the $X$-stabilizers, $\MCC_1$ produces the logical $X$ operators, $\MCC_1^{\perp}$ produces the $Z$-stabilizers, and $\MCC_2^{\perp}$ produces the logical $Z$ operators.
Let $G_{\MCC_1/\MCC_2}$ denote a generator matrix for the quotient group $\MCC_1/\MCC_2$ that represents the ``pure'' logical $X$ operators that do not have any $X$-stabilizer component.
In other words, the rows of $G_{\MCC_1/\MCC_2}$ give the generators of logical $X$ operators for the CSS code.
Let $\mathcal{U}_{\text{Enc}}$ denote an encoding unitary for the code.
Then, the encoded state of $k$ Bell pairs is~\cite{Calderbank-physreva96,Nielsen-2010}
\begin{align}
\left( (\mathcal{U}_{\text{Enc}})_{\text{A}} \otimes (\mathcal{U}_{\text{Enc}})_{\text{B}} \right) & \left( \dket{\Phi_k^+}_{\text{AB}} \otimes \dket{00}_{\text{AB}}^{\otimes (n-k)} \right) \nonumber \\
  & = \frac{1}{\sqrt{2^k}} \sum_{x \in \mathbb{F}_2^k} \mathcal{U}_{\text{Enc}} \left( \dket{x}_{\text{A}} \dket{0}_{\text{A}}^{\otimes (n-k)} \right) \otimes \mathcal{U}_{\text{Enc}} \left( \dket{x}_{\text{B}} \dket{0}_{\text{B}}^{\otimes (n-k)} \right) \\
\label{eq:encoded_bell_pairs}
  & = \frac{1}{\sqrt{2^k}} \sum_{x \in \mathbb{F}_2^k} \left[ \frac{1}{\sqrt{|\MCC_2|}} \sum_{y \in \MCC_2} \dket{xG_{\MCC_1/\MCC_2} \oplus y}_{\text{A}} \right] \otimes \left[ \frac{1}{\sqrt{|\MCC_2|}} \sum_{y' \in \MCC_2} \dket{xG_{\MCC_1/\MCC_2} \oplus y'}_{\text{B}} \right].
\end{align}

For the other direction, we start with $\dket{\Phi_n^+}_{\text{AB}}$ and then apply the projector $\Pi_{\text{CSS}}$ for the code on Alice's qubits.
By the Bell state matrix identity (Section~\ref{sec:bell_state_identity}), this means that we are effectively simultaneously applying $\Pi_{\text{CSS}}^T = \Pi_{\text{CSS}}$ on Bob's qubits as well.
Here, the transpose has no effect because the stabilizer generators for CSS codes are purely $X$-type or purely $Z$-type, and only such operators appear in the expression for the code projector~\eqref{eq:stabilizer_projector}.
Let $G_2$ and $G_1^{\perp}$ represent generator matrices for the codes $\MCC_2$ and $\MCC_1^{\perp}$, respectively.
Then, we have
\begin{align}
\Pi_{\text{CSS}} = \prod_{u \in \text{rows}(G_2)} \frac{I_N + E(u,0)}{2} \cdot \prod_{v \in \text{rows}(G_1^{\perp})} \frac{I_N + E(0,v)}{2} \eqqcolon \Pi_X \cdot \Pi_Z.
\end{align}
For any $z \in \mathbb{F}_2^n$, since $E(0,v) \dket{z} = (-1)^{zv^T} \dket{z}$, we have $(I_N + E(0,v)) \dket{z} = 2 \dket{z}$ if $zv^T = 0$ and $(I_N + E(0,v)) \dket{z} = 0$ otherwise.
This implies that $\Pi_Z \dket{z} = \dket{z}$ or $0$ depending on whether $z \in \MCC_1$ or not, respectively.
Similarly, it is easy to check that $\Pi_X \dket{z} = \frac{1}{|\MCC_2|} \sum_{y \in \MCC_2} \dket{z \oplus y}$.
Putting these together, we observe that
\begin{align}
\left( \left( \Pi_{\text{CSS}} \right)_{\text{A}} \otimes \left( \Pi_{\text{CSS}} \right)_{\text{B}} \right) \dket{\Phi_n^+} & = \frac{1}{\sqrt{2^n}} \sum_{z \in \mathbb{F}_2^n} \Pi_X \Pi_Z \dket{z}_{\text{A}} \otimes \Pi_X \Pi_Z \dket{z}_{\text{B}} \\
  & = \frac{1}{\sqrt{2^n}} \sum_{z \in \MCC_1} \Pi_X \dket{z}_{\text{A}} \otimes \Pi_X \dket{z}_{\text{B}} \\
  & = \frac{1}{\sqrt{2^n}} \sum_{z \in \MCC_1} \frac{1}{|\MCC_2|} \sum_{y \in \MCC_2} \dket{z \oplus y}_{\text{A}} \otimes \frac{1}{|\MCC_2|} \sum_{y' \in \MCC_2} \dket{z \oplus y'}_{\text{B}} \\
  & = \frac{1}{\sqrt{2^n}} \sum_{x \in \mathbb{F}_2^k} \sum_{y'' \in \MCC_2} \frac{1}{|\MCC_2|} \sum_{y \in \MCC_2} \dket{(x G_{\MCC_1/\MCC_2} \oplus y'') \oplus y}_{\text{A}} \otimes \frac{1}{|\MCC_2|} \sum_{y' \in \MCC_2} \dket{(x G_{\MCC_1/\MCC_2} \oplus y'') \oplus y'}_{\text{B}} \\
  & = \frac{1}{\sqrt{2^n}} \sum_{x \in \mathbb{F}_2^k} |\MCC_2| \frac{1}{|\MCC_2|} \sum_{y \in \MCC_2} \dket{x G_{\MCC_1/\MCC_2} \oplus y}_{\text{A}} \otimes \frac{1}{|\MCC_2|} \sum_{y' \in \MCC_2} \dket{x G_{\MCC_1/\MCC_2} \oplus y'}_{\text{B}} \\
\label{eq:projected_bell_pairs}
  & = \frac{1}{\sqrt{2^n}} \sum_{x \in \mathbb{F}_2^k} \frac{1}{\sqrt{|\MCC_2|}} \sum_{y \in \MCC_2} \dket{x G_{\MCC_1/\MCC_2} \oplus y}_{\text{A}} \otimes \frac{1}{\sqrt{|\MCC_2|}} \sum_{y' \in \MCC_2} \dket{x G_{\MCC_1/\MCC_2} \oplus y'}_{\text{B}}.
\end{align}
This state must be normalized by the square root of the probability that we get the all $+1$ syndrome, which corresponds to the subspace of the considered CSS code.
It can be checked that all syndromes are equally likely, so the probability is $1/2^{n-k}$.
Dividing~\eqref{eq:projected_bell_pairs} by $1/\sqrt{2^{n-k}}$, we arrive at exactly the same state in~\eqref{eq:encoded_bell_pairs}.
This establishes that when CSS stabilizer measurements are performed on $n$ Bell pairs, the resulting code state corresponds to $k$ logical Bell pairs.

\section{Proof of Theorem~\ref{thm:ghz_stabilizer_measurement}}
\label{sec:proof_ghz_stabilizer_measurement}

Using the discussion before the statement of the theorem, we will calculate $\ghzmap{I_N}$ and $\ghzmap{E(a,b)}$ to establish the result.
Recollect that $\dketbra{0}_{n=1} = \frac{I + Z}{2}$ and hence $\dketbra{0}^{\otimes n} = \frac{1}{2^n} \sum_{v \in \mathbb{F}_2^n} E(0,v)$.
Then, using Lemma~\ref{lem:Eab}, we have
\begin{align}
\ghzmap{I_N} & = \sum_{x \in \mathbb{F}_2^n} \dketbra{x} \otimes \dketbra{x} \\
  & = \sum_{x \in \mathbb{F}_2^n} \left[ E(x,0) \dketbra{0}^{\otimes n} E(x,0) \right]^{\otimes 2} \\
  & = \sum_{x \in \mathbb{F}_2^n} \left[ E(x,0) \cdot \frac{1}{2^n} \sum_{v \in \mathbb{F}_2^n} E(0,v) \cdot E(x,0) \right]^{\otimes 2} \\
  & = \frac{1}{2^{2n}} \sum_{x \in \mathbb{F}_2^n} \left[ \sum_{v \in \mathbb{F}_2^n} (-1)^{xv^T} E(0,v) \right] \otimes \left[ \sum_{w \in \mathbb{F}_2^n} (-1)^{xw^T} E(0,w) \right] \\
  & = \frac{1}{2^{2n}} \sum_{x \in \mathbb{F}_2^n} \sum_{z \in \mathbb{F}_2^{2n}} (-1)^{[x,x] z^T} E(0,z) \quad (\text{where}\ z = [v,w]) \\
  & = \frac{1}{2^{2n}} \sum_{z \in \mathbb{F}_2^{2n}} E(0,z) \cdot \left( \sum_{x \in \mathbb{F}_2^n} (-1)^{[x,x] z^T} \right) \\
  & = \frac{1}{2^{2n}} \sum_{z \in \mathbb{F}_2^{2n}} E(0,z) \cdot 2^n \mathbb{I}\left(z \perp [x,x] \ \forall \ x \in \mathbb{F}_2^n \right) \\
  & = \frac{1}{2^n} \sum_{z' \in \mathbb{F}_2^n} E([0,0],[z',z']) \\
  & = \bigotimes_{i=1}^n \frac{\left( I_N + E([0,0],[e_i,e_i]) \right)}{2},
\end{align}
where $e_i \in \mathbb{F}_2^n$ is the standard basis vector with $1$ in the $i$-th position and zeros elsewhere.
Note that $E([0,0],[e_i,e_i])_{\text{BC}} = Z_{\text{B}_i} \otimes Z_{\text{C}_i}$ is the GHZ stabilizer $I_{\text{A}} Z_{\text{B}} Z_{\text{C}}$ on the $i$-th triple of qubits between A, B and C~\eqref{eq:ghz_n_stabilizers}.
Next, we proceed to calculate $\ghzmap{E(a,b)}$ using a similar approach.
\begin{align}
\ghzmap{E(a,b)} & = \sum_{x,y \in \mathbb{F}_2^n} \dbra{x} E(a,b) \dket{y} \dketbra{x,x}{y,y} \\
  & = \sum_{x,y \in \mathbb{F}_2^n} \dbra{x} \imath^{ab^T} (-1)^{by^T} \dket{y \oplus a} \dketbra{x,x}{y,y} \\
  & = \sum_{x \in \mathbb{F}_2^n} \imath^{ab^T} (-1)^{b (x \oplus a)^T} \dketbra{x,x}{x \oplus a, x \oplus a} \\
  & = \sum_{x \in \mathbb{F}_2^n} \imath^{-ab^T} (-1)^{b x^T} \left[ E(x,0) \cdot \dketbra{0}^{\otimes n} \cdot E(x \oplus a, 0) \right]^{\otimes 2} \\
  & = \sum_{x \in \mathbb{F}_2^n} \imath^{-ab^T} (-1)^{b x^T} \left[ E(x,0) \cdot \frac{1}{2^n} \sum_{v \in \mathbb{F}_2^n} E(0,v) \cdot E(x \oplus a, 0) \right]^{\otimes 2} \\ 
  & = \frac{1}{2^{2n}} \sum_{x \in \mathbb{F}_2^n} \imath^{-ab^T} (-1)^{b x^T} \left[ E(a,0) \cdot \sum_{v \in \mathbb{F}_2^n} (-1)^{xv^T + av^T} E(0,v) \right]^{\otimes 2} \quad (\text{Lemma~\ref{lem:Eab}(c)}) \\
  & = E(a,0)^{\otimes 2} \cdot \frac{1}{2^{2n}} \sum_{x \in \mathbb{F}_2^n} \imath^{-ab^T} (-1)^{b x^T} \sum_{z \in \mathbb{F}_2^{2n}} (-1)^{[x \oplus a, x \oplus a] z^T} E(0,z) \\
  & = E([a,a],[0,0]) \sum_{z \in \mathbb{F}_2^{2n}} \imath^{-ab^T} (-1)^{[a,a] z^T} E(0,z) \cdot \left( \frac{1}{2^{2n}} \sum_{x \in \mathbb{F}_2^n} (-1)^{[x,x] (z + [b,0])^T} \right) \\
  & = E([a,a],[0,0]) \cdot \frac{1}{2^n} \sum_{z' \in \mathbb{F}_2^{n}} \imath^{-ab^T} (-1)^{z'a^T + z'a^T + ab^T} E([0,0],[z' \oplus b, z']) \\
  & = E([a,a],[0,0]) \cdot \frac{\imath^{ab^T}}{2^n} \sum_{z' \in \mathbb{F}_2^{n}} E([0,0],[b,0]) \,  E([0,0],[z', z']) \\
  & = \imath^{-ab^T} E([a,a],[b,0]) \cdot \imath^{ab^T} \cdot \ghzmap{I_N} \quad (\text{Lemma~\ref{lem:Eab}(b)}) \\
  & = \left( E(a,b) \otimes E(a,0) \right) \cdot \ghzmap{I_N}.
\end{align}
Thus, when Alice's measurement applies the projector $M = \frac{I_N + \varepsilon E(a,b)}{2}$, Bob's and Charlie's qubits experience the projector
\begin{align}
\ghzmap{M^T} & = \frac{\ghzmap{I_N} + \varepsilon (-1)^{ab^T} \ghzmap{E(a,b)}}{2} \\
  & = \frac{\left( I_N \otimes I_N \right) \cdot \ghzmap{I_N} + \varepsilon (-1)^{ab^T} \left( E(a,b) \otimes E(a,0) \right) \cdot \ghzmap{I_N}}{2} \\
  & = \frac{ \left( I_N \otimes I_N + \varepsilon (-1)^{ab^T} E(a,b) \otimes E(a,0) \right) }{2} \cdot \bigotimes_{i=1}^n \frac{\left( I_N + E([0,0],[e_i,e_i]) \right)}{2}.
\end{align}
Since the second term, $\ghzmap{I_N}$, only corresponds to already existing stabilizers $Z_{\text{B}_i} \otimes Z_{\text{C}_i}$ for $n$ copies of the GHZ state, the only new measurement corresponds to the Pauli operator $\varepsilon (-1)^{ab^T} E(a,b) \otimes E(a,0)$. \hfill \IEEEQEDhere

\section{Implementation Details of the GHZ Distillation Protocol}
\label{sec:ghz_implementation}

In this appendix we provide additional details regarding the implementation of our protocol.
Specifically, we discuss how to identify the appropriate diagonal Clifford to be applied on qubits C, and explain the reasoning behind Algorithm~\ref{algo:logical_paulis_ghz_msmt} that generates logical Pauli operators for any stabilizer code.
The discussion related to Algorithm~\ref{algo:logical_paulis_ghz_msmt} will also clarify some aspects of the distillation protocol.

\subsection{Diagonal Clifford on qubits C}
\label{sec:diagonal_clifford}

A diagonal Clifford unitary on $n$ qubits can be described using an $n \times n$ binary symmetric matrix $R$ as~\cite{Dehaene-physreva03,Rengaswamy-tqe20} 
\begin{align}
U_R = \sum_{v \in \mathbb{F}_2^n} \imath^{vRv^T \bmod 4} \dketbra{v} = \text{diag}\left( \{ \imath^{vRv^T \bmod 4} \}_{v \in \mathbb{F}_2^n} \right).
\end{align}
Its action on a Pauli matrix $E(a,b)$ is given by~\cite{Rengaswamy-pra19,Rengaswamy-tqe20} 
\begin{align}
U_R \, E(a,b) \, U_R^{\dagger} & = E(a, b+aR) \\
  & = E(a, (b \oplus aR) + 2(b \ast aR)) \\
  & = \imath^{2 a (b \ast aR)^T} E(a, b \oplus aR) \\
  & = (-1)^{a (b \ast aR)^T} E(a, b \oplus aR),
\end{align}
where $b \ast aR$ is the entrywise product of the two binary vectors.
It is well-known that any diagonal Clifford operator can be formed using the phase gate, $P$, and the controlled-$Z$ gate, CZ, defined as
\begin{align}
P = 
\begin{bmatrix}
1 & 0 \\
0 & \imath
\end{bmatrix} \quad , \quad
\text{CZ} = 
\begin{bmatrix}
I_2 & 0 \\
0 & Z
\end{bmatrix} =
\begin{bmatrix}
1 & 0 & 0 & 0 \\
0 & 1 & 0 & 0 \\
0 & 0 & 1 & 0 \\
0 & 0 & 0 & -1
\end{bmatrix}.
\end{align}
Set $n=2$.
Then, the $R$ matrices for $P_1 = (P \otimes I), P_2 = (I \otimes P),$ and CZ are respectively 
$$ R_{P_1} = \begin{bmatrix} 1 & 0 \\ 0 & 0 \end{bmatrix}, 
R_{P_2} = \begin{bmatrix} 0 & 0 \\ 0 & 1 \end{bmatrix}, \text{ and }
R_{\text{CZ}} = \begin{bmatrix} 0 & 1 \\ 1 & 0 \end{bmatrix}. $$
This generalizes naturally to more than $2$ qubits.
The diagonal entries of $R$ describe which qubits get acted upon by $P$, and the pairwise off-diagonal entries describe which pairs of qubits get acted upon by CZ~\cite{Rengaswamy-tqe20}.

In our protocol, from Theorem~\ref{thm:ghz_stabilizer_measurement} we observed that Alice's stabilizers of the form $E(a_i,b_i)$, with $a_i \neq 0$, become the stabilizer $E(a_i,0)$ for Charlie.
This also means that any logical $X$ operator of the form $E(c_j,d_j)$ would have transformed into $E(c_j,0)$ (e.g., see last row of Step (2) in Table~\ref{tab:ghz_protocol} and compare it to the last row of Step (3)). 
Therefore, the purpose of the diagonal Clifford on qubits C is to convert the stabilizers $E(a_i,0)$ back into $E(a_i,b_i)$ and the logical $X$ operators $E(c_j,0)$ back into $E(c_j,d_j)$.
Given the above insight into diagonal Clifford operators, we want to find a binary symmetric matrix $R$ such that $U_R E(a_i,0) U_R^\dagger = E(a_i,b_i)$ for all $i=1,2,\ldots,r_X$ (using notation in Algorithm~\ref{algo:logical_paulis_ghz_msmt}) and $U_R E(c_j,0) U_R^\dagger = E(c_j,d_j)$ for all $j=1,2,\ldots,k$.
Thus, we need a feasible solution $R$ for $\{ a_i R = b_i,\ i=1,2,\ldots,r_X \}$ and $\{ c_j R = d_j,\ j=1,2,\ldots,k \}$.

We solve this system of linear equations on a binary symmetric matrix as follows.
First, let $A$ be the matrix whose rows are $\{ a_i \}$ and $\{ c_j \}$, and let $B$ be the matrix whose rows are $\{ b_i \}$ and $\{ d_j \}$.
Then, we have the system $A R = B$.
We recall the vectorization property of matrices, which implies that 
\begin{align}
\text{vec}(QUV) = (V^T \otimes Q) \text{vec}(U).
\end{align}
Here, vectorization of a matrix is the operation of reading the matrix entries columnwise, top to bottom, and forming a vector (e.g., this is done through the command \texttt{U(:)} in MATLAB).
Setting $Q=A, U=R, V=I$, we get $(I \otimes A) \text{vec}(R) = \text{vec}(B)$, which is a standard linear algebra problem for the unknown vector $\text{vec}(R)$.
However, we desire a binary \emph{symmetric} matrix $R$.
We impose this constraint as $(I - W) \text{vec}(R) = 0$, where $0$ denotes the all-zeros vector of length $n^2$, and $W$ is the permutation matrix which transforms $\text{vec}(Q)$ into $\text{vec}(Q^T)$ for any matrix $Q$.
In summary, we obtain the desired $R$ (or equivalently the diagonal Clifford $U_R$) by solving
\begin{align}
\text{find} \quad R \quad \text{s.t.} \quad (I \otimes A) \text{vec}(R) & = \text{vec}(B), \nonumber \\
  (I - W) \text{vec}(R) & = 0 .
\end{align}
Since $R$ is symmetric, it has $n(n+1)/2$ degrees of freedom, which accounts for the second constraint.
The matrix $A$ has $r_X + k < n$ rows and the Kronecker product with $I$ results in $n(r_X + k)$ constraints on $n(n+1)/2$ variables.
It remains to be shown if there is always a feasible solution for any valid $A$ and $B$.
Note that $[A,B]$ represents a matrix whose rows are stabilizers and logical $X$ operators.
This means any pair of rows must be orthogonal with respect to the symplectic inner product, which implies that $AB^T + BA^T = 0$.
Thus, a given $A$ and $B$ is valid if and only if $AB^T$ is symmetric.

\subsection{Logical Paulis from GHZ Measurements}
\label{sec:logical_paulis_ghz_msmt}

The procedure in Algorithm~\ref{algo:logical_paulis_ghz_msmt} to determine logical $X$ and $Z$ generators of a stabilizer code is inspired by the stabilizer measurements on Bell or GHZ states, viewed through the lens of the stabilizer formalism for measurements (Section~\ref{sec:stabilizer_formalism}).
Though the algorithm could have been constructed just using measurements on Bell states, we preferred GHZ states because there can be an additional non-trivial sign for the logical $X$ operators due to an odd number of subsystems in the GHZ state.
Of course, logical operators obtained using GHZ states will also apply to the Bell protocol since a negative sign on an even number of subsystems (A and B in Bell states) leads to an overall positive sign for $\overline{X}_A \overline{X}_B$ and $\overline{Z}_A \overline{Z}_B$.

We have a code $\mathcal{Q}(S)$ defined by its stabilizer group $S = \langle \varepsilon_i E(a_i,b_i) \, ; \, i = 1,2,\ldots,r=n-k \rangle$.
Define the $r \times (2n+1)$ stabilizer (or parity-check) matrix $H'$ whose $i$-th row is $[a_i,b_i,\ \varepsilon_i]$.
First, we bring the first $2n$ columns of the stabilizer (or parity-check) matrix of the code to the following standard form:
\begin{align}
H_{1:2n} = 
\begin{bmatrix}
0 & H_Z \\
H_1 & H_2
\end{bmatrix}.
\end{align}
Here, the $r_Z$ rows of $H_Z$ form all generators for the purely $Z$-type stabilizers of the code.
The bottom part of the matrix is such that the $r_X \times n$ matrix $H_1$ has full rank ($r_X + r_Z = r = n-k$).
While performing row operations on the initial parity-check matrix $H'$, one has to account for the Pauli multiplication rule in Lemma~\ref{lem:Eab}(b), and not simply perform binary sums of (the first $2n$ columns of the) rows, i.e., the last column of $H$ must be updated to reflect changes in signs.

Next, we simulate the creation of $n$ GHZ states by creating a $2n \times (6n+1)$ GHZ stabilizer matrix $S_{\text{GHZ}}$, whose first $n$ rows are $[0,0,0,\ e_i,e_i,0,\ +1]$ and the second $n$ rows are $[e_i,e_i,e_i,\ 0,0,0,\ +1]$.
This matrix is the same as Step (0) of Table~\ref{tab:ghz_protocol}, but we have omitted the middle section since the measurements on subsystem $A$ trivially commute with entries $I_{\text{A}} Z_{\text{B}} Z_{\text{C}}$ of this section.
Now, we use the stabilizer formalism for measurements (Section~\ref{sec:stabilizer_formalism}) to simulate measurements of the rows of $H$ on subsystem A of $S_{\text{GHZ}}$.
Clearly, the stabilizers from $[0,H_Z]$ commute with the first $n$ rows, so these will only replace $r_Z$ rows in the bottom half of $S_{\text{GHZ}}$.
The stabilizers from $[H_1,H_2]$ will necessarily anticommute with at least one of the first $n$ rows of $S_{\text{GHZ}}$, and these $r_X$ rows get replaced.
This can be established by counting the dimension of purely $Z$-type operators with which each row of $[H_1,H_2]$ can commute, one after the other.
Crucially, the stabilizer formalism guarantees that all rows of the evolved $S_{\text{GHZ}}$ remain linearly independent and always commute.

The $(n-r_X)$ non-replaced rows within the first $n$ rows can be divided into two types. 
Before we simulate any stabilizer measurements, the first $n$ rows have standard basis vectors $e_i$ for the $Z$-parts of A and B. 
These can be rewritten such that we have $r_Z$ rows of the form $[0,0,0,\ z,z,0,\ +1]$, where $z$ corresponds to rows of $H_Z$, all of which are linearly independent by assumption of the standard form.
Since these correspond to code stabilizers (on A as well as B), the measurement of rows of $[H_1,H_2]$ will not replace these.
After the measurements, when $r_Z$ rows in the bottom half have been replaced by $[0,0,0,\ z,0,0,\ \varepsilon_z]$, we can multiply with the corresponding rows of the top half, i.e., $[0,0,0,\ z,z,0,\ +1]$, to produce purely $Z$-type stabilizers on subsystem B, which later define Bob's code.
These $Z$-operators on B in the top half form the first type of $r_Z$ rows.
The remaining $(n-r_X)-r_Z = k$ rows form the second type, and they have to form logical $Z_{\text{A}_j} Z_{\text{B}_j}$, for $j=1,2,\ldots,k$, since they commute with all code stabilizers and the columns of subsystem C remain zero.
Thus, the $Z$-component of subsystem A of these $k$ rows produce the logical $Z$ generators of the code, and they always have sign $+1$.

A similar argument applies to the bottom half of the evolved $S_{\text{GHZ}}$ matrix.
The stabilizer measurements from $H_Z$ replace $r_Z$ rows out of the $n$ rows.
The remaining $(n-r_Z)$ rows can again be divided into two types.
The first type of rows give operators that can be rewritten as the BC stabilizers guaranteed by Theorem~\ref{thm:ghz_stabilizer_measurement}.
Specifically, these can be identified by the fact that their A-parts will be linearly dependent on the A-parts of the other rows of the evolved $S_{\text{GHZ}}$ matrix.
Indeed, this is how one can cancel the A-parts of these $r_X$ rows to produce the $r_X$ BC stabilizers corresponding to $[H_1,H_2]$.
The remaining $(n-r_Z)-r_X = k$ rows of the bottom half form the second type, and they have to form logical $X_{\text{A}_j}' X_{\text{B}_j}'$, for $j=1,2,\ldots,k$, since they commute with all code stabilizers and are linearly independent from all other rows.
The A-parts of these rows are used to define the logical $X'$ operators of the code.
The primes on these logical operators indicate that they might not exactly pair up with the corresponding logical $Z$ operators defined earlier.
This is because they are only guaranteed to be logical operators independent of the logical $Z$ operators, but not to be the appropriate pairs $\{ \overline{X}_j \}$ of the previously determined $\{ \overline{Z}_j \}$.

Once these pseudo logical $X$ operators are determined, we can easily find the necessary pairs for the logical $Z$ operators.
Let the logical $Z$ operators be $E(0,f_i), i = 1,2,\ldots,k$, and let these pseudo logical $X$ operators be $\nu_j E(c_j,d_j), j = 1,2,\ldots,k$.
If they are the correct pairs, then we would get $\syminn{[0,f_i]}{[c_j,d_j]} = \delta_{ij}$ for all $i,j \in \{ 1,2,\ldots,k \}$, where $\delta_{ij} = 1$ if $i=j$ and $0$ otherwise.
The symplectic inner product can be expressed as 
$$ \syminn{[0,f_i]}{[c_j,d_j]} = [0,f_i] \, \Omega \, [c_j,d_j]^T,\ \text{where}\ \Omega = \begin{bmatrix} 0 & I_n \\ I_n & 0 \end{bmatrix}. $$
Therefore, if $F$ is the matrix whose rows are $f_i$ and $[C,D]$ is the matrix whose rows are $[c_j,d_j]$, then we need 
$$ [0,F] \, \Omega \, [C,D]^T \eqqcolon T = I_k. $$
If $T \neq I_k$, then we can simply pre-multiply the equation by $T^{-1}$ (mod $2$) to achieve the desired result.
In this case, we redefine the logical $Z$ operators to be given by the rows of $T^{-1}\, [0,F]$.
This completes the reasoning behind Algorithm~\ref{algo:logical_paulis_ghz_msmt}.

\end{document}